\documentclass[manuscript]{aastex}

\slugcomment{}

\shorttitle{}
\shortauthors{}

\begin{document}

\title{The accretion rate independence of horizontal branch oscillation in \object{XTE J1701-462}}

\author{Zhaosheng Li\altaffilmark{1,2}}
\email{lizhaosheng@pku.edu.cn}

\author{Li Chen\altaffilmark{1}}
\email{chenli@bnu.edu.cn}

\author{Jinlu Qu\altaffilmark{3}}
\email{qujl@ihep.ac.cn}

\author{Qingcui Bu\altaffilmark{1}}
\and
\author{Dehua Wang\altaffilmark{1}}
\author{Renxin Xu\altaffilmark{2}}
\email{r.x.xu@pku.edu.cn}
\altaffiltext{1}{Department of Astronomy, Beijing Normal University, Beijing 100875, China}
\altaffiltext{2}{School of Physics and State Key Laboratory of Nuclear Physics and Technology, Peking University, Beijing 100871, China}
\altaffiltext{3}{Laboratory for Particle Astrophysics, Institute of High Energy Physics, CAS, Beijing 100049,  China}

\begin{abstract}
We study the temporal and energy spectral properties of the unique neutron star low-mass X-ray binary \object{XTE J1701-462}.
After assuming the HB/NB vertex as a reference position of accretion rate,
the horizontal branch oscillation (HBO) of the HB/NB vertex is roughly 50 Hz. It indicates that the HBO is
independent with the accretion rate or the source intensity. The spectral analysis shows
$R_{\rm{in}}\propto\dot{M}_{\rm{Disk}}^{2.9\pm0.09}$ in the HB/NB vertex and
$R_{\rm{in}}\propto\dot{M}_{\rm{Disk}}^{1.7\pm0.06}$ in the NB/FB vertex, which implies that different accretion rate
may be produced in the HB/NB vertex and the NB/FB vertex.
The Comptonization component could be fitted by
constrained broken power law (CBPL) or nthComp. Different with \object{GX 17+2}, the frequencies of HBO positively correlate with
the inner disk radius, which contradict with the prediction of Lense-Thirring precession model.
\object{XTE J1701-462}, both in the Cyg-like phase and in the Sco-like phase, follows a positive correlation between
the break frequency of broad band noise and the characteristic frequency of HBO, which is called the W-K relation.
An anticorrelation  between the frequency of HBO and photon energy is observed.
Moreover, the rms of HBO increases with photon energy till $\sim \rm 10 ~\rm keV$.
We discuss the possible origin of HBO from corona in \object{XTE J1701-462}.

\end{abstract}


\section{Introduction}

Low mass X-ray Binary (LMXB) is composed of a compact object (neutron star or black hole) accreting matter from its low mass companion star ($\lesssim1M_{\odot}$).  According to the X-ray spectral proporties and rapid timing variabilities, the neutron star (NS) LMXBs are usually classified as the Z sources and the atoll sources. They are named after the shapes which display in X-ray color-color diagram (CD) or hardness-intensity diagrams (HID). The Z sources produce approximate Eddington luminosities with soft X-ray spectra, whereas atoll sources produce a lower luminosities in the range $\sim 0.001-0.5 L_{\rm Edd}$ \citet{}.  A typical Z source track shows three branches, from top to bottom, which are usually called the horizontal branch (HB), the normal branch (NB), and the flaring branch (FB; \citet{hasi89}), respectively. For atoll sources, the three branches are called  extreme island, island, and banana state.  Based on the orientation of branches, six typical Z sources are further divided into the Cyg-
like Z sources (\object{Cyg X-2, GX 5-1}, and \object{GX 340+0}) with a horizontal HB (``Z"-shaped tracks) and the Sco-like Z sources (\object{Sco X-1}, \object{GX 17+2}, and \object{GX 349+2}) with a vertical HB (``$\nu$"-shaped tracks).

The black hole (BH) and NS LMXBs show many similarities in their timing behaviors. Low-frequency quasi-periodic oscillations (LF-QPOs) are observed in Z sources, atoll sources and BH LMXBs. In general, the centroid frequencies of LF-QPOs are $\sim$1-70 Hz. The type C, B and A LF-QPOs in BH LMXBs were considered as corresponding to HBOs, NBOs, FBOs of Z sources \citep{cas05}, respectively.  Moreover, the W-K relation, the strong correlation between the centroid frequency of LF-QPO and the break frequency in power density spectral, was identified in BH LMXBs, accreting millisecond pulsars and atoll sources \citep{wij99,bel02,str05}. Z sources show a similar but slightly shifted relation. These similar characteristics suggest that LF-QPOs are likely produced from the same physical mechanism in LMXBs.  Lense-Thirring precession was introduced to interpret HBOs in NS LMXBs as well as type C LF-QPOs in BH LMXBs \citep{ste98,ste99}.
In this model, the LF-QPOs were arisen from the misalignment between the compact star's spin axis and the rotational axis of the inner accretion disk \citep{len18,bar75}. \citet{ing09} discussed the possible origin of HBO from hot inner flow precession.

The evolution of accretion rate $\dot{m}$  is studied from the HID of LMXB because it contains the variation of X-ray spectrum and radiation intensity. In BH LMXBs,  $\dot{m}$  increases in the following direction: the Low Hard State (LHS) -- the Hard Intermediate State (HIMS) -- the Soft Intermediate State (SIMS) -- the High-Soft State (HSS) \citep{kal06,bel06,bel11}. Meanwhile, the type C LF-QPO was only observed in the HIMS in BH LMXBs. The frequency of type C QPO positively correlates with accretion rate and energy flux \citep{bel05}. However, the evolution of accretion rate $\dot{m}$ in the NS LMXBs is still controversial in the Cyg-like and Sco-like Z sources. The ascending trend of accretion rate is not always consistent with the source intensity increasing direction in the HIDs.  According to multi-wavelength campaigns of the classic Z source Cyg X-2, accretion rate  monotonically increases as HB-NB-FB \citep{has90,vrt90}. Based on the boundary layer emission model, \citet{gil05} also found $\dot{m}$
increasing from the HB to the FB in GX 340+0.  However, Church and co-workers \citep{chu06, chu10} applied the extend ADC model for the Cyg-like Z sources and suggested an opposite direction, i.e., $\dot{m}$ increases from the FB/NB vertex to the HB/NB vertex. \citet{hom02} considered that $\dot{m}$ maintains constant along Z tracks. For classical Z sources, the frequency of HBO increased from HB to HB/NB vertex. So, three distinct correlations between the accretion rate and the frequency of HBO were proposed, that is, positive correlation, anticorrelation and non-correlation.

The referred sources in the above works were analyzed either in the Cyg-like source or in the Sco-like source. The unique Z source, \object{XTE J1701-462},  switched from a Cyg-like Z source to a Sco-like Z source at high luminosity and  from Z source to atoll source at low luminosity, which was observed by the Rossi X-ray Timing Explorer (RXTE) during its 2006-2007 outburst. The secular change of \object{XTE J1701-462} was driven by the accretion rate variation. \citet{lin09} studied the spectra evolutions of \object{XTE J1701-462} prudently, and suggested that the accretion rate $\dot{m}$ maintains constant on the NB and FB of Z sources. While on the HB, the ${L_{\rm{MCD}}-T_{\rm{MCD}}^{4/3}}$  correlation biased from the constant $\dot{m}$ line because the disk component encountered a Comptonization upscattering. The constant $\dot{m}$ should be satisfied after the Comptonization component accounted. In \object{GX 17+2}, the constant $\dot{m}$ was also established \citep{lin12}. \citet{hom10} indicated
that the accretion rate was invariant in the Z sources and the oscillation of accretion rate $\dot{m}$ produced the Z tracks. However, \citet{ding11} concluded that the accretion rate of the disk follow  $\dot{M}_{\rm{Disk}} \propto R^{7/2}_{\rm{in}}$ after considering the magnetic field effect during the accretion.

In previous works, the relation between the spectra parameters and the characteristics of timing variability was not utilized to study the accretion rate variation. \object{XTE J1701-462} provides us a great opportunity to understand the temporal variabilities varying with the accretion rate evolution while the NS LMXB source transited from the Cyg-like, via the Sco-like Z source, to an atoll source. \citet{lin12} indicated two exactly opposite disk radius-HBO frequency relations in \object{GX 17+2} when the cutoff power law was replaced by nthComp. In this paper, we will provide a model independent method to study the HBOs' behaviors with decreasing accretion rate.

In Sec. 2, we analyze the public archive data of \object{XTE J1701-462}. In Sec. 3, we study the  X-ray spectra and timing variabilities of the HB/NB vertices and the NB/FB vertices, and then investigate the energy dependence of the HBO. The discussions and conclusions are displayed in Sec. 4 \& Sec. 5, respectively.

\section{Observations}

During its 2006-2007 outburst, \object{XTE J1701-462} experienced from a Z source to an atoll source in the X-ray spectral and timing behaviours collected by the the Proportional Counter Array (PCA) instrument on board of \emph{RXTE} \citep{rem06}.

\subsection{Data Selection and Reduction}

\begin{figure}
\epsscale{0.8}
\plotone{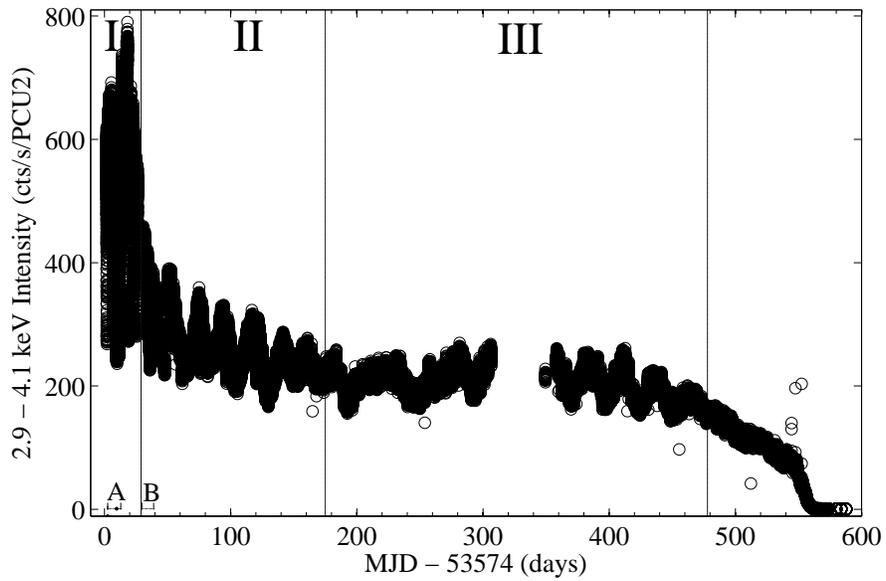}
\caption{The 16 s light curve of XTE J1701-462 in 2.9-4.1 keV during its 2006-2007 outburst. Following \citet{lin09}, three stages were divided. Stage I belongs to the Cyg-like Z source, stage II-III belong to the Sco-like Z source. The ``A" and ``B" mark the time intervals for our detailed temporal and spectral analysis. \label{fig1}}
\end{figure}

\begin{figure}
\epsscale{0.8}
\plotone{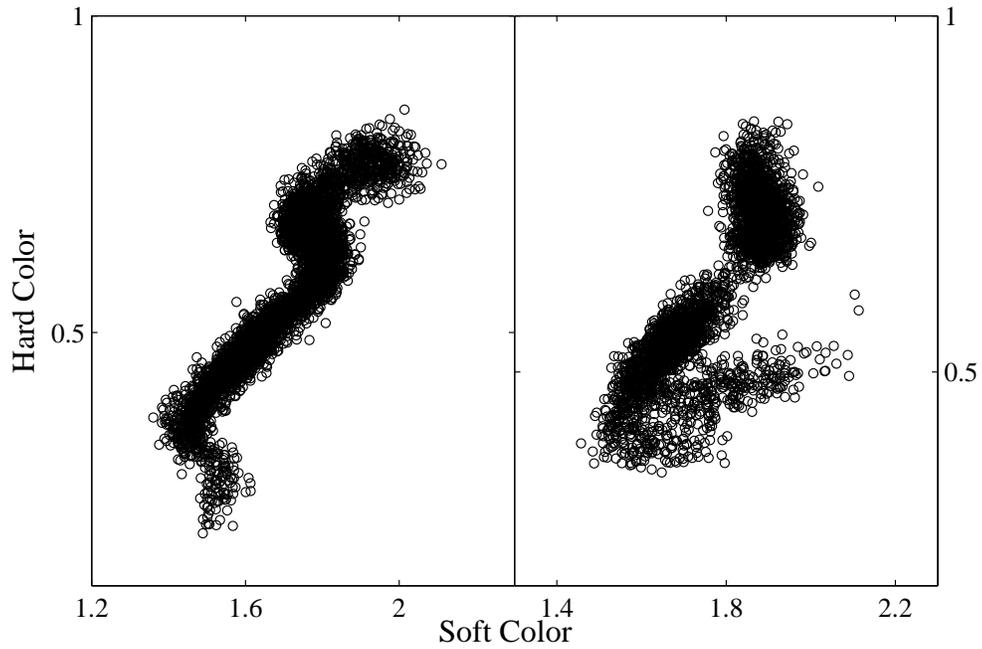}
\caption{The CDs for Intervals A and B in Table~\ref{tbl-1}. Each dot represents 16 s data from PCU2 with background-subtracted data.  The soft
color is defined as 4.5-7.4 keV/2.9-4.1 keV (channels 10-17/6-9) count rates ratio and the hard color is defined as 10.2-18.1 keV/7.8-9.8 keV (channels 24-43/18-23) count rates ratio. \label{fig2}}
\end{figure}

\begin{figure}
\epsscale{0.6}
\plotone{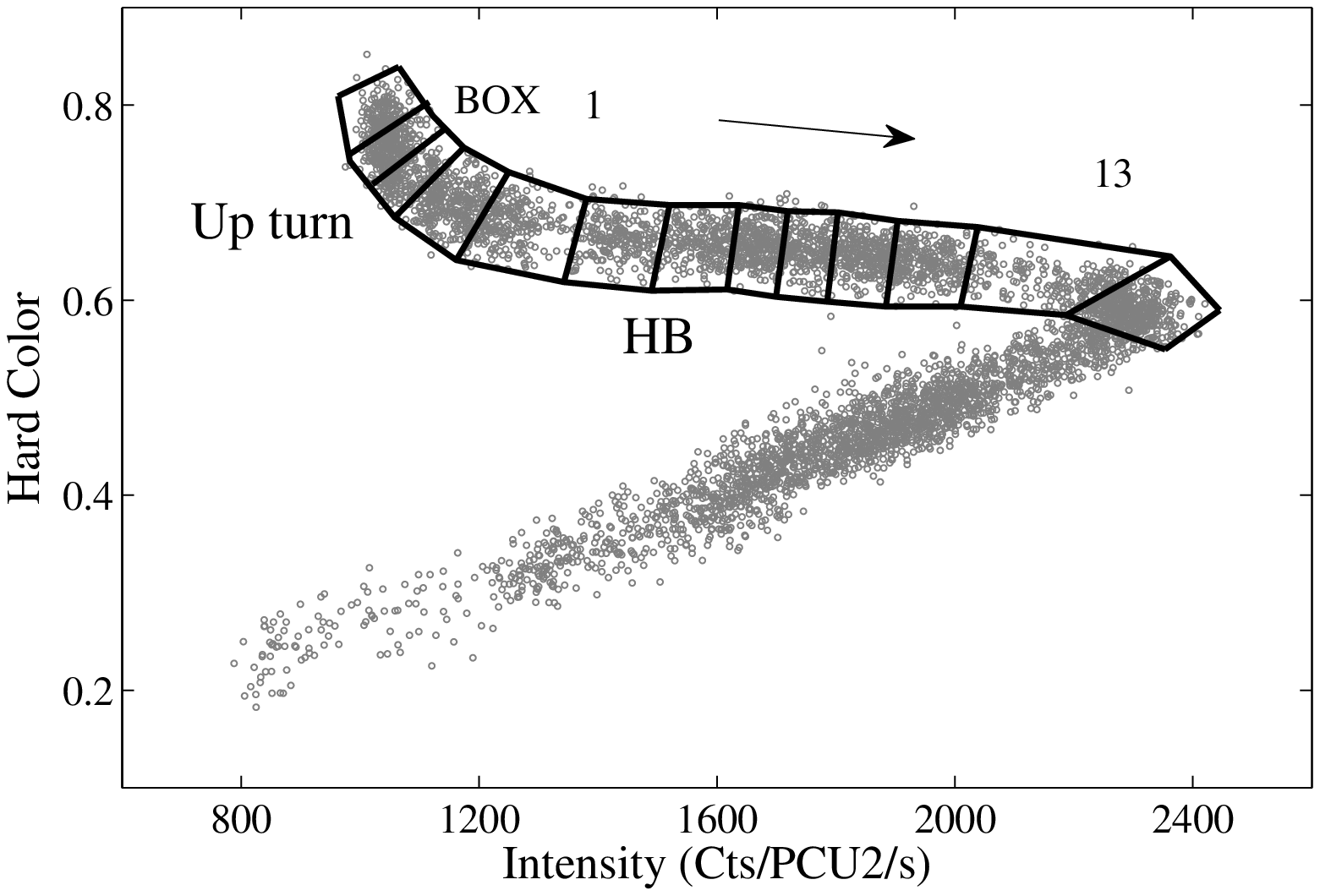}
\plotone{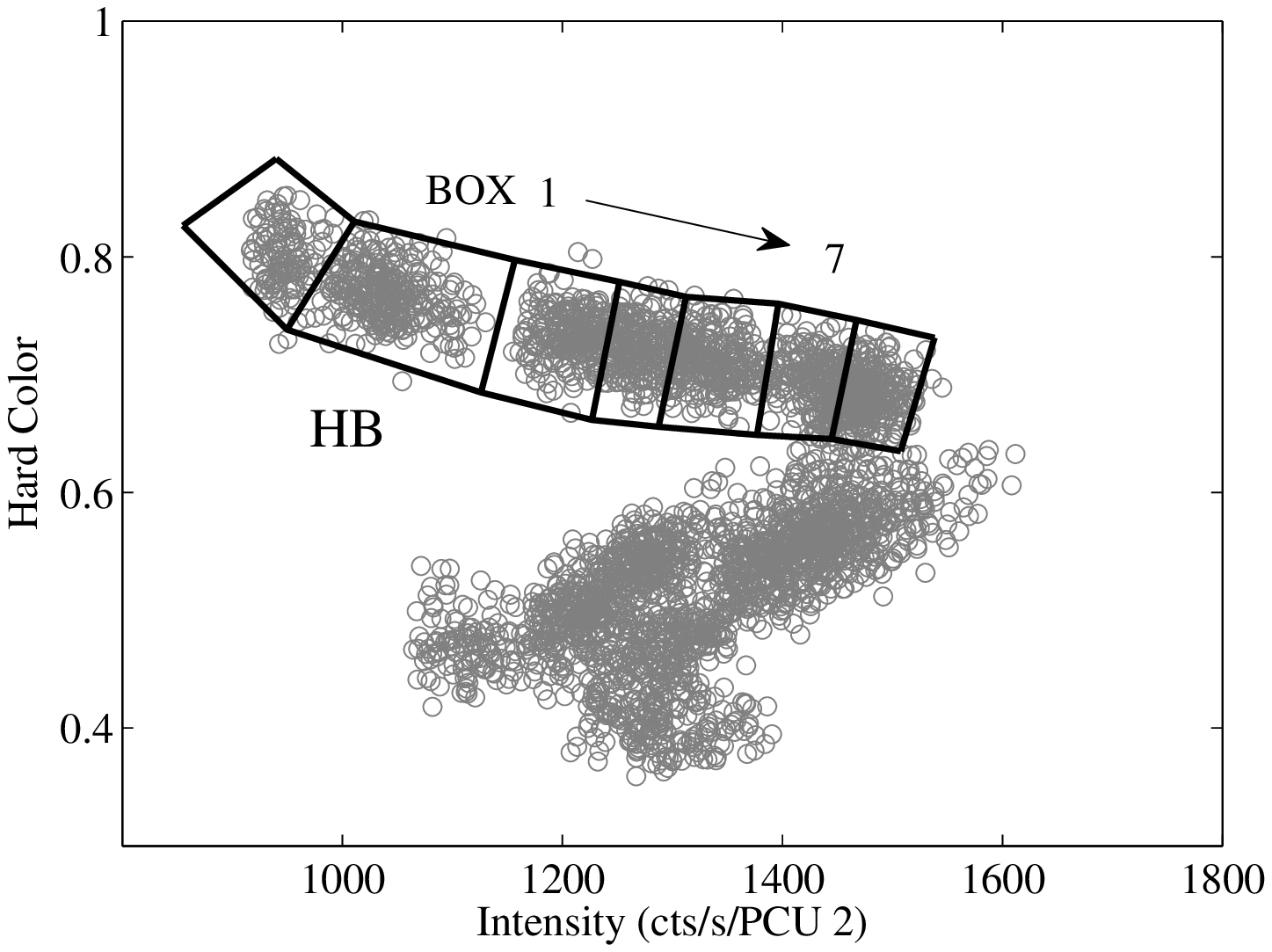}
\caption{The HIDs for Intervals A and B in Table~\ref{tbl-1}. Two divided HBs are labeled as Box 1-13 and Box 1-7 for Intervals A and B, respectively. The intensity is the photon count rates covering the energy range 2.9-18.1 keV (channels 6-43). \label{fig3}}
\end{figure}

We used the Heasoft 6.12 package to extract background-subtracted light
curves with a 16 s time resolution from the ``Standard 2" mode data in Figure~\ref{fig1}. Only data collected by PCU 2 were used. We applied the following standard data selection criteria, a source elevation of $>10^\circ$, a pointing offset of $<0^\circ.01$ and a South Atlantic Anomaly exclusion time of 30 minutes. In order to obtain the CCDs/HIDs from the light curves, we define the soft color as 4.5-7.4 keV/2.9-4.1 keV (channels 10-17/6-9) count rate ratios, and the hard color as 10.2-18.1 keV/7.8-9.8 keV (channels 24-43/18-23) count rate ratios, then define the intensity as count rate covering the energy range 2.9-18.1 keV (channels 6-43). The selected time intervals and observations are list in Table~\ref{tbl-1}. We constructed two CDs/HIDs from the observation of \object{XTE J1701-462} in Figure~\ref{fig2} and Figure~\ref{fig3}, Interval A for the Cyg-like phase and Interval B for the Sco-like phase. Suggested by \citet{lin09}, we also reproduced their HB/NB and NB/FB vertex in Stages I-III as the
reference region for estimating accretion rate. Each point represents  960 s duration in Figure~\ref{fig4}.

\subsection{Timing Analysis}

To investigate the HBOs for the Intervals A and B and the HB/NB vertex, we extracted the light curves of channels 0-13, 14-35 and 36-149 from SB\_125us\_0\_13\_1s, SB\_125us\_14\_35\_1s and E\_125us\_64M\_36\_1s,  respectively. We merged these three energy bands light curves to obtain channel 0-149 light curves with 1/512 s time resolution. When the source counts became lower, we extracted the channel 0-149 light curves from E\_125us\_64M\_0\_1s mode directly. No dead time correction or background substraction were applied. In order to study the energy dependence of HBOs, we extracted the light curves PCA channels 0-10, 11-13, 14-16, 17-22, 23-35 with 1/512 s resolution from B\_2ms\_8B\_0\_35\_Q. The choice of energy bands and their centroid energies are listed in Table~\ref{tbl-2}.  We created  power density spectra (PDSs) by using 16 s data segments for the full channel 0-149 and all sub-channel light curves. Therefore, the Fourier frequency is in the range 1/16-256 Hz. A multi-Lorentzian model is used to
fit the PDS. In Figure~\ref{fig5}, we show the PDS with frequency-times-power representation ($\nu P_{\nu}$) of Interval A. In order to obtain the best fitting result, four Lorentzian components are needed \citep{bel02,li13}. We obtain three characteristic frequencies of PDS, which are the centroid frequency of HBO ($\nu_{\rm HBO}$), the break frequency of low frequency noise ($\nu_b$) and $\nu_c=(\nu_{\rm HBO}^2+\Delta^2/4)^{1/2}$ (the frequency where the maximum of $\nu P_{\nu}$ occurs, where $\Delta$ is the Full-Width-Half-Maximum (FWHM) of the Lorentzian component), respectively.

\subsection{Energy Spectral Analysis}

We obtained the Stages I-III vertices and intervals A and B spectra with 16 s resolution from Standard 2 model data and  added a $0.6 \%$ systematic error to each channel. We also extracted the HEXTE data from Cluster B to study the high energy emission mechanism simultaneously. For the HEXTE data, we rebinned the spectra to obtain high S/N ratio. In the Z stage of \object{XTE J1701-462}, \citet{lin09} fitted the spectral with the model \textbf{wabs(bbodyrad+diskbb+gaussian+CBPL)} (in Xspec, hereafter Model I), which accounted for the absorbed black body from the NS surface, the multi-color disk, the iron line, the Comptonized component, respectively.  Suggested by \citet{lin09}, we fixed the hydrogen column density and the iron line width at $N_{\rm{H}}=2\times10^{22}~{\rm cm^{-2}}$ and 0.3 keV respectively. For the constrained broken power law (CBPL) model, the break energy is fixed at 20 keV and the low energy photon index is set as 2.5. We assume a 70 degree inclination and the source distance is 8.8 kpc
 with $15\%$ uncertainty\citep{lin09b}.

Lin et al. (2012) provided opposite $R_{\rm{in}}-\nu_{\rm{HBO}}$ relations when the CBPL was replaced by the nthcomp in \object{GX 17+2}, i.e., \textbf{wabs(bbodyrad+diskbb+gaussian+nthcomp)} (in XSPEC, hereafter Model II). Therefore, we attempted to study the  $R_{\rm{in}}-\nu_{\rm{HBO}}$ relation when these two different models were accounted for the Comptonized component in \object{XTE J1701-462}. We utilized the XSPEC procedure to fit the HB of Intervals A and B by using Model I and Model II simultaneously. During the energy spectral fitting, we fixed $kT_{\rm{e,nthcomp}}$ at 6.0 keV and $\Gamma_{\rm{nthcomp}}$ at 2.5.
Nevertheless, the vertices of Stages I-III could be well fitted by the  model \textbf{wabs(bbodyrad+diskbb+gaussian)}, where the Comptonized component could be ignored.

From the above model parameters, the apparent NS radius was derived from $Norm_{\rm{BB}}=(R_{\rm{NS,~km}}/D_{\rm{10kpc}})^2$, where the $Norm_{\rm {BB}}$ is the normalization of Blackbody component.  It is noticeable that the inner radius from Model II in \citet{lin12} (i.e., marked as $R_{\rm{MCDPS}}$ in their paper) accounted the MCD component as well as the nthComp component. We also attempted to account the high energy photon from nthComp to obtain the refined inner disk radius. However, the inner disk radius did not vary significiantly after adding the nthComp to MCD. Hereafter, for both Model I and Model II, the inner disk radius $R_{\rm{in}}$ was deduced from $Norm_{\rm{MCD}}=(R_{\rm{MCD,~km}}/D_{\rm{10kpc}})^2\cos(i)$, where the $Norm_{\rm{MCD}}$ and $i$ are the normalization of MCD component and inclination angle, respectively. The individual component flux was integrated from the fitting model in the energy range $\sim$3-60 keV. The luminosity was obtained from flux by multiplying $4\pi D^2$.

Figure~\ref{fig6-1}-\ref{fig6-4} showed the unfolded spectra of the HB in Intervals A and B. We listed the spectral and timing parameters in Tabel~\ref{tbl-3} -~\ref{tbl-6}. All quoted errors are 90\% confidence level. The fitting results were accounted 15\% uncertainty of the distance to the source.

\section{Results}

\subsection{The Spectral States of the HB/NB and NB/FB Vertex }

In Figure~\ref{fig4}, the HB/NB and NB/FB vertex of the Z source followed two straight lines in HID. In Figure~\ref{fig7}, we showed the black body component ($L_{\rm{BB}}$) and the multi-color disk component ($L_{\rm{MCD}}$) as a function of total luminosity ($L_{\rm{tot}}$).  We obtained similar results as Figure 16 in \citet{lin09}.  For the HB/NB vertices, both $L_{\rm{BB}}$ and $L_{\rm{MCD}}$ increased with $L_{\rm{tot}}$ enhancement. For the NB/FB vertices, $L_{\rm{MCD}}$ performed a similar behavior. However, $L_{\rm{BB}}$ descends at the beginning of Stage II. We presented the disk accretion rate versus the inner disk radius in Figure \ref{fig8}. For the HB/NB vertices, the relation ${\rm{log}}{\dot{M}_{\rm{Disk}}}=(2.90\pm0.09){\rm{log}}R_{\rm{in}}-(4.19\pm0.42)$ was obtained. However, a more gradual relation was shown for the NB/FB vertices, ${\rm{log}}{\dot{M}_{\rm{Disk}}}=(1.74\pm0.06){\rm{log}}R_{\rm{in}}-(2.89\pm0.06)$. The obviously discrepancy of slop indicates that the accretion disk
transfers from a slim disk in HB/NB vertex to a standard disk in NB/FB vertex \citep{lin12}.

\subsection{The HBO Evolution with Decreasing Accretion Rate}

\citet{lin09} suggested that the disk accretion rate of HB/NB vertex could be traced by using $2(L_{\rm MCD}+L_{\rm CBPL})R_{\rm in}/GM$ in place of  $2L_{\rm MCD}R_{\rm in}/GM$ to take into account the Comptonization emission in the HB. As we mentioned above, the $L_{\rm CBPL}$ component was ignored because of its faintness. Hereafter, we adopted the disk accretion rate as $\dot{M}_{\rm{Disk}}=L_{\rm{MCD}}R_{\rm{in}}/2GM$, where we assumed the 1.4$M_{\odot}$ for  the mass of neutron star and $\dot{M}_{\rm{Edd}}=3.24\times10^{-8}M_{\odot}\rm{yr}^{-1}$ for the Eddington accretion rate. The maximum $\nu_{\rm{HBO}}$ is $57.2 \pm1.8 \rm{Hz}$ which located at the Cyg-like stage. The total luminosity/disk accretion rate-$\nu_{\rm{HBO}}$ relation of HB/NB vertex were displayed in Figure~\ref{fig9}. With decreasing accretion rate, the correlation coefficient between the total luminosity and the frequency of  HBO is merely 0.36. We used ${\rm{log}}({\nu_{\rm{HBO}}})=a{\rm{log}}(L_{\rm{tot}})+b$ to fit the data
from Stage 1 and 2 displayed in Figure~\ref{fig9}, and the least squared method was applied. Then we obtained ${\rm{log}}({\nu_{\rm{HBO}}})=(0.073^{+0.165}_{-0.017}){\rm{log}}(L_{\rm{tot}})-1.11^{+2.36}_{-4.58}$.  We also discuss whether a constant model can fit better than the above model. We denote the above model as model 1, and ${\rm{log}}({\nu_{\rm{HBO}}})=\bar{b}$ as model 2. An approach to this question is using F-test. There are $n$ data points (n=23 in our works) to estimate parameters of model 1 and 2. We can calculate $F$ statistic, which is
$F=(SSR_2-SSR_1)/(SSE_2/(n-2))$,
where, $SSR_i$ is the regression sum of squares of model $i$, and $SSE_2$ is the residual sum of squares of model 2. Assuming the null hypothesis that model 1 does not provide a significantly better fitting than model 2, the statistic $F$ will have F-distribution with  degree of freedom of $(1,n-2)$. For a given false-rejection probability (eg., $\alpha=0.05$), the null hypothesis is accepted if the statistic $F$ calculated from our data is less than the critical value of the $F$ distribution, $F_{\alpha}(1,n-2)$. From our data, $F$ equals to 2.54, which is less than $F_{0.05}(1,21)=4.32$. It means that model 1 does not provide a significantly better fit than model 2. So, a constant model fits the data well enough.  It means that the total luminosity as well as the disk accretion rate are independent with the HBO. In Figure~\ref{fig10}, we also show that the inner disk radius has no correlation to the $\nu_{\rm{HBO}}$ at the HB/NB vertex.

\subsection{The $\nu_{\rm{HBO}}-R_{\rm{in}}$ Relation}

We fitted the HB of Intervals A and B with Model I and II simultaneously. Two models provided nearly the same results, which show positive $\nu_{\rm{HBO}}-R_{\rm{in}}$ relation in  Figure~\ref{fig8}. Within these two models, \citet{lin12} offered two distinct opposite $\nu_{\rm{HBO}}-R_{\rm{in}}$ relations in \object{GX 17+2}. However, it was not appeared in our results. The presumable reason is that the Comptonized component is relatively weak on HB in \object{XTE J1701-462}. Since the maximum of  the Comptonized component occupies only 20\% of the multi-color disk component in Table~\ref{tbl-3} of our fitting results as well as in \citet{lin09} and \citet{ding11}, it could not strongly effect the  $\nu_{\rm{HBO}}-R_{\rm{in}}$ relation.

\subsection{The Upper Limit of NS Radius in XTE J1701-462}

From the energy spectral fitting by Model I, the Box 1 of Interval A had minimum inner radius, which was roughly $11.5\pm2.3 \rm km$ with 15\% distance uncertainty accounted. Causality, the radius of NS should be smaller than the minimum inner radius in accretion disk system. Therefor, we utilized the minimum inner radius of accretion disk to constrain the upper radius of NS. The deduced inner radius depends on the inclination angle, which could not be directly measured in \object{XTE J1701-462} so far. However, the lack of dips/eclipses in the light curve of \object{XTE J1701-462} provided an upper limit of the inclination angle to a value close to 75 degrees \citep{lin09}. Here, we plotted the upper limit of NS radius for a wide range of inclination angle in Figure~\ref{fig11}.

\subsection{The W-K Relation of XTE J1701-462}

We computed the characteristic frequencies for the HB in Intervals A and B and the HB/NB vertex in Stages I-II. The results accompany with data from \citet{wij99} displayed in Figure~\ref{fig14}. At low frequency ($\nu_{\rm{HBO}}<35$ Hz), the W-K relation of \object{XTE J1701-462} locates in the atoll sources region. However, at high frequency ($\nu_{\rm{HBO}}>35$ Hz), the W-K relation of \object{XTE J1701-462} happens to locating between the Z sources and the atoll sources. That is, \object{XTE J1701-462} follows a linear correlation between the break frequency and the HBO, without offset appearing.

\subsection{The Flux Dependence of HBO}

We studied the HBO-total flux dependence in both Cyg-like phase and Sco-like phase. The total flux was integrated in the energy range $\sim$3-60 keV. In Figure~\ref{fig13}, the positive correlations between the frequency of HBO and the total flux were displayed. Compared with the Sco-like phase, the Cyg-like phase showed more complicated correlation, that is, the frequency of HBO increases gradually at lower flux and smoothly at higher flux.

\subsection{The Energy Dependence of HBO and its rms}

In order to investigate the energy dependence of HBO in Interval A, we extracted six sub-bands and one full band lighcurves, and then fitted their PDSs. When the $\nu_{\rm{HBO}}$ is larger than 30 Hz, we could not obtain high signal to noise ratio QPOs in channel 0-10, 11-13 and 36-149. In our analysis, only upper HB regions (i.e., $\nu_{\rm{HBO}} < 30~ \rm{Hz}$) were returned. The centroid frequency of full band and six subbands were derived. In left panels of Figure~\ref{fig15}, the QPO decreases with increasing photon energy. By using the bootstrap procedure, the correlation coefficients between the HBO frequency and energy are $-0.32\pm0.70$, $-0.85\pm0.06$, $-0.77\pm0.21$, $-0.77\pm0.22$, $-0.18\pm 0.43$ for Boxes 1-5 in Interval A, respectively. All Boxes show negative HBO-enenrgy correlations, but the Boxes 1 and 5 have relatively large errors. The HBO-energy relation in \object{XTE J1701-462} is distinct from the black hole transit XTE J1550-564, in which displayed more complicated correlations
between type C QPO frequency and photon energy \citep{li13b}. For each subband, the rms of HBO are also computed, which positively correlate with photon energy in five lower energy bands and slightly drop at highest energy band in right panels of Figure~\ref{fig15}.

\section{Discussion}

The unique transient NS LMXB \object{XTE J1701-462} emitted  from near Eddington luminosity to quiescence. The accretion rate $\dot{m}$ played a leading role in the whole process. For \object{XTE J1701-462}, the NB/FB vertex was considered as reference position to trace accretion rate. However, we could not confirm the strength of accretion rate only from the source intensity in a single Z track. The spectral state should be analyzed simultaneously. Homan and co-workers
\citep{hom02,lin09,lin12} argued that the accretion rate maintains constant along tracks in Z stage of \object{XTE J1701-462}. \citet{ding11} considered the consequence of magnetic field during accretion and derived that $\dot{M}_{\rm{Disk}}\propto R_{\rm{in}}^{7/2}$. In this paper, we simultaneously studied the timing properties and spectral states on the relative regions (the HB/NB vertex and the  NB/FB vertex) when the accretion rate decreased.

\begin{figure}
\epsscale{0.8}
\plotone{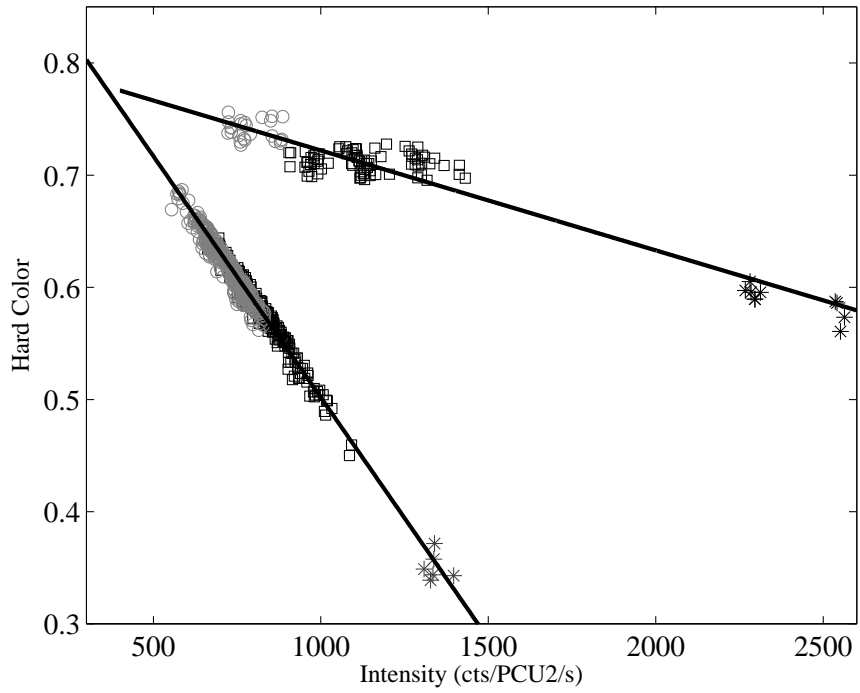}
\caption{The HB/NB vertices (upper line) and NB/FB vertices (lower line) for Stages I-III in \citet{lin09}, which are represented as star, square and circle respectively. The time resolution is 960 s. \label{fig4}}
\end{figure}

\begin{figure}
\epsscale{1.}
\plotone{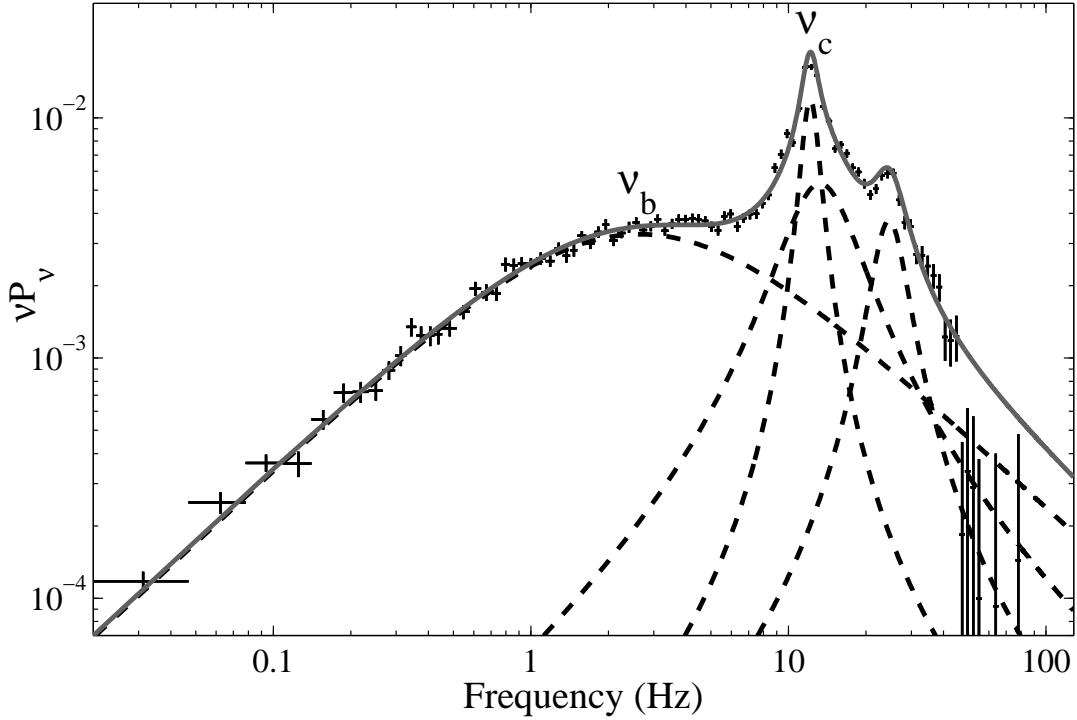}
\caption{The PDS of up turn in Interval A. Four Lorentzian components are presented as dash lines. The gray curve displays the best fitting result. $\nu_{\rm{b}}$ and $\nu_{\rm{c}}$ are labeled as the maximum of broad band noise and characteristic frequency of HBO in $\nu P_{\nu}$ representation, respectively. It should be noticed that $\nu_{\rm{c}}=(\nu_{\rm HBO}^2+\Delta^2/4)^2$, where, $\nu_{\rm HBO}$ is centroid frequency of HBO and $\Delta$ is the full-width-half-maximum of HBO.  $\nu_{\rm{c}}$ is the frequency where $\nu P_{\nu}$ reach its local maximum.  \label{fig5}}
\end{figure}

\begin{figure}


\includegraphics[width=3.25cm,angle=270]{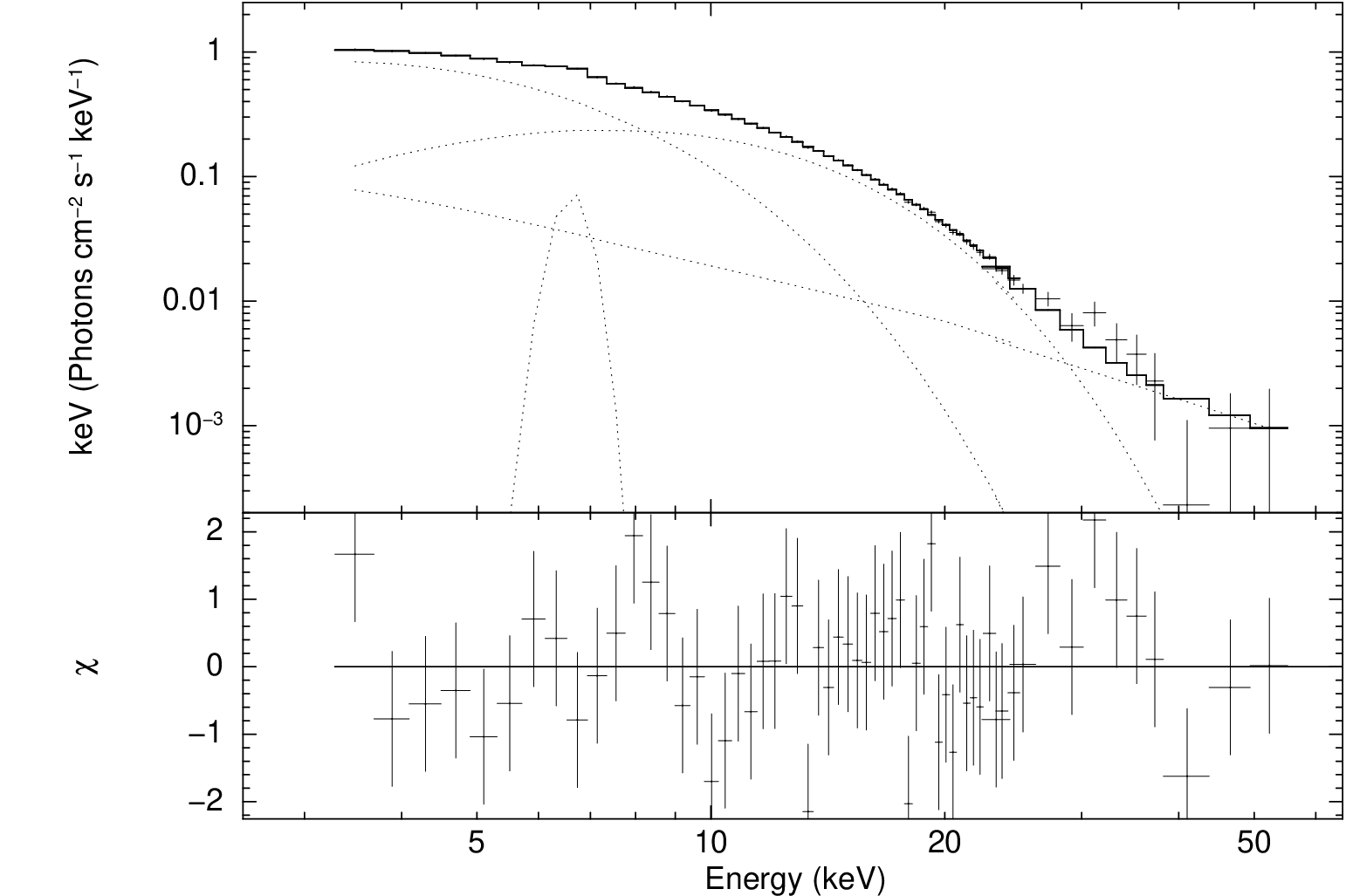}
\includegraphics[width=3.25cm,angle=270]{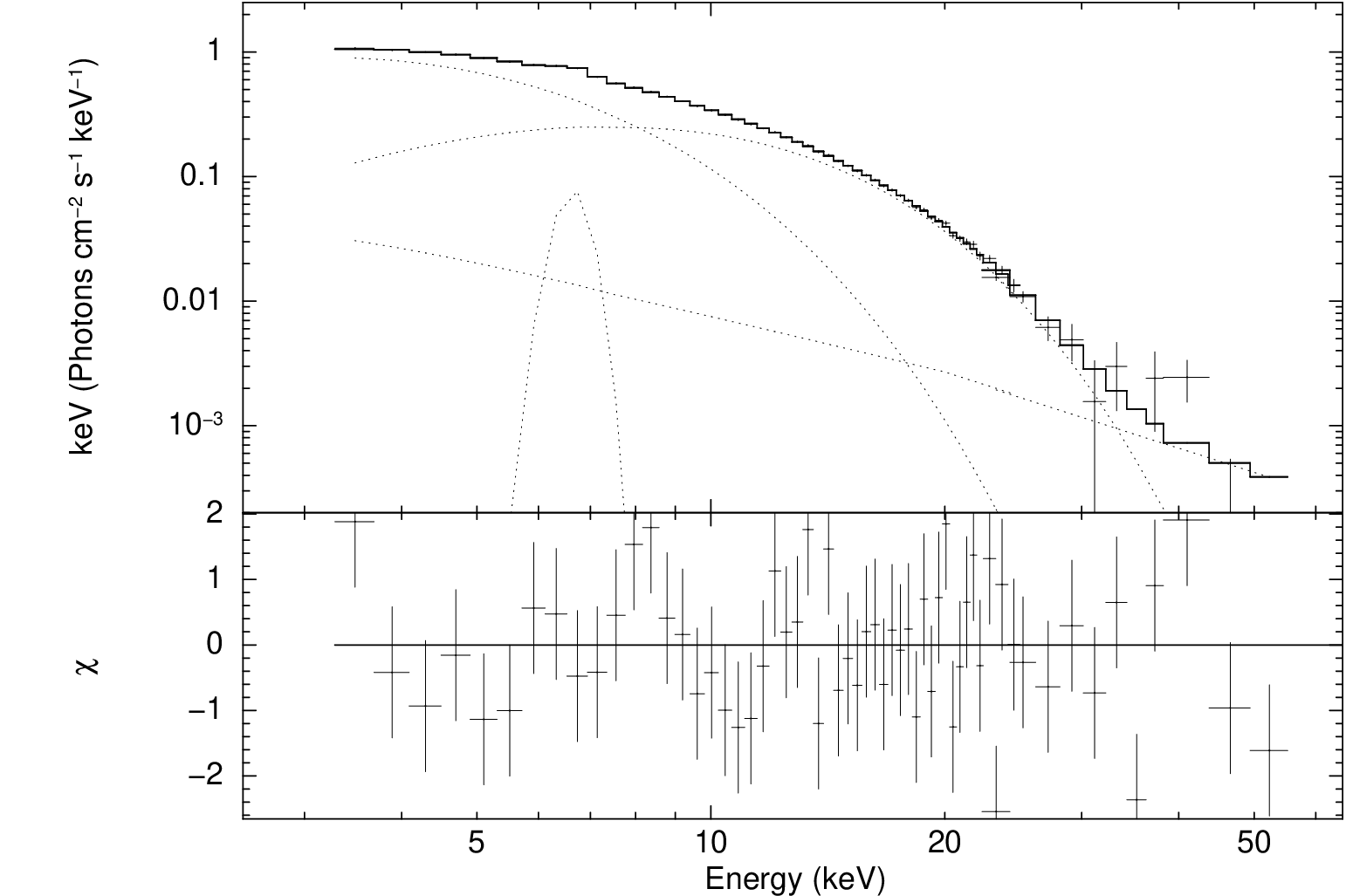}
\includegraphics[width=3.25cm,angle=270]{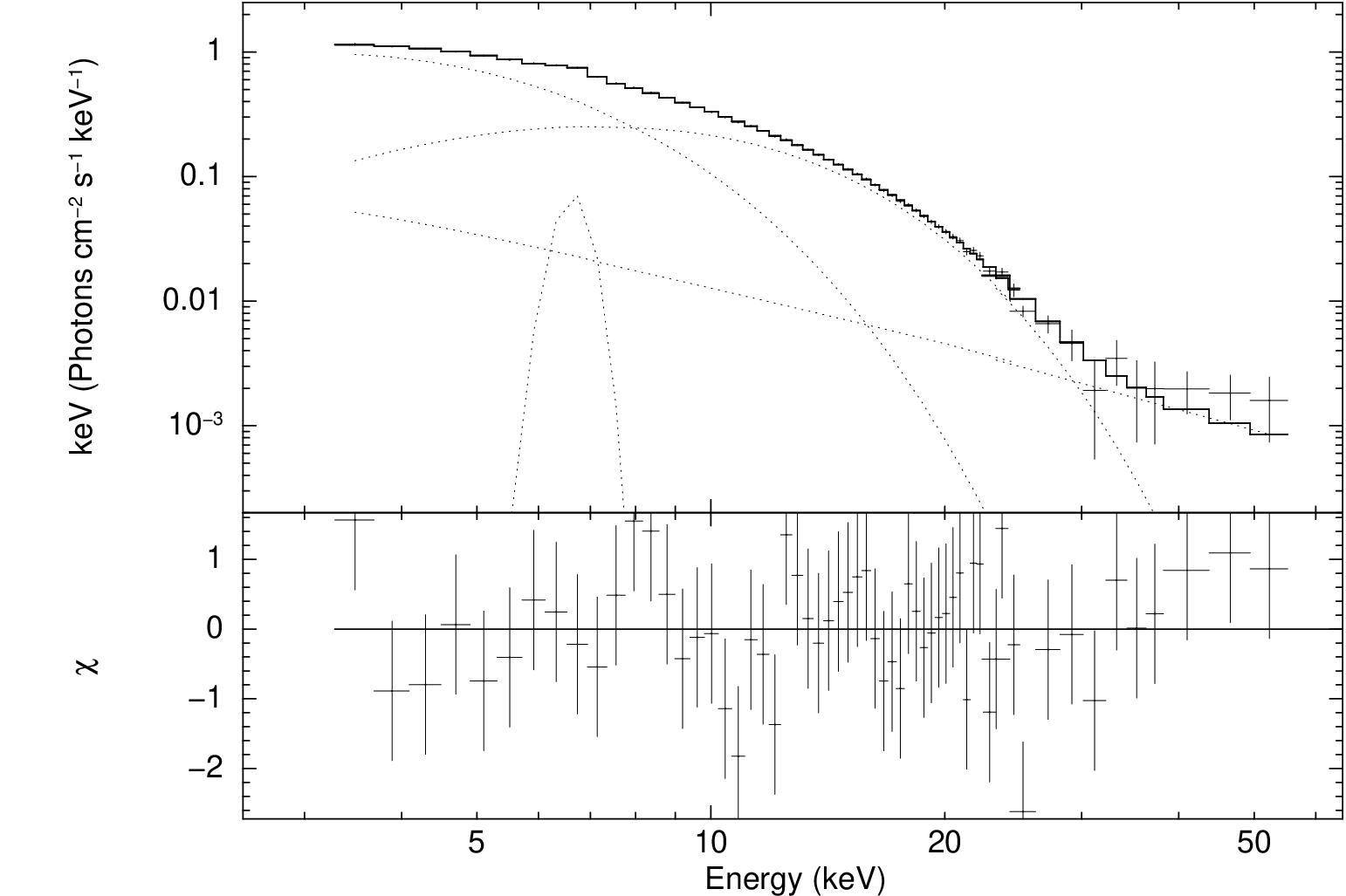}

\includegraphics[width=3.25cm,angle=270]{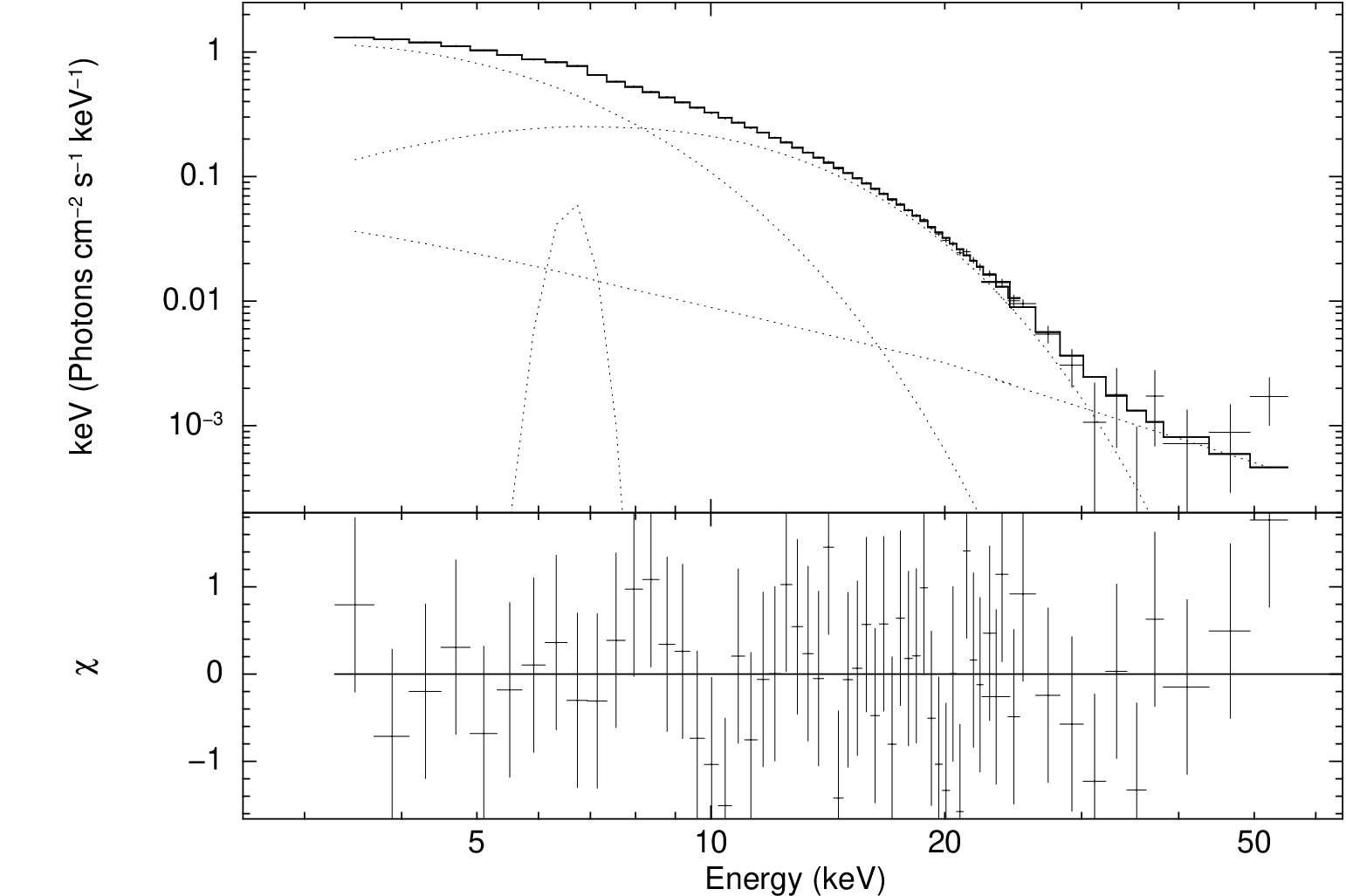}
\includegraphics[width=3.25cm,angle=270]{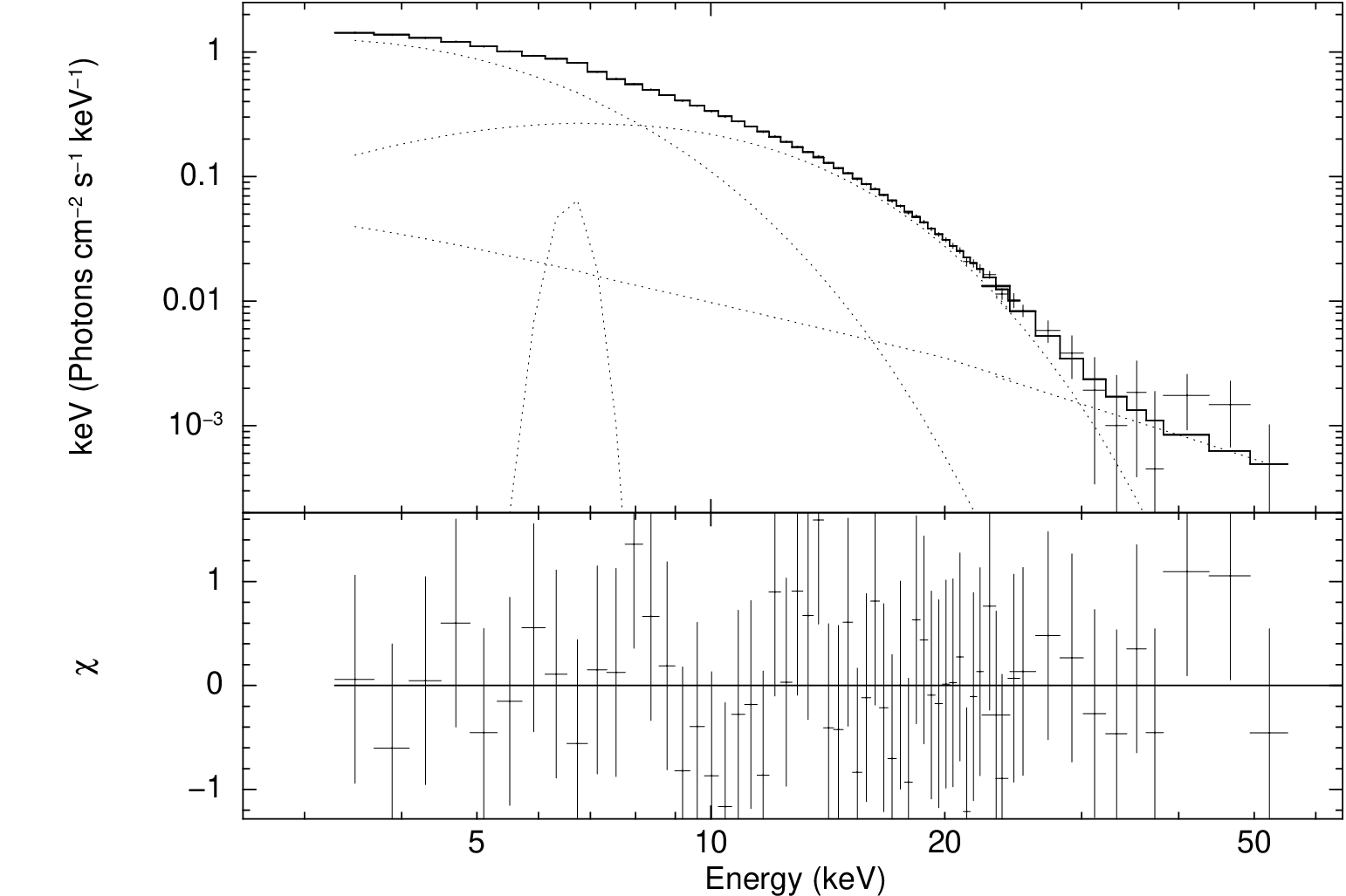}
\includegraphics[width=3.25cm,angle=270]{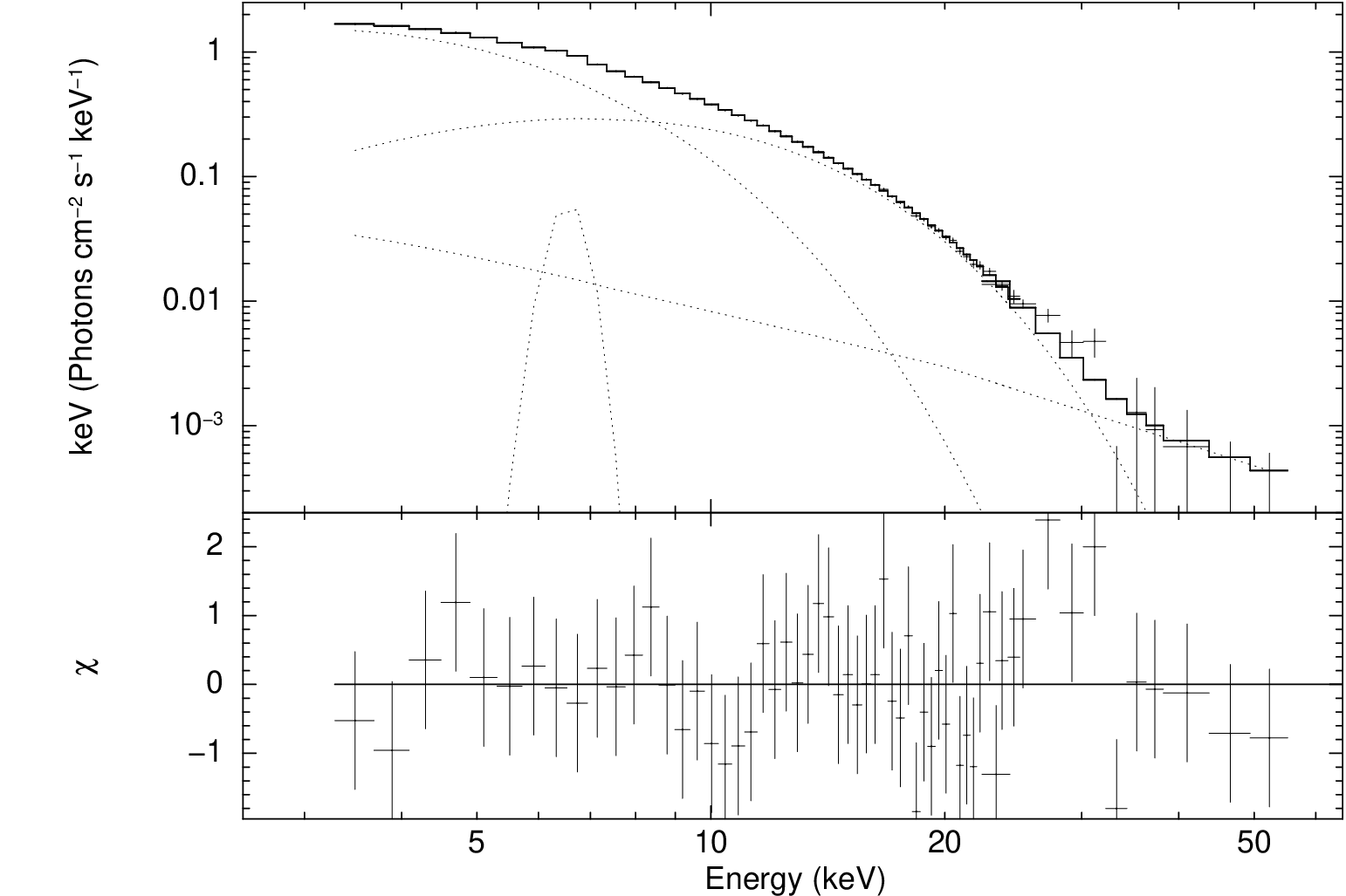}

\includegraphics[width=3.25cm,angle=270]{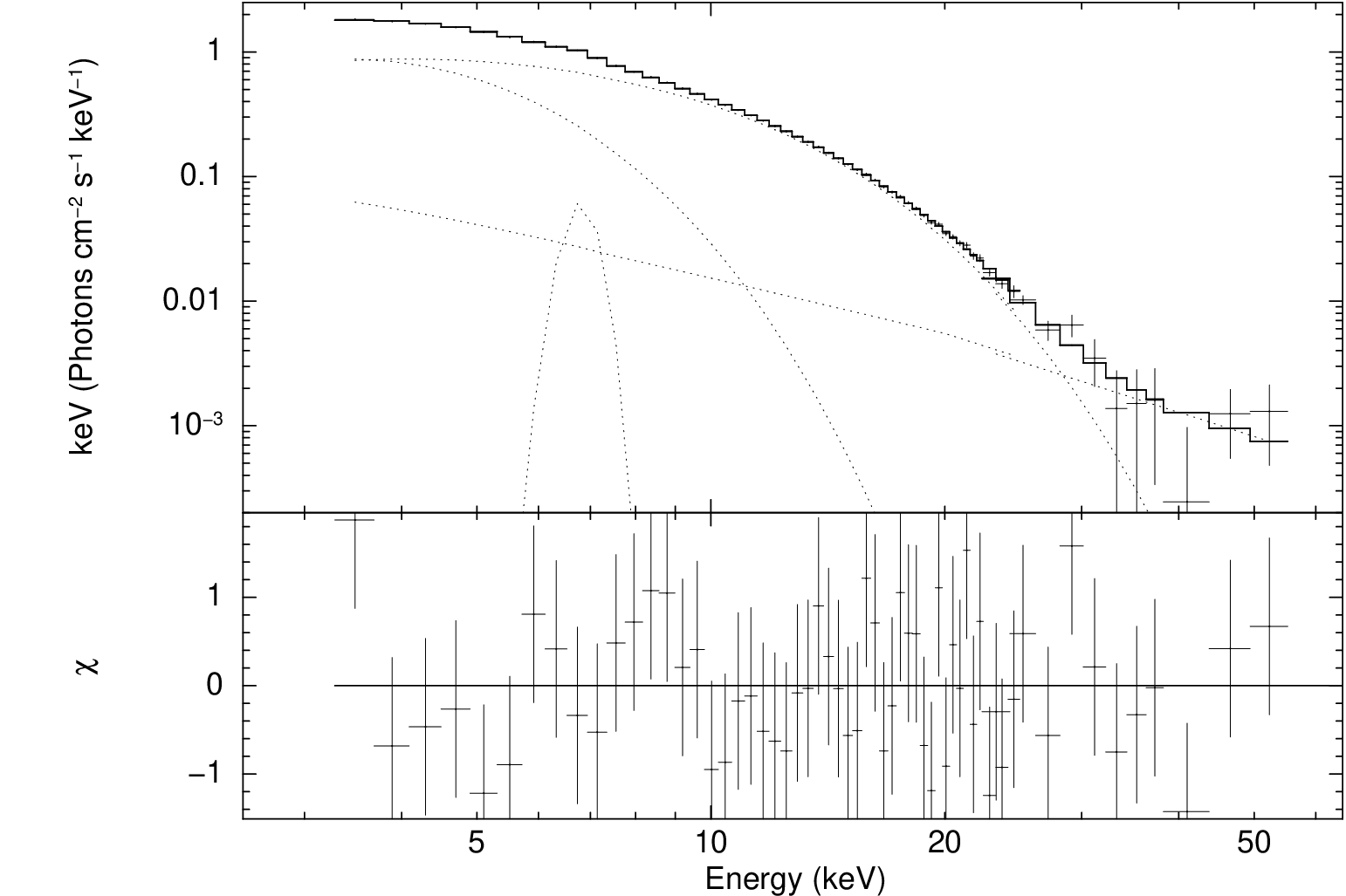}
\includegraphics[width=3.25cm,angle=270]{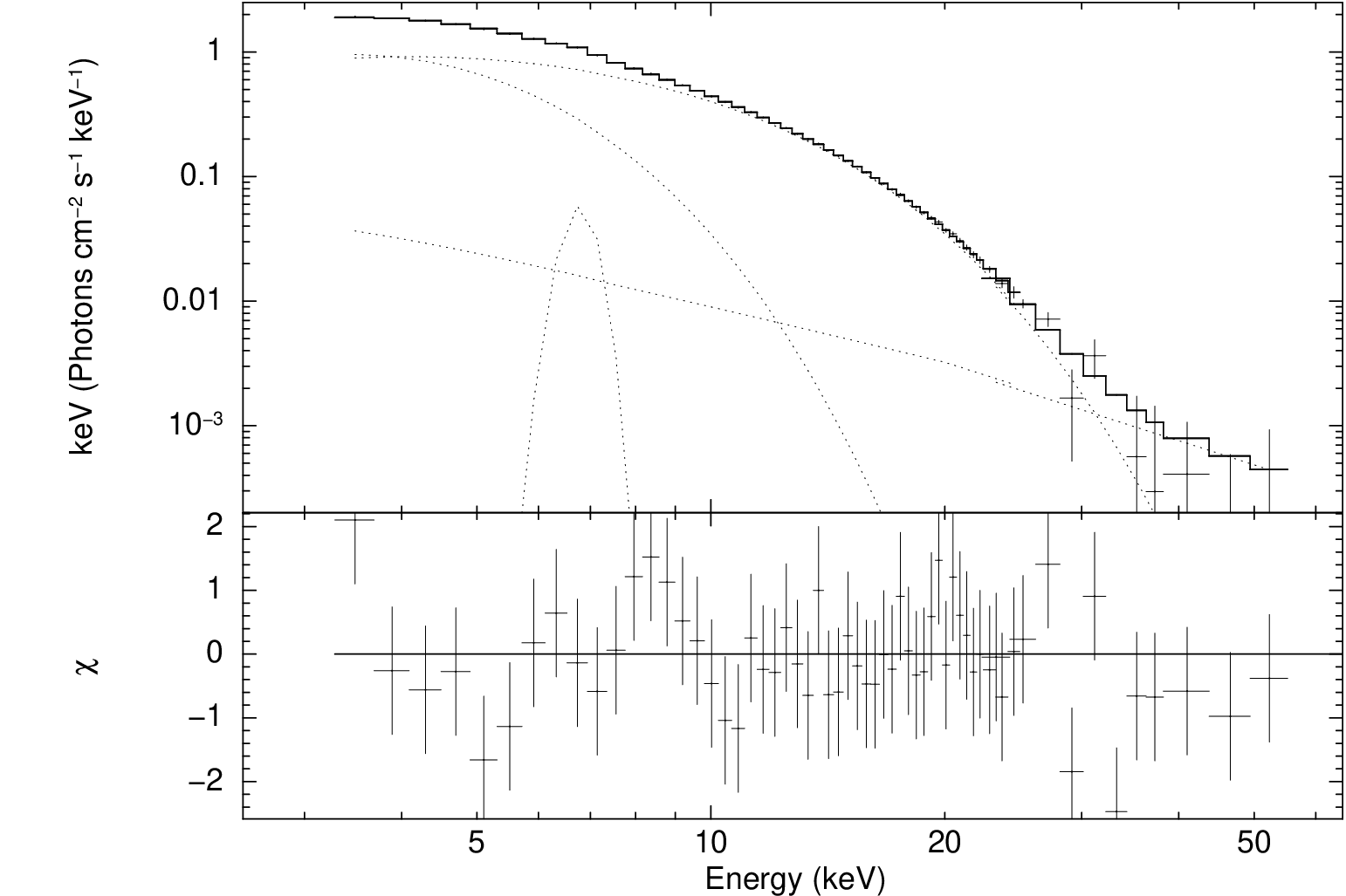}
\includegraphics[width=3.25cm,angle=270]{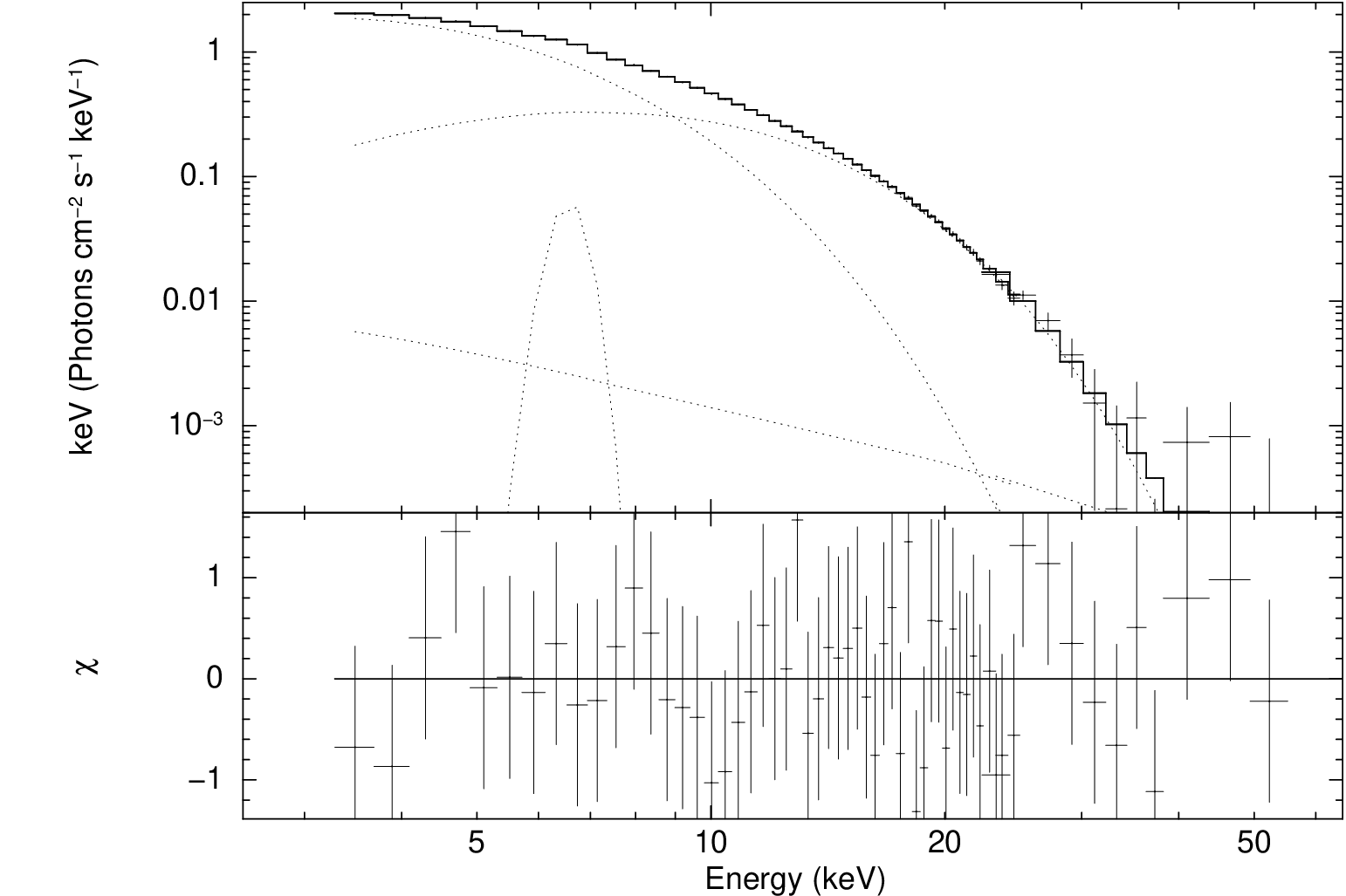}

\includegraphics[width=3.25cm,angle=270]{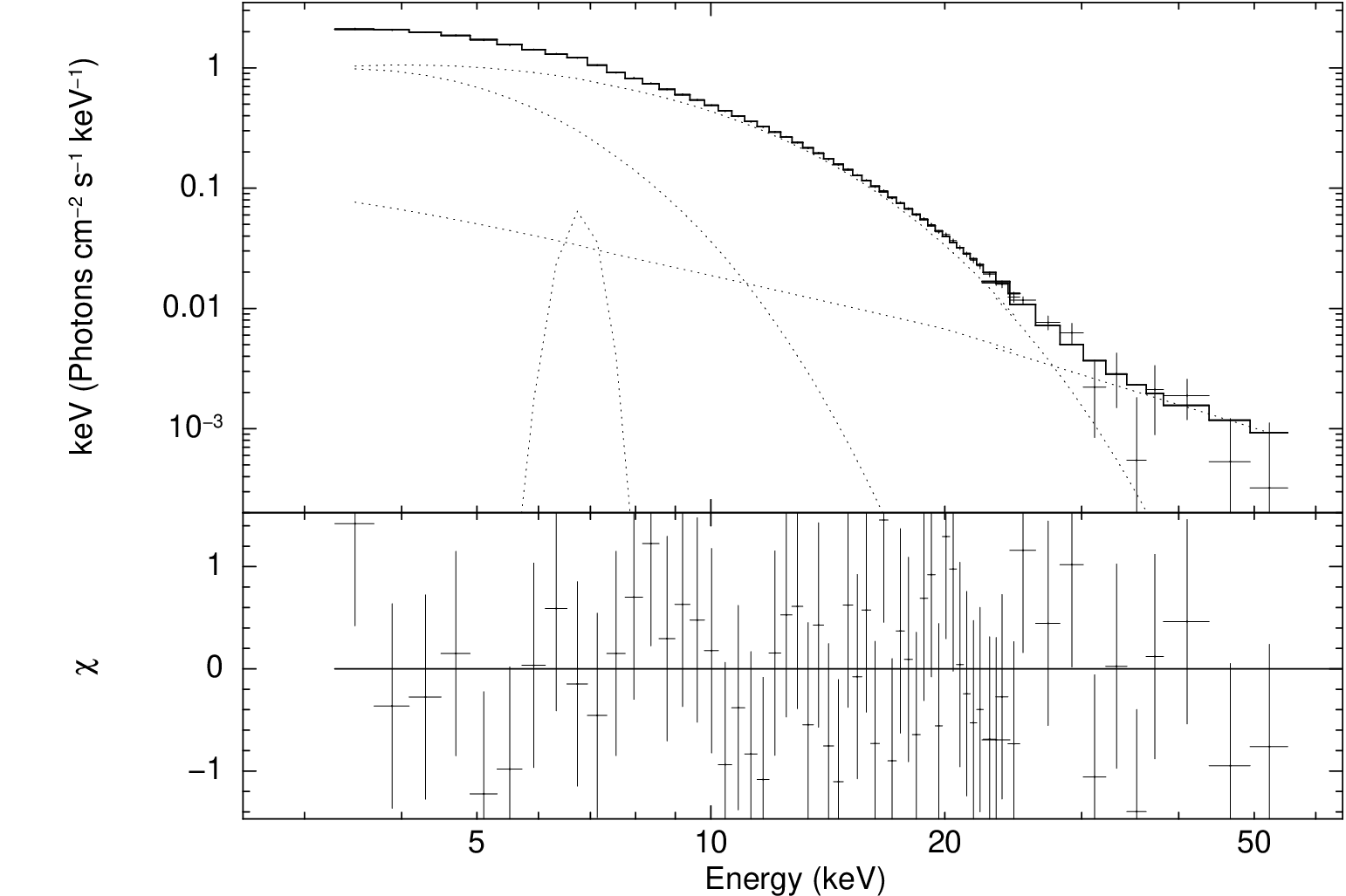}
\includegraphics[width=3.25cm,angle=270]{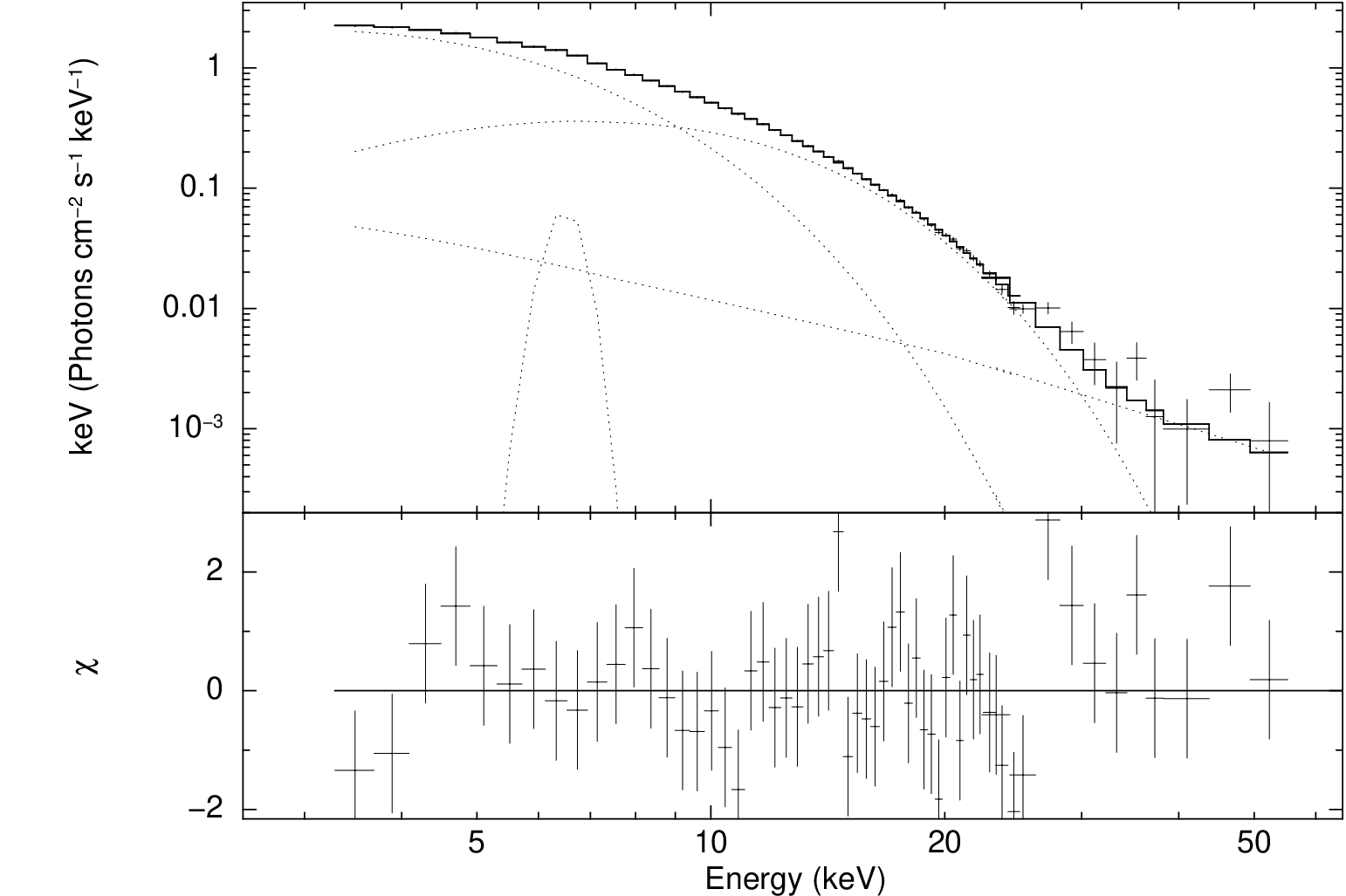}
\includegraphics[width=3.25cm,angle=270]{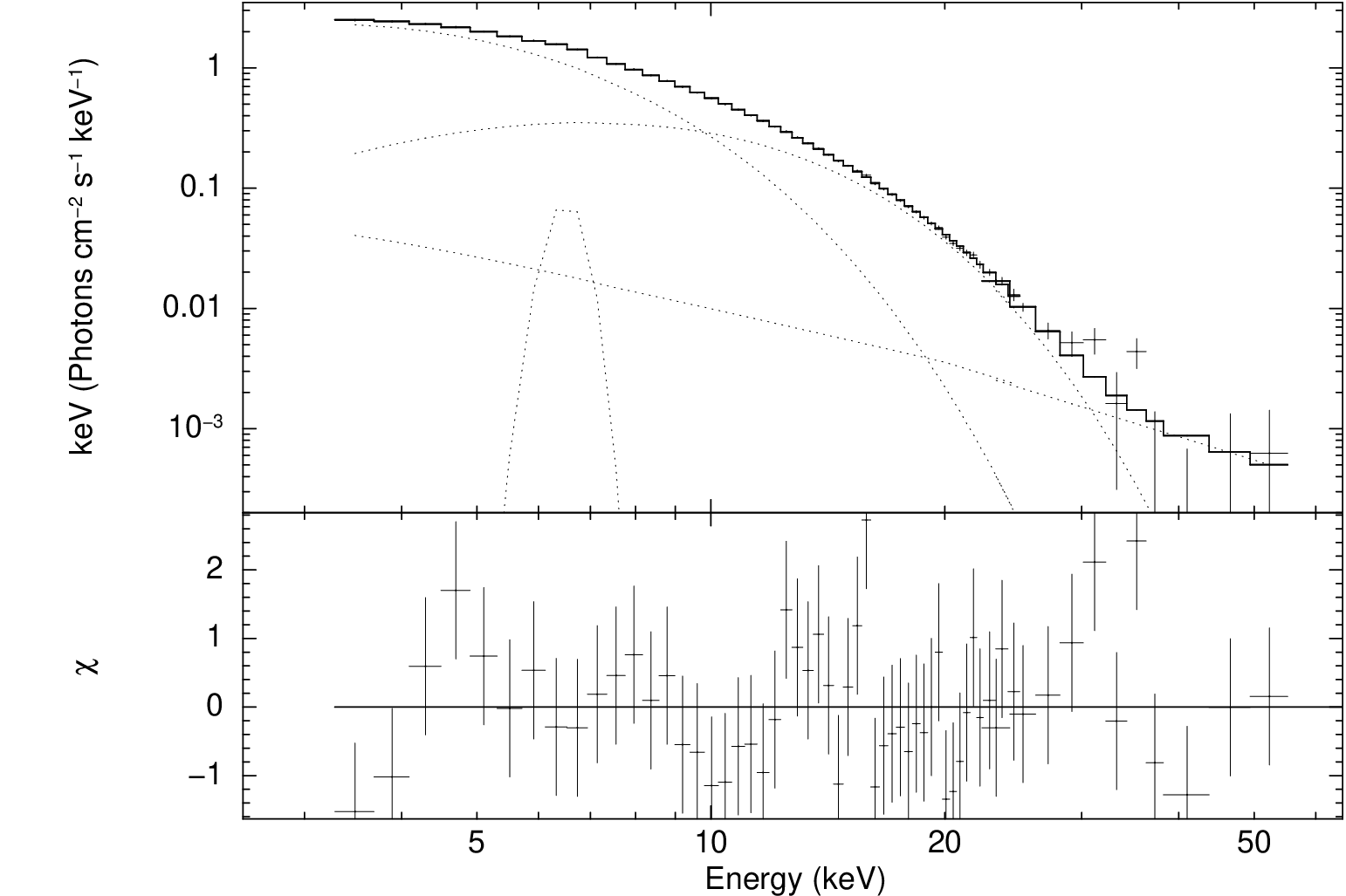}

\includegraphics[width=3.25cm,angle=270]{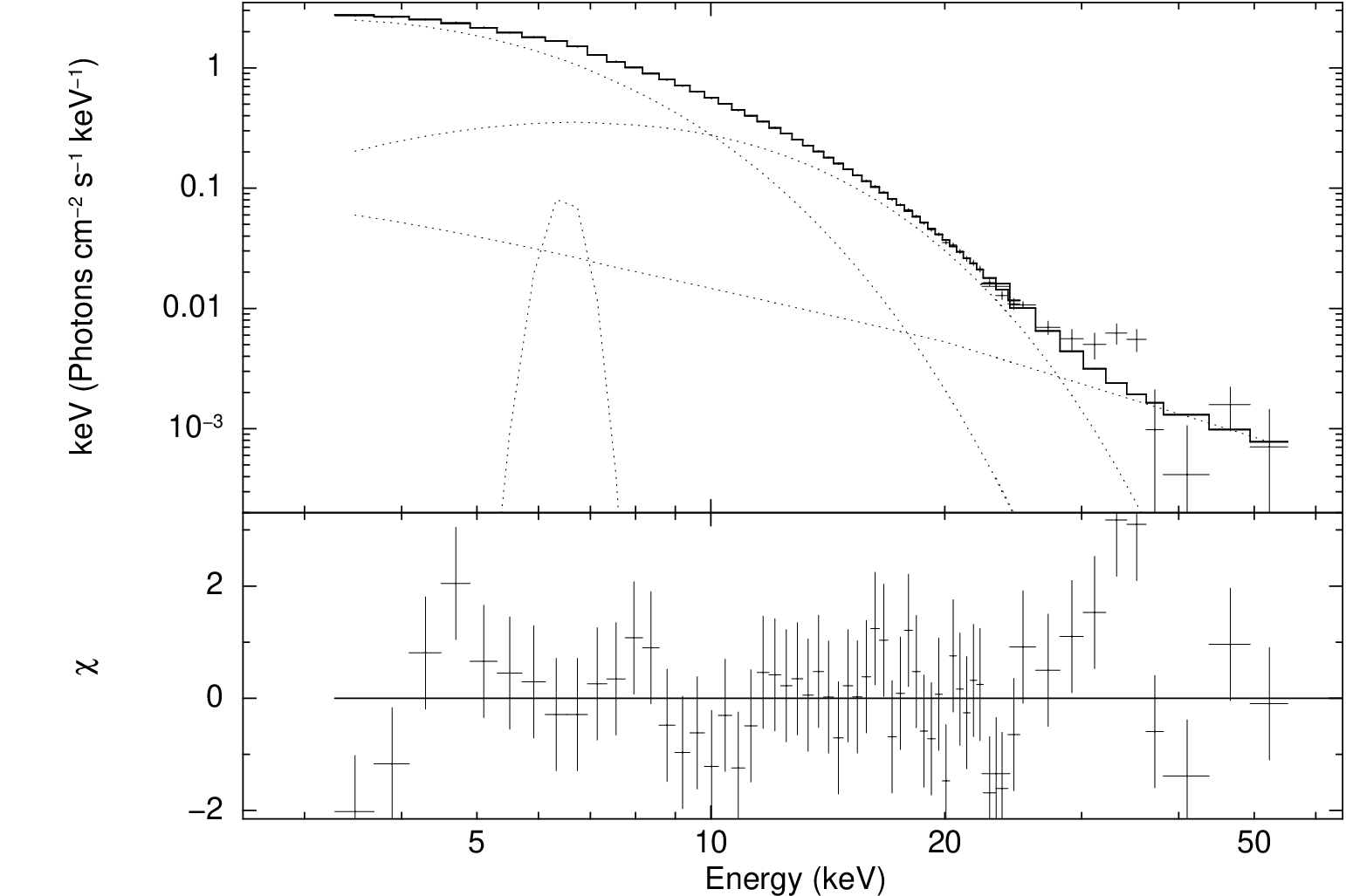}

\caption{The unfolded spectral of upper HB  in Interval A (Box 1-13, from left-to-right and top-to-bottom) fitted by Model I.    \label{fig6-1}}
\end{figure}

\begin{figure}


\includegraphics[width=3.25cm,angle=270]{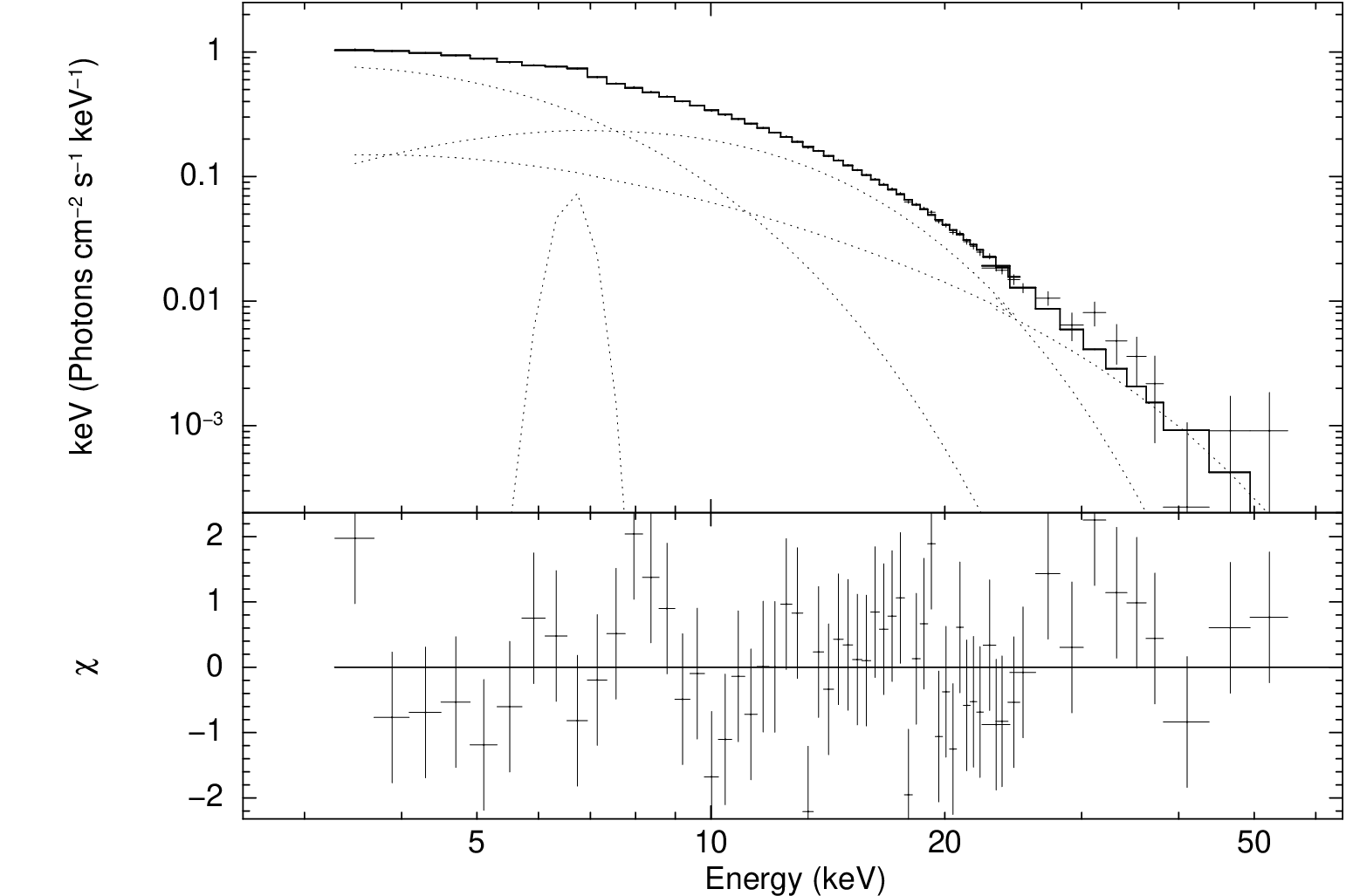}
\includegraphics[width=3.25cm,angle=270]{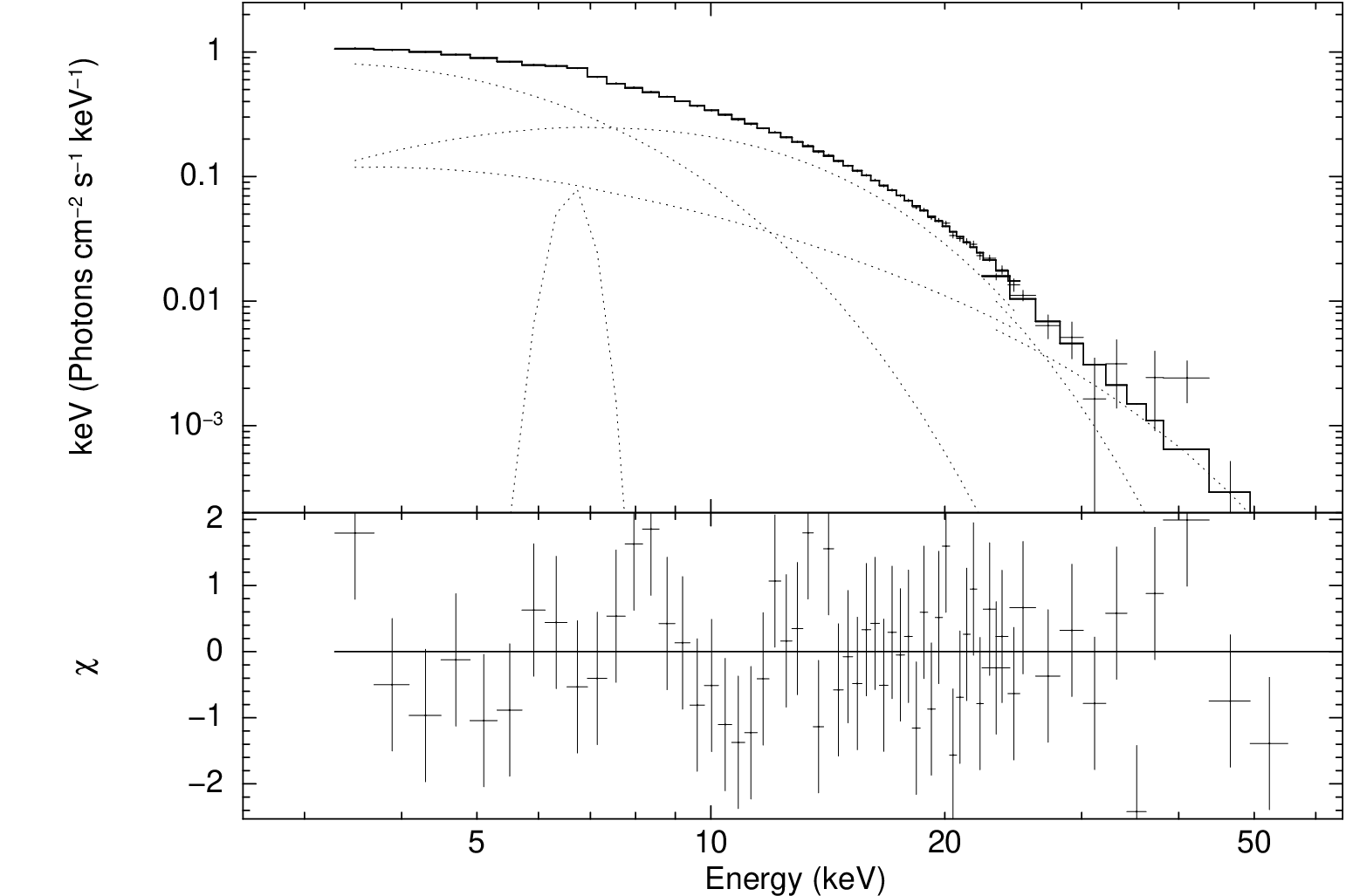}
\includegraphics[width=3.25cm,angle=270]{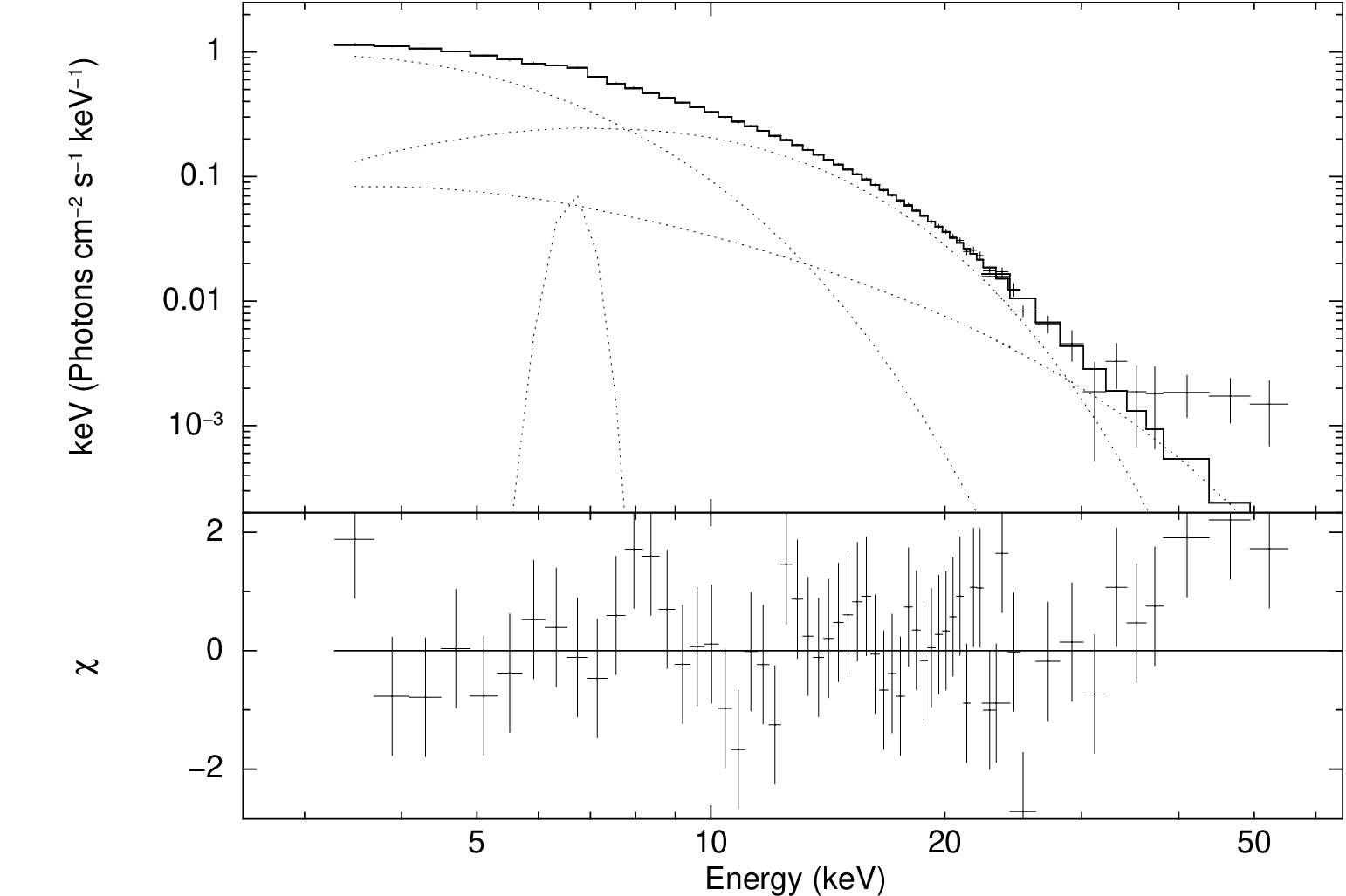}

\includegraphics[width=3.25cm,angle=270]{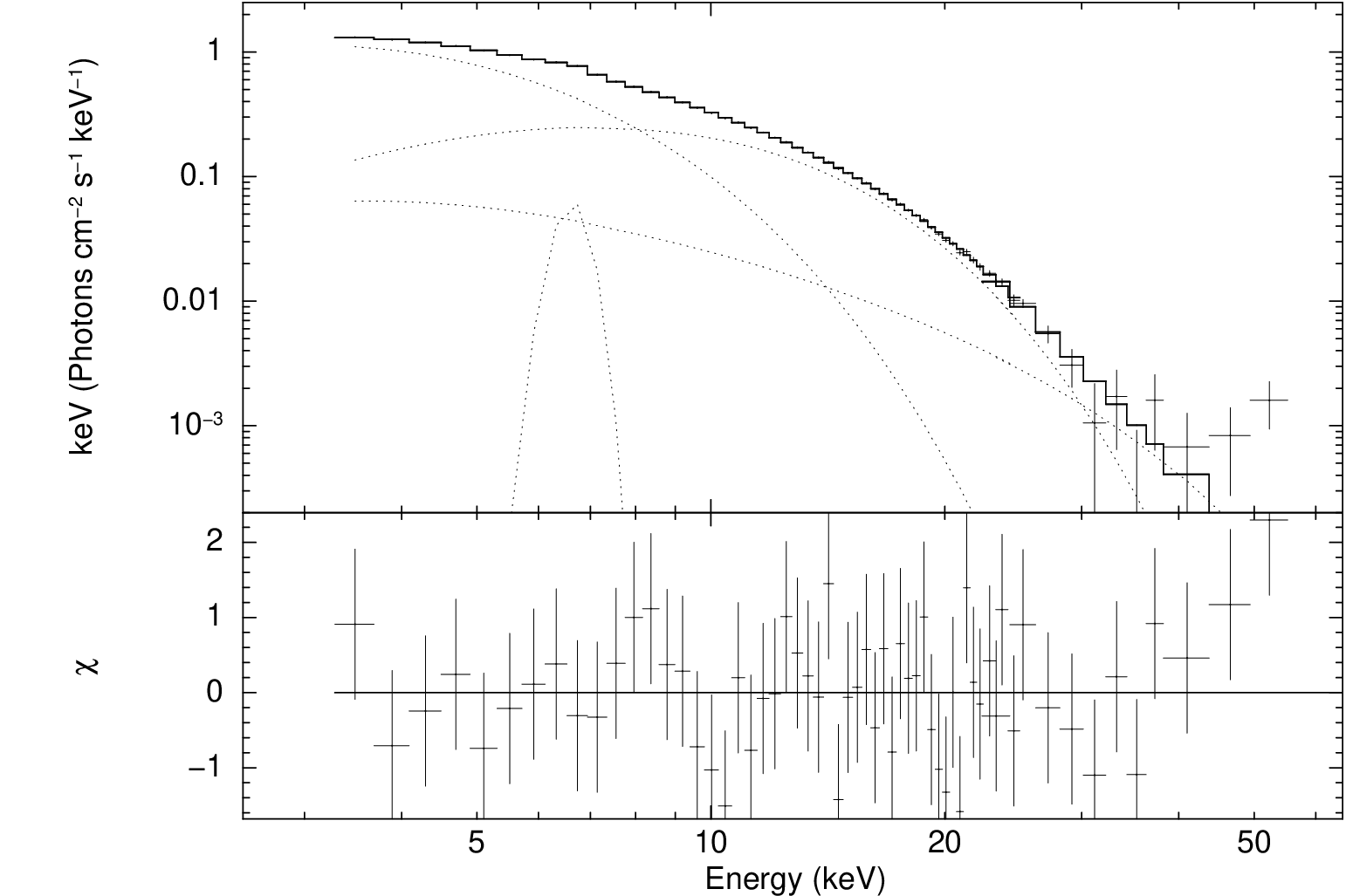}
\includegraphics[width=3.25cm,angle=270]{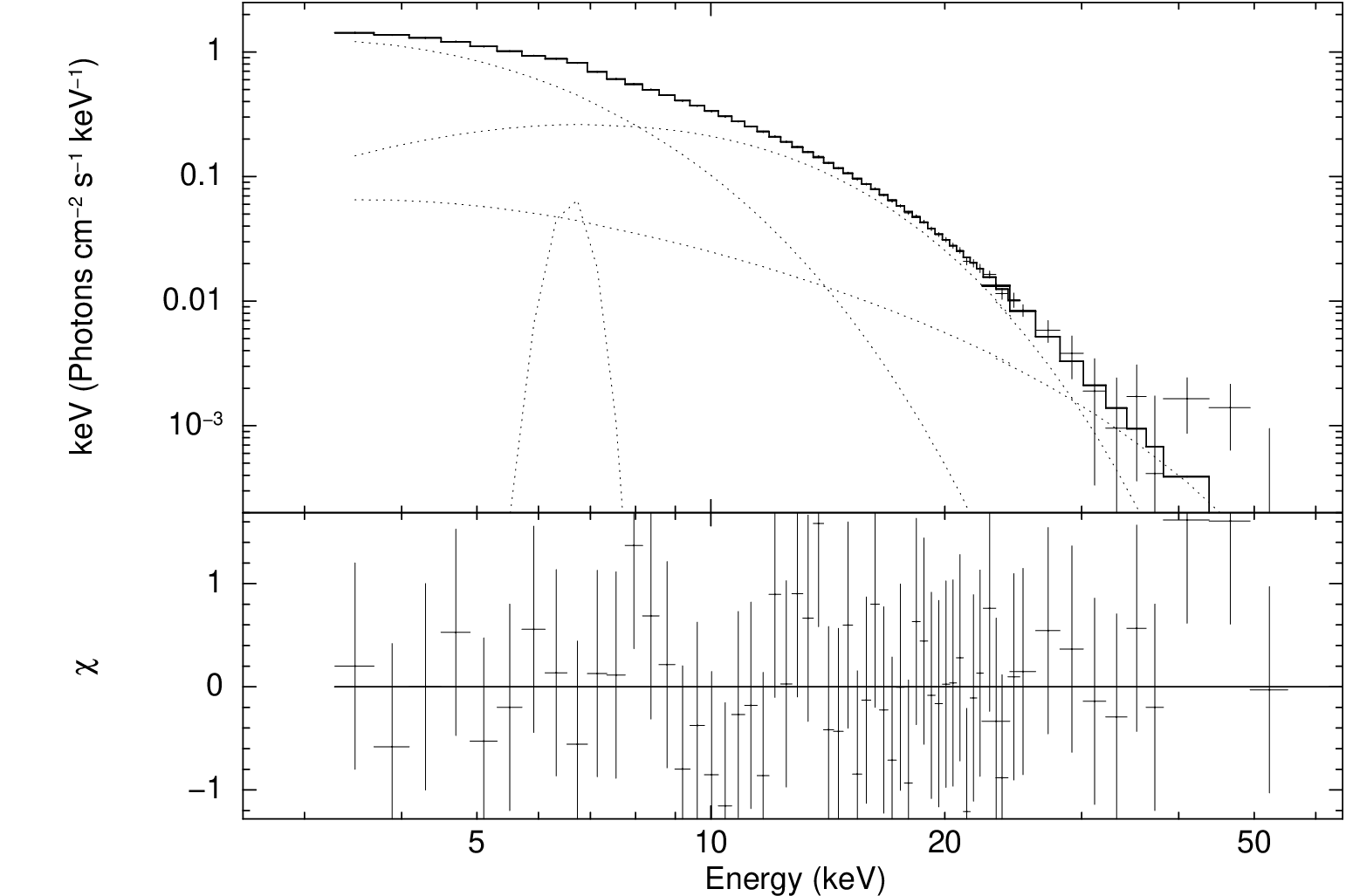}
\includegraphics[width=3.25cm,angle=270]{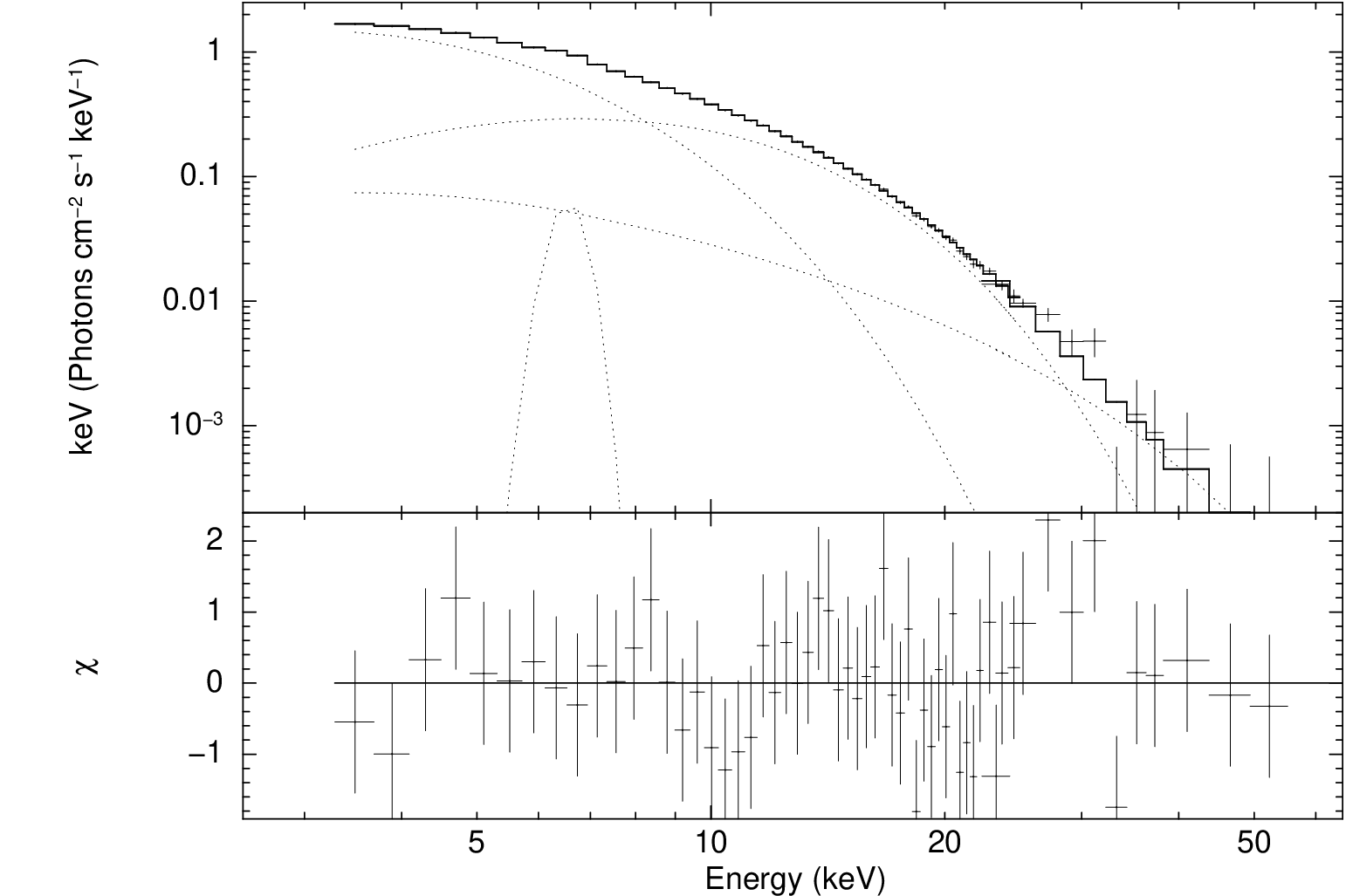}

\includegraphics[width=3.25cm,angle=270]{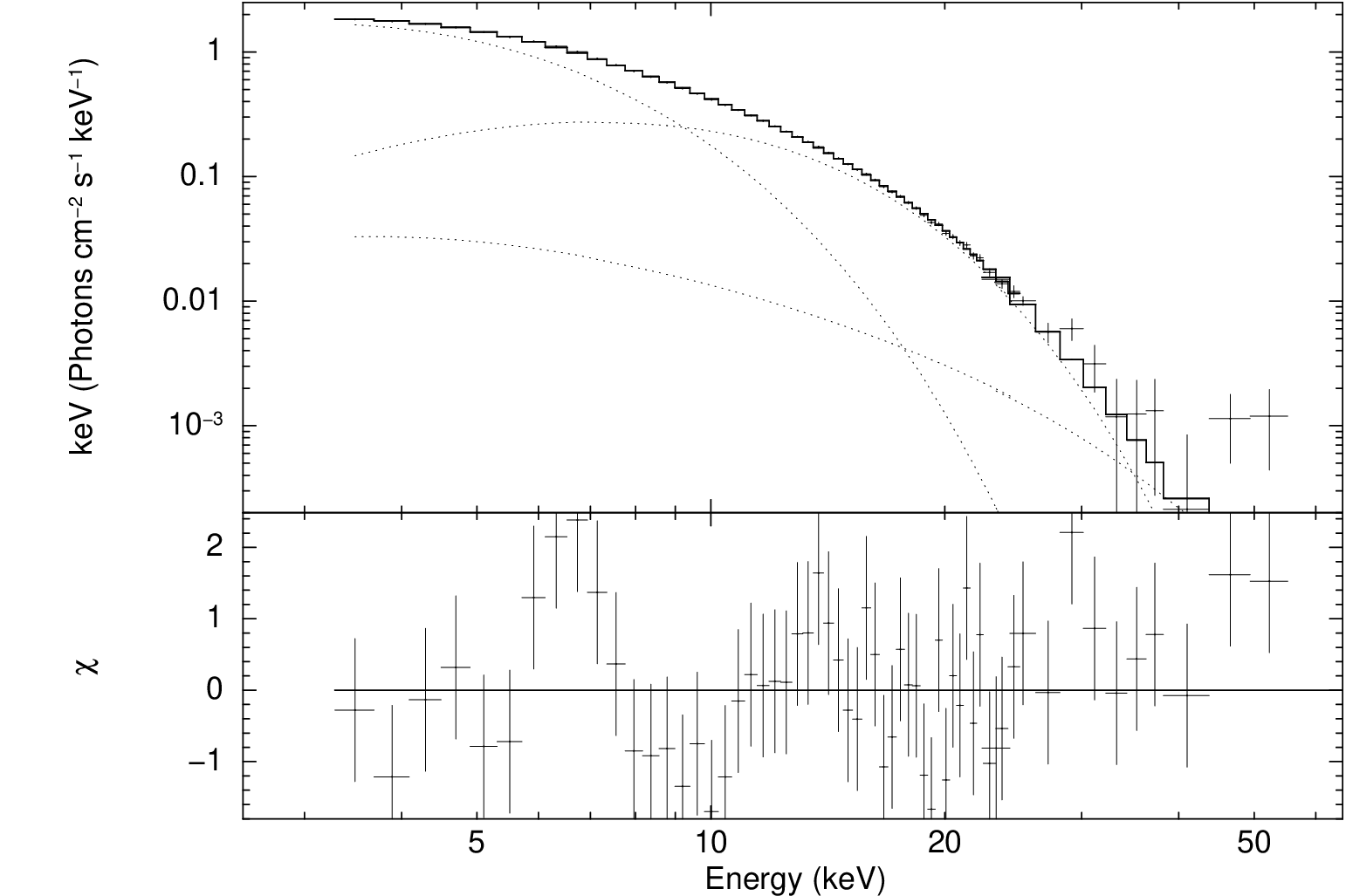}
\includegraphics[width=3.25cm,angle=270]{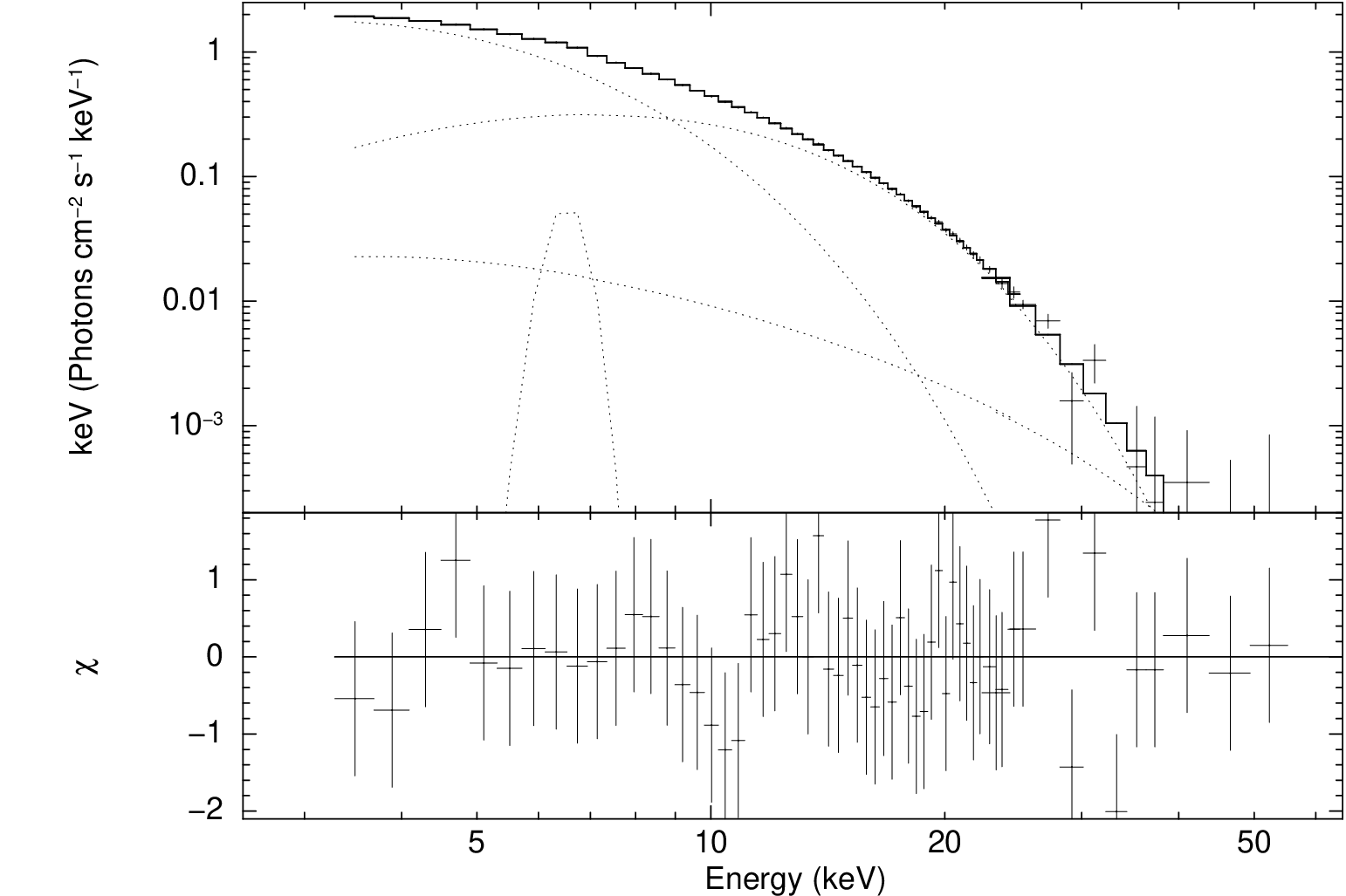}
\includegraphics[width=3.25cm,angle=270]{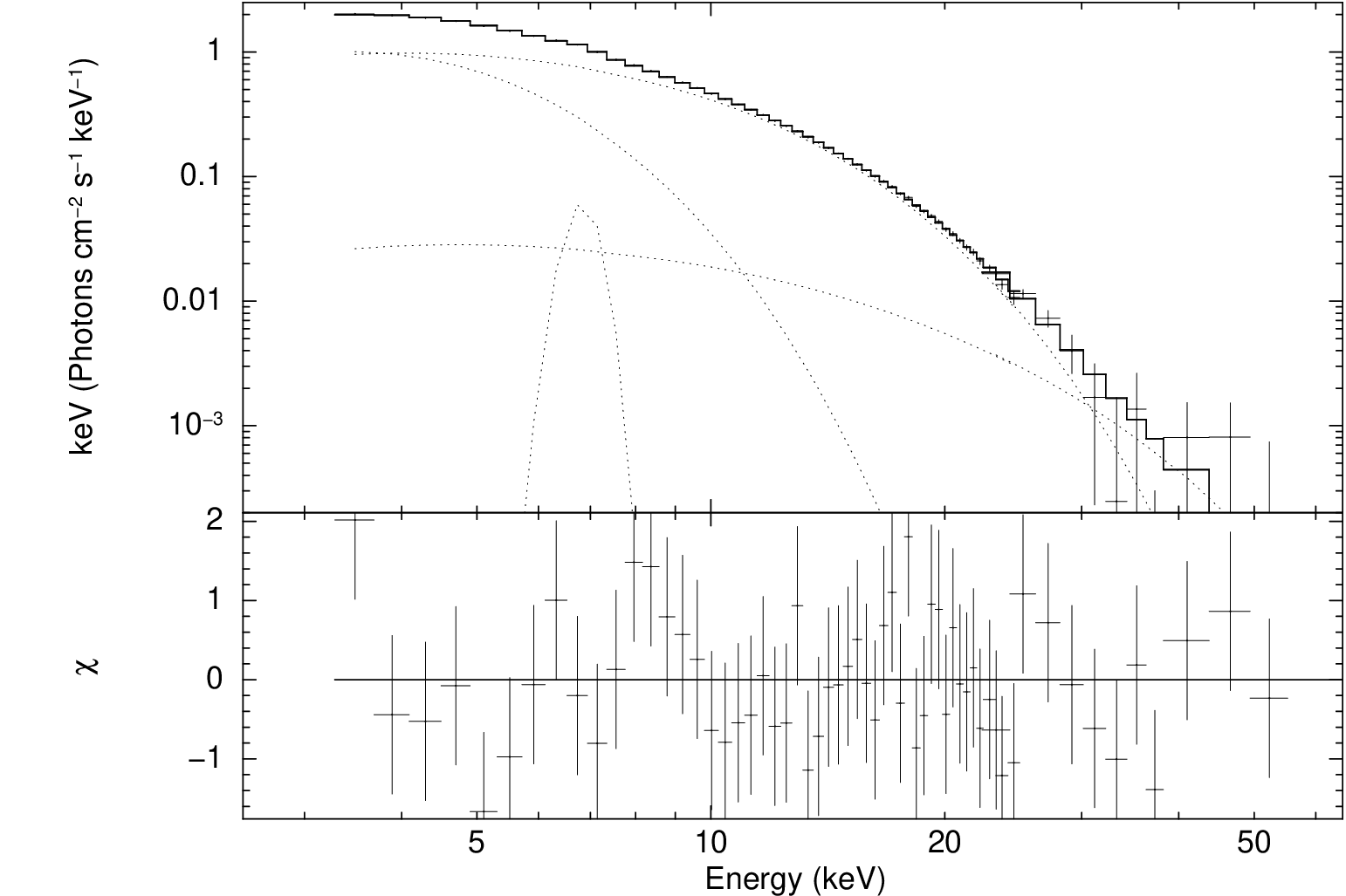}

\includegraphics[width=3.25cm,angle=270]{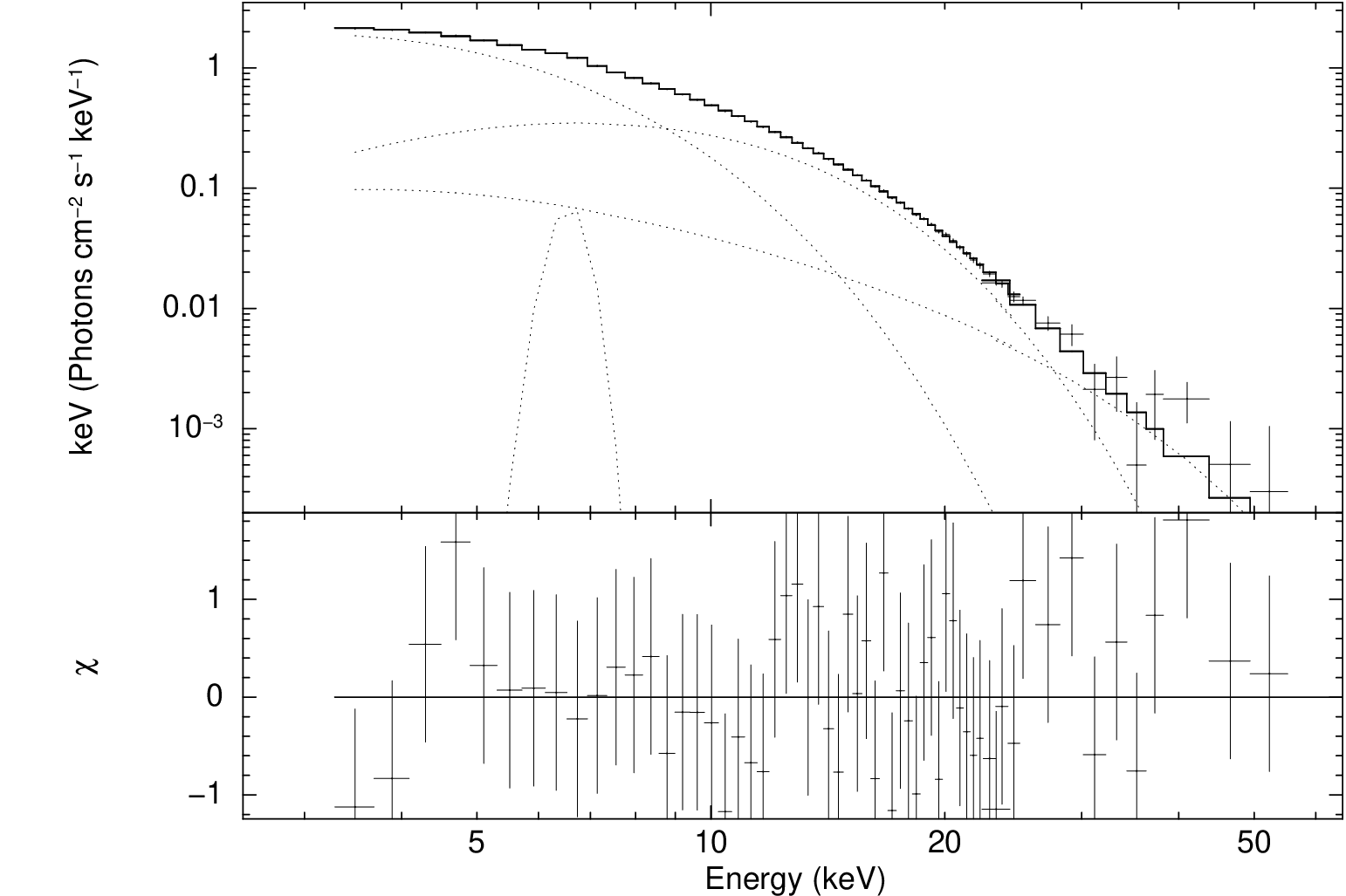}
\includegraphics[width=3.25cm,angle=270]{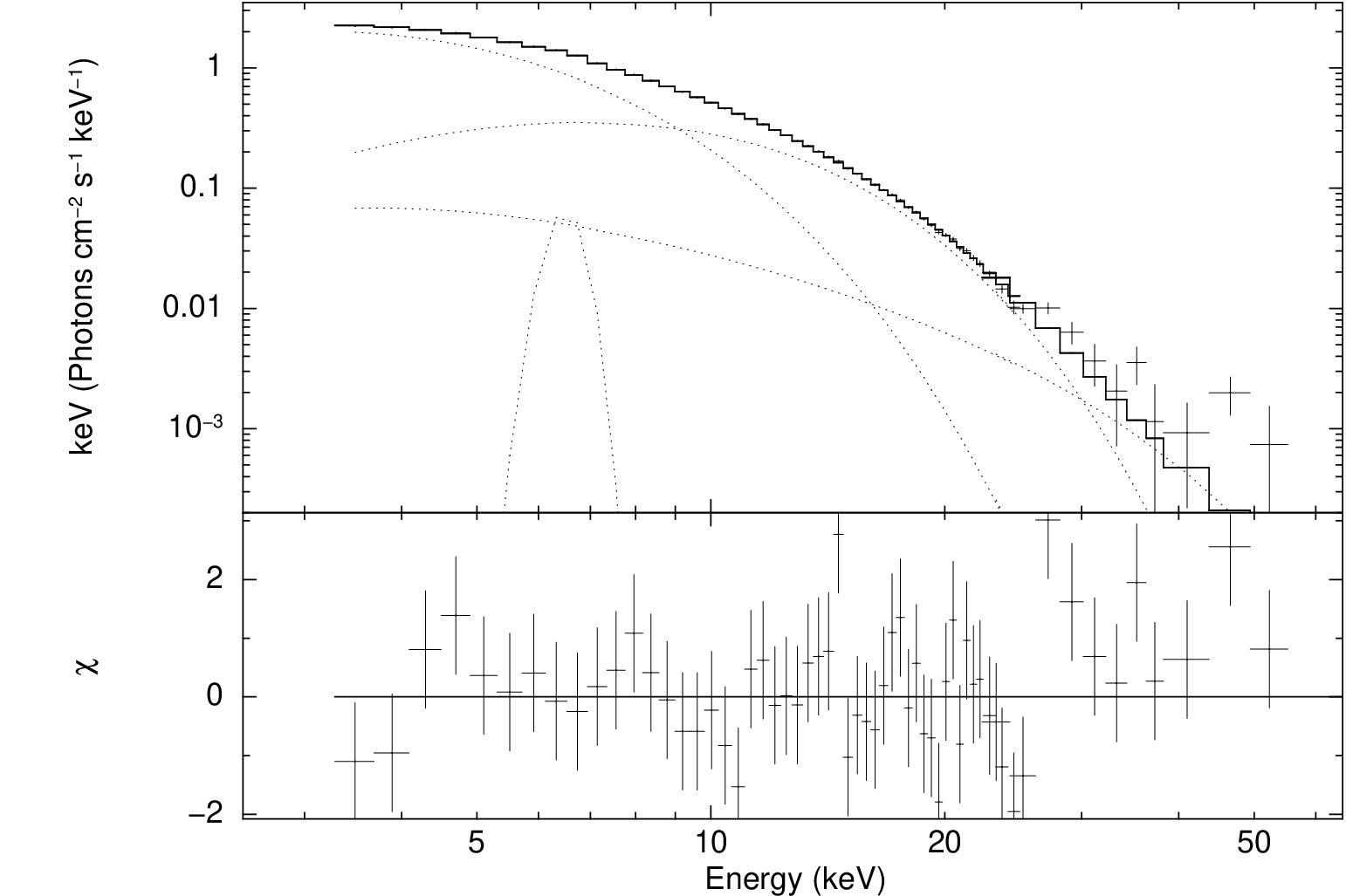}
\includegraphics[width=3.25cm,angle=270]{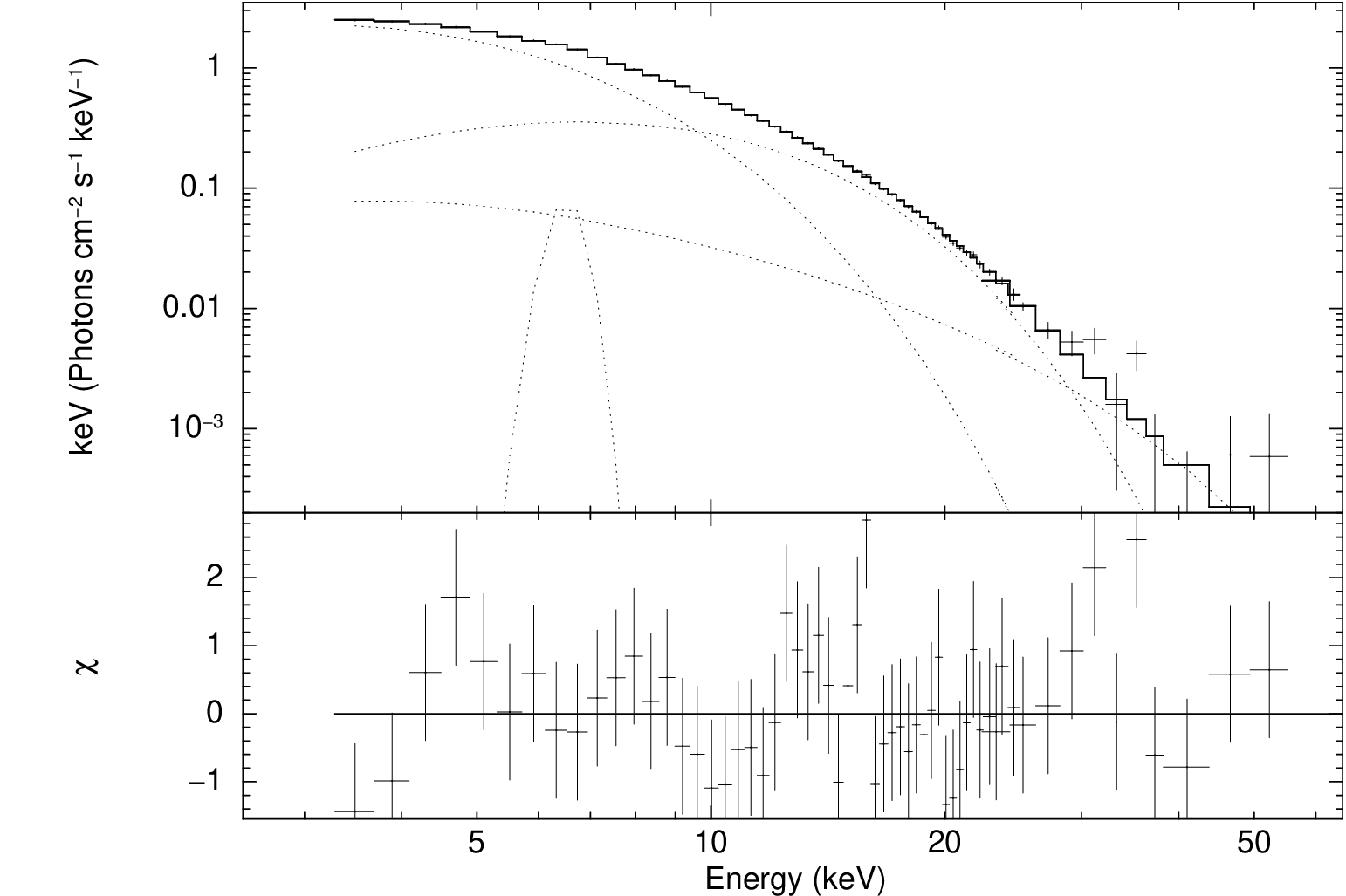}

\includegraphics[width=3.25cm,angle=270]{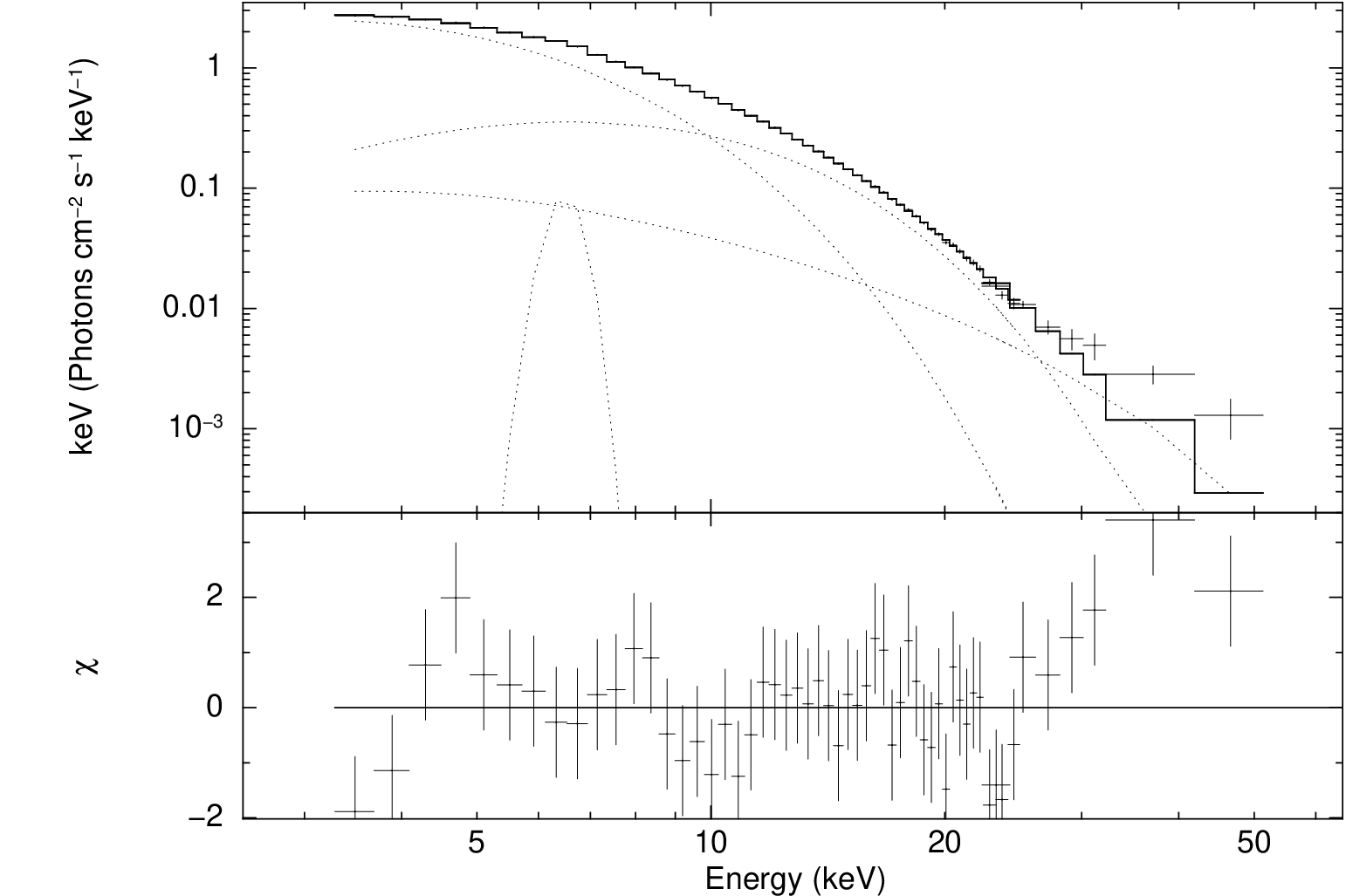}

\caption{The unfolded spectral of upper HB  in Interval A (Boxes 1-13) fitted by Model II.    \label{fig6-2}}
\end{figure}

\begin{figure}


\includegraphics[width=3.25cm,angle=270]{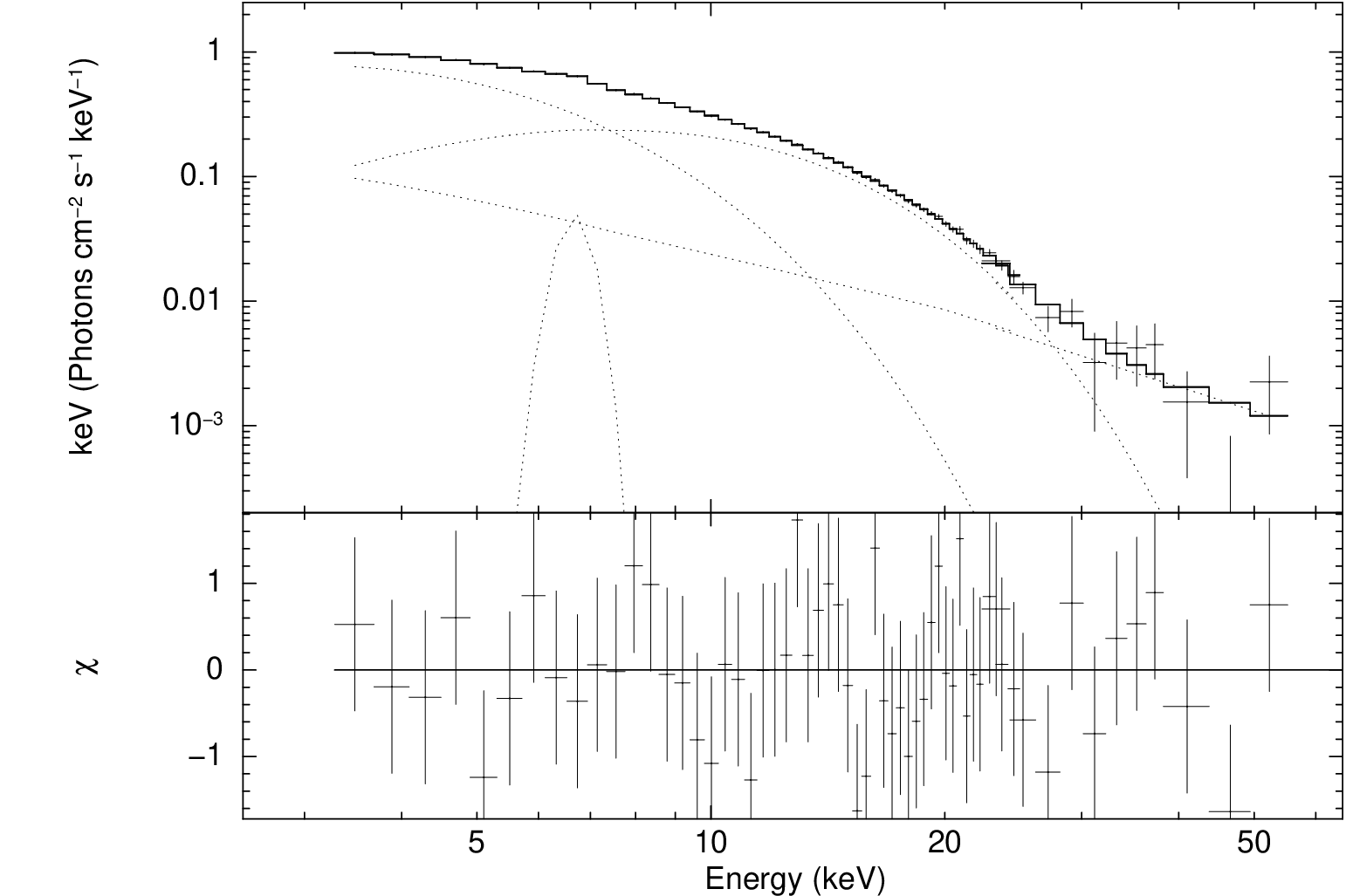}
\includegraphics[width=3.25cm,angle=270]{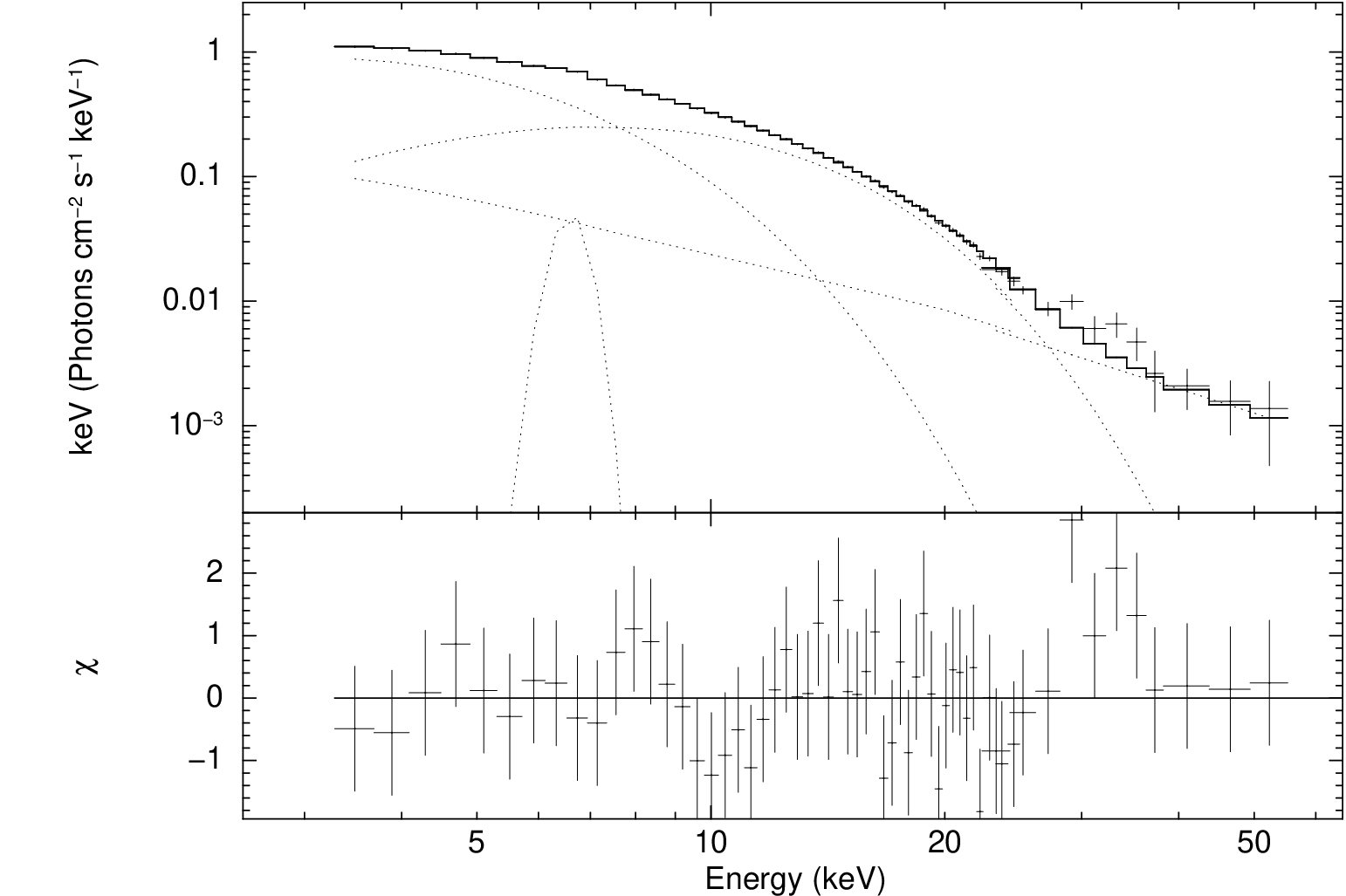}
\includegraphics[width=3.25cm,angle=270]{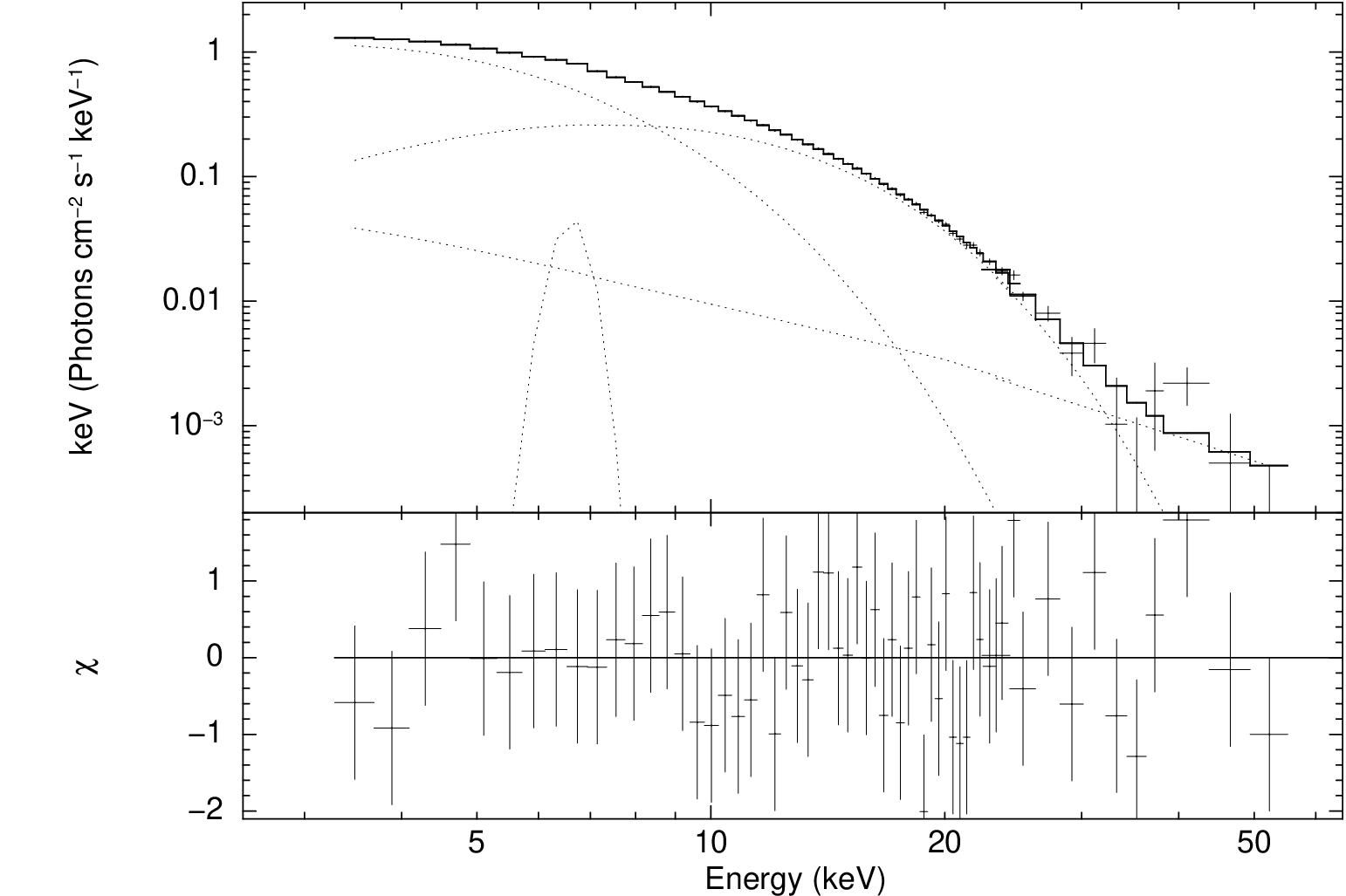}
\includegraphics[width=3.25cm,angle=270]{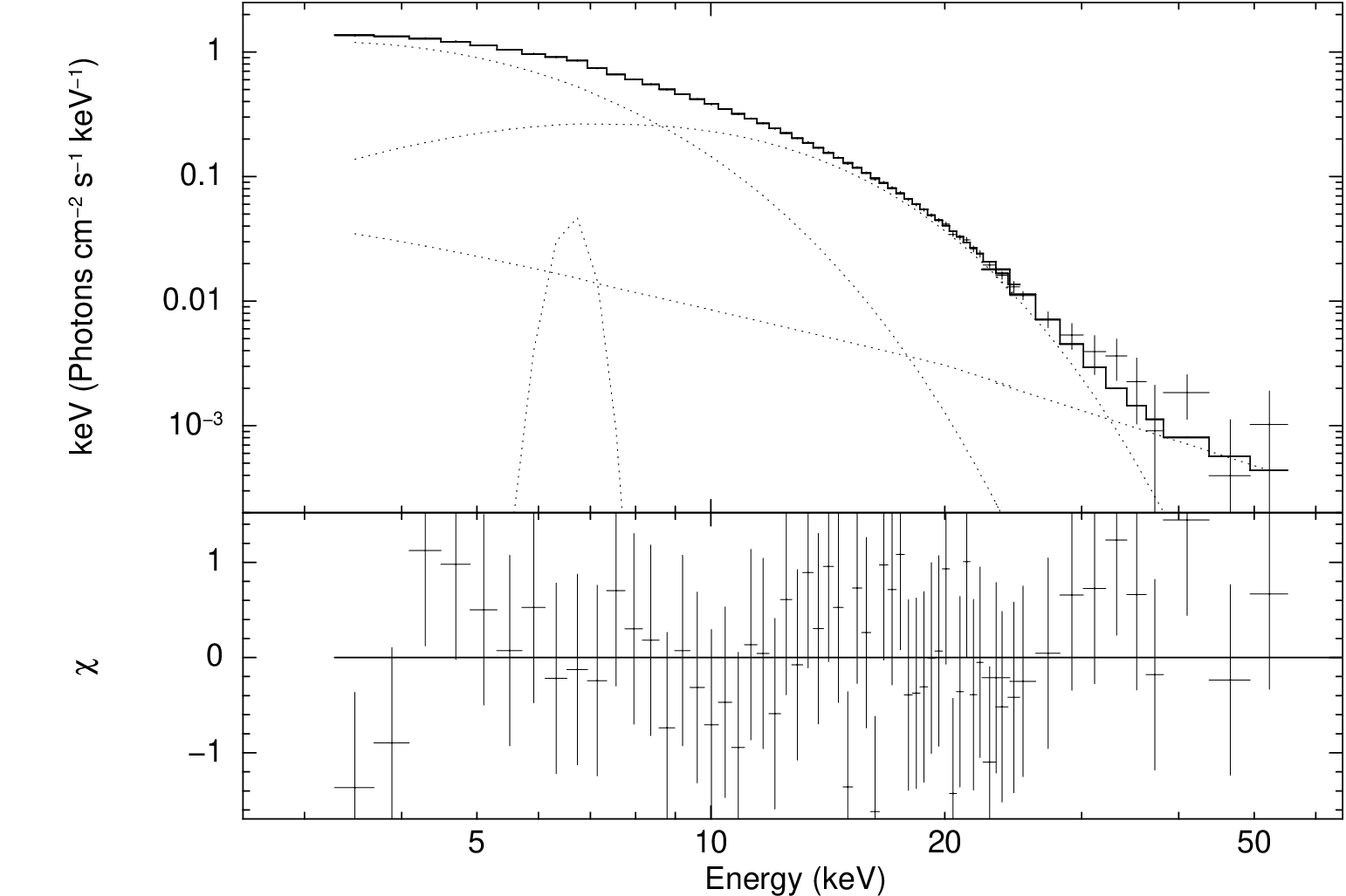}
\includegraphics[width=3.25cm,angle=270]{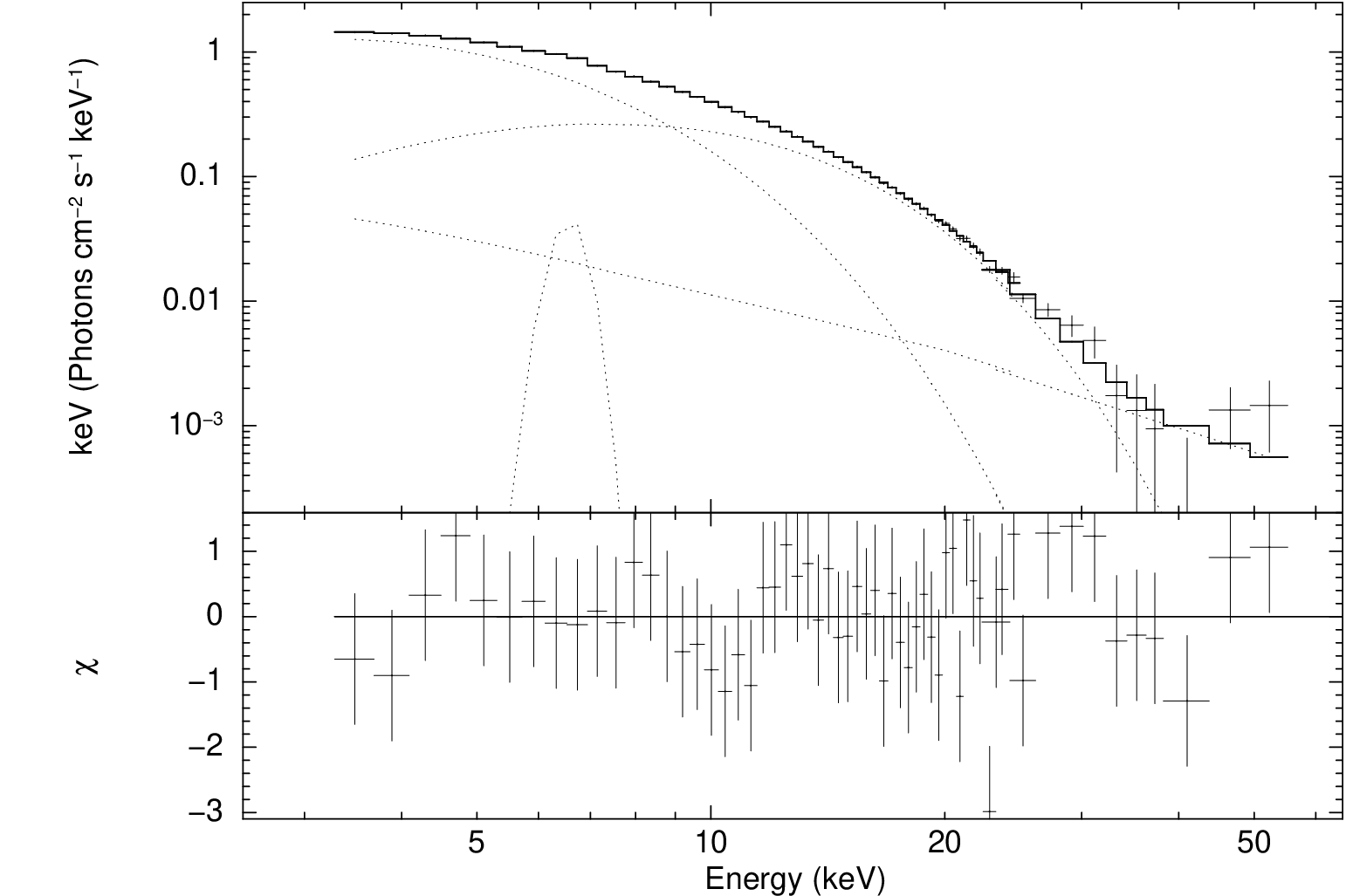}
\includegraphics[width=3.25cm,angle=270]{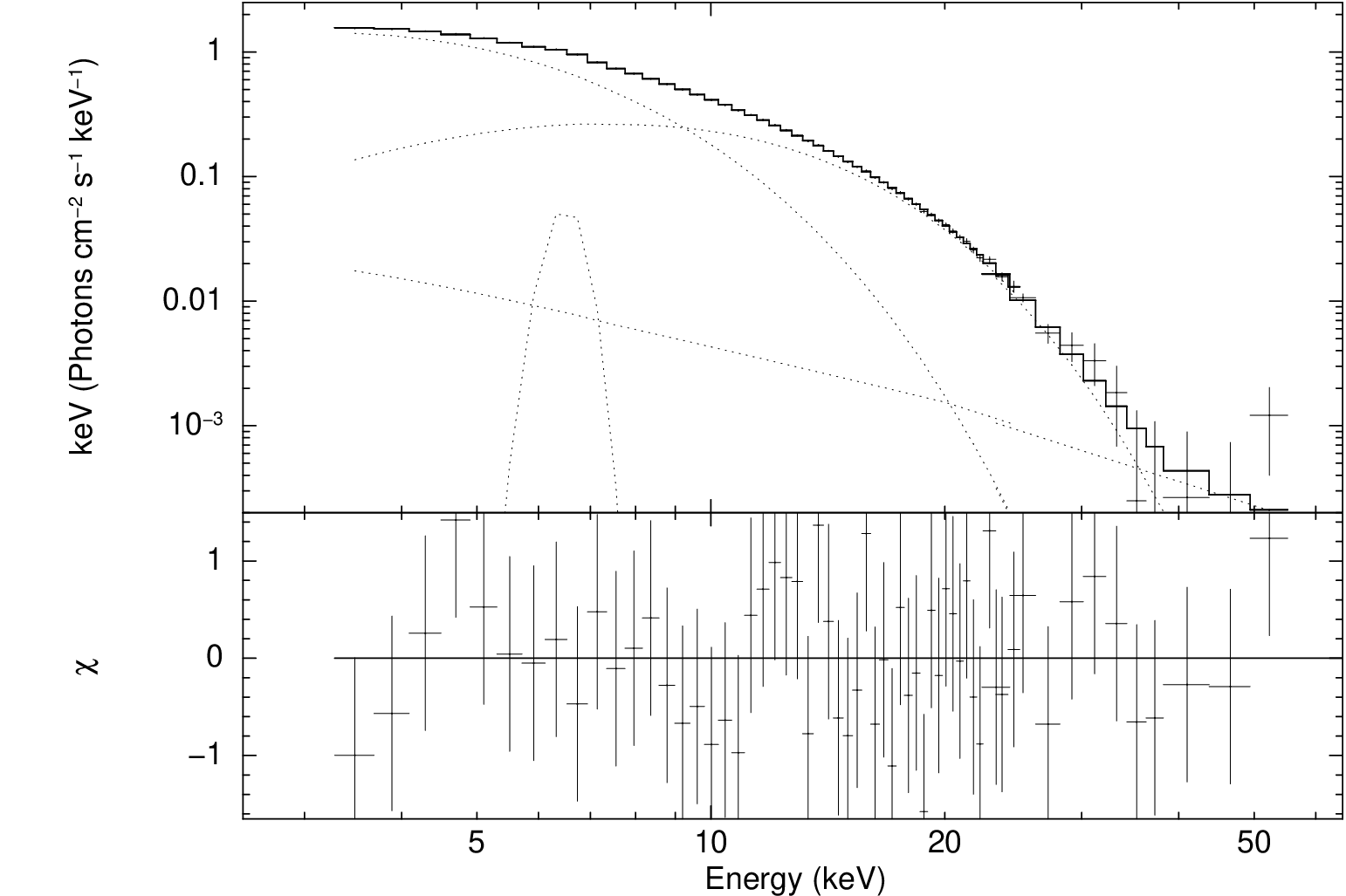}
\includegraphics[width=3.25cm,angle=270]{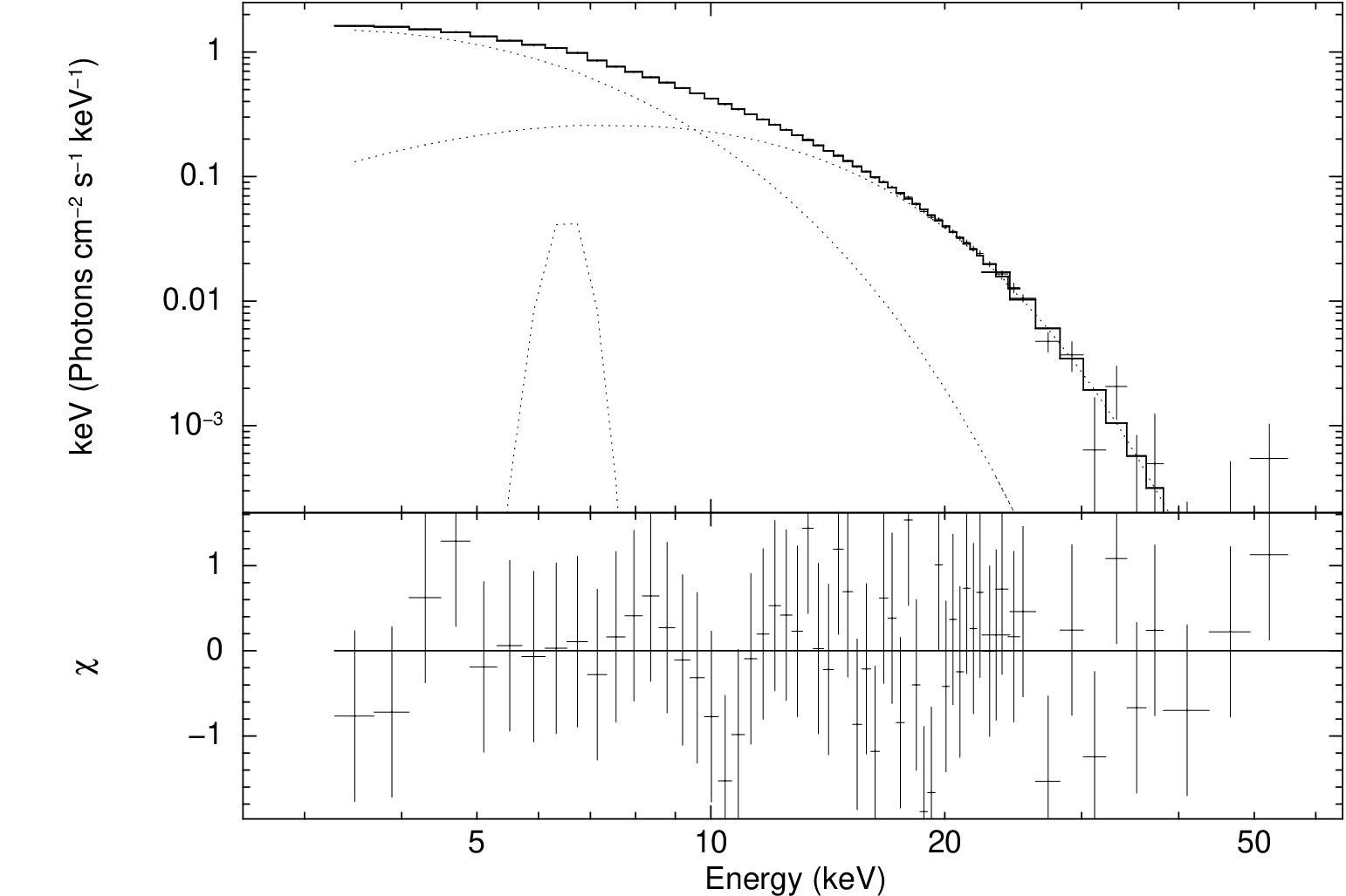}

\caption{The unfolded spectral of upper HB in Interval B (Boxes 1-7, from left-to-right and top-to-bottom) fitted by Model I.    \label{fig6-3}}
\end{figure}

We implemented a detailed temporal and spectral analysis of the HB/NB vertex and the NB/FB vertex in Figure~\ref{fig4}. The HB/NB vertex and the NB/FB vertex trace two distinct lines in HID. Their energy spectra can be well fitted by\textbf{ wabs(diskbb+bbodyrad+gauss)} without Comptonization component. The inner radius and the apparent radius of NS can be deduced from normalizations of \textbf{diskbb} and \textbf{bbodyrad} respectively. The HB/NB vertices display ${\dot{M}_{\rm{Disk}}\propto R^{2.9}_{\rm{in}}}$ relation, where the NB/FB vertices show a flatter trend, that is, ${\dot{M}_{\rm{Disk}}\propto R^{1.7}_{\rm{in}}}$.
The HB/NB vertices have a slightly steeper ${\dot{M}_{\rm{Disk}}-R_{\rm{in}}}$ relation
compared with the NB/FB vertices, which indicate the accretion disk evolves from a slim disk to a standard disk \citep{lin12} because of decreasing accretion rate. Since the disk accretion rate is proportional to $L_{\rm MCD}R_{\rm in}$, the disk accretion rate decreases with the luminosity decaying for the HB/NB vertices. The HB/NB vertices of Stages I and II present near 50 Hz HBOs with luminosity variation. We firstly conclude that the frequency of HBO is independent of the accretion rate or the source intensity, which might be applied in other Z sources. This conclusion is obtained without prior model restriction.

The radius-mass relation of NS can be constrained from Type I x-ray burst\citep{sal12}, kHz QPO\citep{zhang09}, X-ray emission of quiescent LMXBs in globular clusters\citep{gui13,lat13}, emission lines\citep{cot02} and so on. In \object{XTE J1701-462}, the inner disk radius depends on the inclination angle of the binary system. Although, the inclination angle of \object{XTE J1701-462} has not measured directly until now, we still plot the inner radius versus inclination angle in Figure~\ref{fig11}. When the accretion disk moves closest to the NS, the deduced inner radius must larger than the NS radius. This figure shows the upper limit of NS radius. For 70 degree of inclination angle, the upper limit of NS radius is $11.7\pm2.3$ km, which covers the radius confidence intervals from Type I x-ray burst \citep{sal12} and quiescent LMXBs\citep{gui13}.

The W-K relation between break frequency and LF-QPO was initially found in the BHs and atoll sources, as well as  in Z sources but with a slightly shift \citep{wij99}. It spanned nearly three orders of magnitude in frequency. Recently, the accreting millisecond pulsars were also found following this correlation \citep{bel02,str05}.  The W-K relation implies that the similar physical mechanism likely happens in both BH-LMXBs and NS-LMXBs. The W-K relation in BH-LMXBs, NS-LMXBs and accreting millisecond pulsars demonstrate that the intrinsic magnetic field and the solid surface are probably not the reason for it. The W-K relation in \object{XTE J1701-462} may constrain the potential physics process. \object{XTE J1701-462}, in the Cyg-like phase, Sco-like phase and HB/NB vertex, follows the W-K relation without shifted like other typical Z sources or dropped at high break frequency like atoll sources. More interesting, the low characteristic frequencies locate in the atoll sources region. As mentioned above,
the accretion rate in HB/NB vertex decreases towards lower source intensity in Figure~\ref{fig4}. The well correlated W-K relation of HB/NB vertex in \object{XTE J1701-462} indicates that the characteristic frequencies on the HB are also not determined by the accretion rate. Some other mechanisms, e.g., the propagation of accretion flow instabilities \citep{str05}, are presumably responsible for it.

The origin of LF-QPO is still puzzling. The Lense-Thirring precession has been suggested as the explanation of the HBO in Z sources, the Low frequency QPO in atoll sources and the typ-C QPO in BH LMXBs. The $\sim$1 Hz QPO in the dipping/eclipsing NS-LMXB was considered as a likely signature of Lense-Thirring \citep{hom12}. The Lense-Thirring precession predicted a $\nu\propto R^{-3}$ relation \citet{ste98} arose in \object{GX 17+2} when the Comptonized component was fitted by nthCompt \citep{lin12}. We investigated the $\nu_{\rm{HBO}}-R_{\rm{in}}$ relation in Z stage of \object{XTE J1701-462}. Whether the Comptonized component fitted by the cutoff power law or nthCompt, the nearly identical $\nu_{\rm{HBO}}-R_{\rm{in}}$ relations were obtained in Figure~\ref{fig12}. That is the $\nu_{\rm{HBO}}$ positively correlate with the inner disk radius $R_{\rm{in}}$. The possible reason of dissimilar with \object{GX 17+2} is that the relative weak Comptonized component in \object{XTE J1701-462} unlikely affected the
fitting results of inner disk radius.

In BH LMXBs, the energy dependence of Type C QPO shows complicated correlations, i.e., \object{GRS 1915+105}\citep{qu10}, \object{XTE J1550}\citep{li13a} and \object{H1743-322}\citep{li13b}.  In Figure~\ref{fig15}, we studied the energy dependence of HBO in \object{XTE J1701-462}. On the HB, the frequency of HPO is anticorrelated with photon energy, which is similar to the BH LMXB GRS 1915+105 with LF-QPOs $<3$~Hz \citep{qu10}. In \object{XTE J1701-462}, the HBO frequency is $\sim$10-55 Hz. We only obtained the HBO frequency-energy relation with HBO  $<30$ Hz. Because at higher HBO frequency, we could not have high S/N PDS to constrain the HBO properly in some sub-bands. If we have high S/N PDSs above 30 Hz in some other NS LMXBs, more complicated HBO frequency-energy relations could probably appear. Although the mechanism of type C QPO frequency-energy relation is still not known yet, the difference between the BH LMXBs and NS LMXBs may due to two reasons. The first is the affection of emission from NS
surface which can cool down the corona comparable to the disk emission. The second is the central mass dependence of characteristic frequencies in the accretion flow as suggested by \citet{sun00}.

In the right panels of Figure~\ref{fig15}, the rms of HBO increases with photon energy, from $\sim 5\%$ in 2-4.5 keV to $\sim 15\%$ in 9.4$-$14.8 keV, which indicates that the higher energy photon provide more contribution of HBO. From our fitting results, the temperature of black body from NS surface is about $2.6~ \rm keV$ and the temperature of multi-color disk is about $1.7~ \rm keV$, which are much less than $\sim 10 ~\rm keV$ of the maximum rms of HBO occurring. The amplitude of HBOs as well as hard color dropped from upturn to HB \citep{li13}, meanwhile, the strengh of the Comptonization emission also became fainter from our spectral fitting results. On the NB, the inverse Compton process turned into undetectable, while HBO disappeared simultaneously. The HB/NB vertex could be considered as a spectral state transition zone, where the spectral switched from a non-thermal emission contained state to a thermal emission completely dominated state. This feature of spectral transition also exhibited 
in GX 17+2. The spectral state change of GX 17+2 occurred on the NB/FB vertex, moreover, the HBO exhibit on both HB and NB. On FB, the Comptonization component became detectable marginally and HBOs disappeared similar as in the case of XTE J1701-462 \citep{lin12}. The non-thermal emission connection of HBO were researched in Cyg X-2 and 4U 1728-34, which the photon indices of Cyg X-2 and 4U 1728-34 increased with HBO \citep{tit05,tit07}. If the normalization of Comptoniztion component did not change significantly, the HBO in Cyg X-2 and 4U 1728-34 were also anti-correlated with Comptonization emission strength.  Based on the above observational evidences, we interpret that the HBO possibly arise in the corona, where the HBO generate in the process of Comptonization emission \citep{yan13}. The frequency and rms of HBO maybe associate with the scale as well as the optical depth of corona region, respectively. This kind of dependence needs further observation constraining.


\section{Conclusion}

We study the HBOs of \object{XTE J1701-462} with decreasing accretion rate in the HB/NB vertex. We conclude that the HBO in \object{XTE J1701-462}, unlike the type C LF-QPOs in BH LMXBs, is independent with accretion rate. In other word, the ascending of HBO is not representing the accretion rate increasing. We also find that the anti-correlation relation between the frequency of HBO and its centroid energy. The energy dependence of HBO implies that the higher QPO produced a disk moves away from the NS. Both the Cyg-like phase and the Sco-like phase follow the W-K relation which are presumably caused by the same mechanism. The derived $R_{\rm in}-\nu_{\rm HBO}$ relations contradict the prediction of Lense-Thirring precession. We conclude that the HBO may origin from the corona.

\acknowledgments

We appreciate the referee's insightful suggestions and comments which improve our works distinctly. Z.S. Li thank S.N. Zhang for critical remarks, Z.B. Li for helpful discussions. This work is supported by the National Natural Science Foundation of China (10778716), the National Basic Research program of China 973 Program 2009CB824800, the National Natural Science Foundation of China (11173024) and the Fundamental Research Funds for the Central Universities. This research has made use of data obtained from the High Energy Astrophysics Science Archive Research Center (HEASARC), provided by NASA's Goddard Space Flight Center.

\clearpage
\begin{deluxetable}{cccccc}
\tabletypesize{\scriptsize}
\tablecaption{The data selection for the CDs/HIDs of Intervals A and B.\label{tbl-1}}
\tablewidth{0pt}
\tablehead{
\colhead{Interval}    & \colhead{Begin of date } & \colhead{Begin of Obs.} &
\colhead{End of date} & \colhead{End of Obs.}   & \colhead{Source Type}   \\
\colhead{}            & \colhead{(DD/MM/YY)}     &  \colhead{}   &
\colhead{(DD/MM/YY)}            & \colhead{}    &  \colhead{}            }
\startdata
  I &   22/01/06 & 91106-01-07-00 &   29/01/06 & 91106-02-02-07 &   Cyg-like \\

 II &   17/02/06 & 91442-01-07-02 &   26/02/06 & 91442-01-03-05 &   Sco-like \\

\enddata
\end{deluxetable}

\begin{deluxetable}{ccc}
\tabletypesize{\scriptsize}
\tablecaption{Six sub-bands used in our analysis. Column 1 is the PCA channels and Column 2 is the corresponding energy range. The centroid energy is shown in Column 3. \label{tbl-2}}
\tablewidth{0pt}
\tablehead{
\colhead{PCA channel}    & \colhead{Energy range (keV) } & \colhead{Centroid energy (keV)}   }
\startdata
0-10       &        2-4.5         &        3.25   \\
11-13      &        4.5-5.7       &        5.1    \\
14-16      &        5.7-6.9       &        6.3    \\
17-22      &        6.9-9.4       &        8.2    \\
23-35      &        9.4-14.8      &        12.1   \\
36-149     &        14.8-65       &        16.5   \\

\enddata
\end{deluxetable}

\clearpage

\begin{deluxetable}{cllllllllc}
\tabletypesize{\scriptsize}
\tablecaption{The spectral fitting results  of Intervals A and B by using Model I. \label{tbl-3}}
\tablewidth{0pt}
\tablehead{

\colhead{Box} & \colhead{$L_{\rm Tot}$}  & \colhead{$R_{\rm NS}$ } & \colhead{$kT_{\rm BB}$} & \colhead{$L_{\rm BB}$ } & \colhead{$R_{\rm in}$  } & \colhead{$kT_{\rm Disk}$} & \colhead{$L_{\rm Disk}$}& \colhead{ $L_{\rm CBPL}$ } & \colhead{$\chi^2/{\rm d.o.f}$}\\

\colhead{}& \colhead{($10^{38}$erg/s)}& \colhead{(km)}& \colhead{(keV) }& \colhead{($10^{38}$erg/s)}& \colhead{(km)}&\colhead{(keV)}& \colhead{($10^{38}$erg/s)}& \colhead{($10^{38}$erg/s)}& \colhead{}
}
\startdata
Interval A &&&&&&&&&\\
\hline
1  &  $1.05\pm0.31  $  &  $ 2.79\pm0.66 $   & $  2.54\pm0.10  $&  $ 0.40\pm0.14 $ &   $ 11.51_{ - 2.35}^{ + 2.41}$ & $ 1.73\pm0.06$ &  $0.56 \pm 0.18 $  &   $ 0.096_{  -0.060 }^{  +0.062 } $  & 51.82/52     \\
2  &  $1.06\pm0.32  $  &  $ 2.94\pm0.69 $   & $  2.51\pm0.10  $&  $ 0.43\pm0.15 $ &   $ 12.42_{ - 2.56}^{ + 2.63}$ & $ 1.67\pm0.06$ &  $0.55 \pm 0.18 $  &   $ 0.084_{  -0.060 }^{  +0.061 } $  & 54.04/52     \\
3  &  $1.08\pm0.32  $  &  $ 2.87\pm0.65 $   & $  2.52\pm0.09  $&  $ 0.42\pm0.14 $ &   $ 14.06_{ - 2.78}^{ + 2.88}$ & $ 1.63\pm0.05$ &  $0.62 \pm 0.19 $  &   $ 0.035_{  -0.033 }^{  +0.045 } $  & 41.82/53     \\
4  &  $1.13\pm0.34  $  &  $ 2.96\pm0.62 $   & $  2.48\pm0.07  $&  $ 0.41\pm0.13 $ &   $ 16.50_{ - 3.11}^{ + 3.19}$ & $ 1.56\pm0.04$ &  $0.69 \pm 0.21 $  &   $ 0.018_{  -0.017 }^{  +0.029 } $  & 35.38/52     \\
5  &  $1.20\pm0.36  $  &  $ 3.16\pm0.68 $   & $  2.42\pm0.07  $&  $ 0.42\pm0.14 $ &   $ 17.97_{ - 3.41}^{ + 3.52}$ & $ 1.53\pm0.04$ &  $0.74 \pm 0.22 $  &   $ 0.029_{  -0.026 }^{  +0.031 } $  & 23.00/52     \\
6  &  $1.38\pm0.41  $  &  $ 3.35\pm0.70 $   & $  2.41\pm0.06  $&  $ 0.46\pm0.15 $ &   $ 19.43_{ - 3.63}^{ + 3.72}$ & $ 1.54\pm0.03$ &  $0.88 \pm 0.27 $  &   $ 0.030_{  -0.025 }^{  +0.028 } $  & 43.19/52     \\
7  &  $1.52\pm0.46  $  &  $ 3.34\pm0.73 $   & $  2.44\pm0.06  $&  $ 0.49\pm0.16 $ &   $ 19.69_{ - 3.68}^{ + 3.78}$ & $ 1.57\pm0.03$ &  $1.00 \pm 0.30 $  &   $ 0.023_{  -0.021 }^{  +0.028 } $  & 36.05/52     \\
8  &  $1.60\pm0.48  $  &  $ 3.32\pm0.67 $   & $  2.47\pm0.09  $&  $ 0.51\pm0.17 $ &   $ 19.87_{ - 3.64}^{ + 3.73}$ & $ 1.59\pm0.04$ &  $1.08 \pm 0.33 $  &   $ 0.003_{  -0.002 }^{  +0.002 } $  & 41.71/52     \\
9  &  $1.69\pm0.51  $  &  $ 3.44\pm0.64 $   & $  2.46\pm0.02  $&  $ 0.53\pm0.18 $ &   $ 20.38_{ - 3.59}^{ + 3.59}$ & $ 1.59\pm0.02$ &  $1.15 \pm 0.35 $  &   $ 0.001_{  -0.001 }^{  +0.004 } $  & 27.20/52     \\
10 &  $1.77\pm0.53  $  &  $ 3.65\pm0.81 $   & $  2.40\pm0.08  $&  $ 0.54\pm0.19 $ &   $ 20.49_{ - 3.80}^{ + 3.95}$ & $ 1.59\pm0.04$ &  $1.18 \pm 0.36 $  &   $ 0.049_{  -0.035 }^{  +0.037 } $  & 32.34/52     \\
11 &  $1.86\pm0.56  $  &  $ 3.80\pm0.83 $   & $  2.38\pm0.05  $&  $ 0.55\pm0.20 $ &   $ 20.87_{ - 3.86}^{ + 3.96}$ & $ 1.60\pm0.03$ &  $1.24 \pm 0.38 $  &   $ 0.037_{  -0.023 }^{  +0.024 } $  & 58.19/52     \\
12 &  $2.06\pm0.62  $  &  $ 3.72\pm0.85 $   & $  2.40\pm0.06  $&  $ 0.55\pm0.21 $ &   $ 21.17_{ - 3.91}^{ + 3.66}$ & $ 1.64\pm0.03$ &  $1.45 \pm 0.44 $  &   $ 0.028_{  -0.020 }^{  +0.022 } $  & 50.64/52     \\
13 &  $2.18\pm0.66  $  &  $ 3.91\pm0.91 $   & $  2.32\pm0.05  $&  $ 0.52\pm0.20 $ &   $ 22.74_{ - 4.16}^{ + 4.27}$ & $ 1.62\pm0.03$ &  $1.56 \pm 0.47 $  &   $ 0.046_{  -0.025 }^{  +0.026 } $  & 45.24/48     \\
\hline
Interval B &&&&&&&&&\\
\hline
1  &  $ 0.97 \pm  0.29  $  &  $      2.77  \pm0.59 $   & $   2.56  \pm 0.06  $&  $  0.40 \pm  0.13   $ &   $    13.08 _{ -   2.65}^{ +   2.75    }$ & $  1.59  \pm 0.05 $ &  $  0.47 \pm 0.15 $  &   $   0.070  _{  -  0.047  }^{  +   0.051} $  &   35.92/52    \\
2  &  $ 1.05 \pm  0.31  $  &  $      2.92  \pm0.59 $   & $   2.51  \pm 0.05  $&  $  0.41 \pm  0.13   $ &   $    14.10 _{ -   2.72}^{ +   2.80    }$ & $  1.58  \pm 0.04 $ &  $  0.54 \pm 0.16 $  &   $   0.076  _{  -  0.039  }^{  +   0.040} $  &   44.40/52    \\
3  &  $ 1.20 \pm  0.36  $  &  $      2.92  \pm0.61 $   & $   2.55  \pm 0.05  $&  $  0.44 \pm  0.15   $ &   $    14.87 _{ -   2.80}^{ +   2.87    }$ & $  1.64  \pm 0.04 $ &  $  0.71 \pm 0.22 $  &   $   0.025  _{  -  0.021  }^{  +   0.022} $  &   37.11/52    \\
4  &  $ 1.26 \pm  0.38  $  &  $      2.95  \pm0.62 $   & $   2.54  \pm 0.05  $&  $  0.45 \pm  0.15   $ &   $    15.04 _{ -   2.82}^{ +   2.89    }$ & $  1.66  \pm 0.03 $ &  $  0.77 \pm 0.23 $  &   $   0.027  _{  -  0.020  }^{  +   0.021} $  &   30.84/52    \\
5  &  $ 1.32 \pm  0.40  $  &  $      2.94  \pm0.64 $   & $   2.54  \pm 0.06  $&  $  0.44 \pm  0.15   $ &   $    15.12 _{ -   2.83}^{ +   2.87    }$ & $  1.68  \pm 0.04 $ &  $  0.83 \pm 0.25 $  &   $   0.027  _{  -  0.021  }^{  +   0.022} $  &   41.46/52    \\
6  &  $ 1.40 \pm  0.42  $  &  $      2.89  \pm0.62 $   & $   2.57  \pm 0.06  $&  $  0.45 \pm  0.16   $ &   $    15.85 _{ -   2.92}^{ +   2.99    }$ & $  1.69  \pm 0.03 $ &  $  0.93 \pm 0.28 $  &   $   0.007  _{  -  0.005  }^{  +   0.014} $  &   29.27/52    \\
7  &  $ 1.44 \pm  0.43  $  &  $      2.80  \pm0.44 $   & $   2.60  \pm 0.06  $&  $  0.45 \pm  0.15   $ &   $    16.19 _{ -   2.84}^{ +   1.71    }$ & $  1.70  \pm 0.06 $ &  $  0.99 \pm 0.30 $  &   $   - $  &   35.51/52    \\
\enddata
\tablenotemark{-}{Each error is quoted 90\% confidence level for the fitting parameters. In Box 7 of Interval B, the CBPL component is too faint to be constrained well.}
\end{deluxetable}

\begin{deluxetable}{cllllllllc}
\tabletypesize{\scriptsize}
\tablecaption{The spectral fitting results of Intervals A and B by using Model II . \label{tbl-4}}
\tablewidth{0pt}
\tablehead{

\colhead{Box} & \colhead{$L_{\rm Tot}$}  & \colhead{$R_{\rm NS}$ } & \colhead{$kT_{\rm BB}$} & \colhead{$L_{\rm BB}$ } & \colhead{$R_{\rm in}$  } & \colhead{$kT_{\rm Disk}$} & \colhead{$L_{\rm Disk}$}& \colhead{ $L_{\rm nthComp}$ } &\colhead{$\chi^2/{\rm d.o.f}$}\\

\colhead{}& \colhead{($10^{38}$erg/s)}& \colhead{(km)}& \colhead{(keV) }& \colhead{($10^{38}$erg/s)}& \colhead{(km)}&\colhead{(keV)}& \colhead{($10^{38}$erg/s)}& \colhead{($10^{38}$erg/s)}& \colhead{}

}
\startdata
Interval A &&&&&&&&&\\
\hline
1  & $ 1.05  \pm  0.31   $   & $   2.93  \pm 0.66   $ & $  2.4561 \pm  0.10  $ & $ 0.37  \pm  0.13   $ &   $ 12.45  _{- 2.35  }^{+ 2.41 } $  & $ 1.62   \pm  0.09  $ & $ 0.47  \pm  0.15  $  & $   0.177 _{-   0.091 }^{+   0.090 } $   &   55.67/52    \\
2  & $ 1.06  \pm  0.32   $   & $   3.00  \pm 0.69   $ & $  2.4637 \pm  0.10  $ & $ 0.40  \pm  0.14   $ &   $ 13.17  _{- 2.56  }^{+ 2.63 } $  & $ 1.60   \pm  0.09  $ & $ 0.49  \pm  0.17  $  & $   0.135 _{-   0.086 }^{+   0.085 } $   &   54.44/52    \\
3  & $ 1.07  \pm  0.32   $   & $   3.06  \pm 0.65   $ & $  2.4187 \pm  0.09  $ & $ 0.38  \pm  0.13   $ &   $ 14.73  _{- 2.78  }^{+ 2.88 } $  & $ 1.55   \pm  0.07  $ & $ 0.52  \pm  0.17  $  & $   0.130 _{-   0.078 }^{+   0.077 } $   &   55.14/53    \\
4  & $ 1.13  \pm  0.34   $   & $   3.06  \pm 0.62   $ & $  2.4294 \pm  0.07  $ & $ 0.39  \pm  0.13   $ &   $ 16.91  _{- 3.11  }^{+ 3.19 } $  & $ 1.53   \pm  0.05  $ & $ 0.64  \pm  0.20  $  & $   0.058 _{-   0.045 }^{+   0.047 } $   &   38.60/52    \\
5  & $ 1.20  \pm  0.36   $   & $   3.23  \pm 0.68   $ & $  2.3853 \pm  0.07  $ & $ 0.41  \pm  0.14   $ &   $ 18.33  _{- 3.41  }^{+ 3.52 } $  & $ 1.50   \pm  0.05  $ & $ 0.70  \pm  0.22  $  & $   0.068 _{-   0.051 }^{+   0.052 } $   &   25.63/52    \\
6  & $ 1.38  \pm  0.41   $   & $   3.48  \pm 0.70   $ & $  2.3544 \pm  0.07  $ & $ 0.44  \pm  0.15   $ &   $ 19.96  _{- 3.63  }^{+ 3.72 } $  & $ 1.50   \pm  0.04  $ & $ 0.82  \pm  0.26  $  & $   0.079 _{-   0.051 }^{+   0.052 } $   &   42.23/52    \\
7  & $ 1.53  \pm  0.46   $   & $   3.50  \pm 0.73   $ & $  2.3837 \pm  0.07  $ & $ 0.47  \pm  0.16   $ &   $ 20.14  _{- 3.68  }^{+ 3.78 } $  & $ 1.54   \pm  0.04  $ & $ 0.94  \pm  0.29  $  & $   0.076 _{-   0.052 }^{+   0.054 } $   &   57.95/52    \\
8  & $ 1.60  \pm  0.48   $   & $   3.41  \pm 0.67   $ & $  2.4459 \pm  0.07  $ & $ 0.50  \pm  0.17   $ &   $ 20.10  _{- 3.64  }^{+ 3.73 } $  & $ 1.57   \pm  0.04  $ & $ 1.05  \pm  0.33  $  & $   0.018 _{-   0.008 }^{+   0.032 } $   &   28.87/52    \\
9  & $ 1.69  \pm  0.51   $   & $   3.47  \pm 0.64   $ & $  2.4554 \pm  0.07  $ & $ 0.53  \pm  0.18   $ &   $ 20.45  _{- 3.59  }^{+ 3.59 } $  & $ 1.59   \pm  0.04  $ & $ 1.13  \pm  0.36  $  & $   0.014 _{-   0.014 }^{+   0.030 } $   &   38.49/52    \\
10 & $ 1.77  \pm  0.53   $   & $   3.83  \pm 0.81   $ & $  2.3433 \pm  0.07  $ & $ 0.52  \pm  0.18   $ &   $ 21.08  _{- 3.80  }^{+ 3.95 } $  & $ 1.56   \pm  0.04  $ & $ 1.10  \pm  0.34  $  & $   0.106 _{-   0.060 }^{+   0.060 } $   &   33.28/52    \\
11 & $ 1.86  \pm  0.56   $   & $   3.80  \pm 0.83   $ & $  2.3683 \pm  0.07  $ & $ 0.53  \pm  0.18   $ &   $ 21.06  _{- 3.86  }^{+ 3.96 } $  & $ 1.59   \pm  0.04  $ & $ 1.21  \pm  0.37  $  & $   0.079 _{-   0.050 }^{+   0.052 } $   &   58.64/52    \\
12 & $ 2.06  \pm  0.62   $   & $   3.88  \pm 0.85   $ & $  2.3494 \pm  0.08  $ & $ 0.54  \pm  0.20   $ &   $ 21.60  _{- 3.91  }^{+ 3.66 } $  & $ 1.62   \pm  0.04  $ & $ 1.39  \pm  0.43  $  & $   0.086 _{-   0.057 }^{+   0.058 } $   &   50.93/52    \\
13 & $ 2.17  \pm  0.65   $   & $   4.04  \pm 0.91   $ & $  2.2803 \pm  0.07  $ & $ 0.49  \pm  0.18   $ &   $ 23.13  _{- 4.16  }^{+ 4.27 } $  & $ 1.60   \pm  0.04  $ & $ 1.51  \pm  0.46  $  & $   0.107 _{-   0.055 }^{+   0.056 } $   &   50.70/48    \\

Interval B &&&&&&&&&\\
\hline
1 & $ 0.97  \pm  0.29  $  & $ 2.85 \pm  0.19  $ & $ 2.44\pm   0.09   $ & $ 0.35 \pm   0.12  $ &$ 14.20  _{-  0.86 }^{+   1.06} $& $  1.46 \pm   0.08 $& $ 0.36  \pm  0.13  $ & $ 0.236  _{-  0.131}^{+    0.126} $ &  37.17/52 \\
2 & $ 1.05  \pm  0.31  $  & $ 2.95 \pm  0.18  $ & $ 2.44\pm   0.07   $ & $ 0.37 \pm   0.12  $ &$ 14.90  _{-  0.73 }^{+   0.87} $& $  1.50 \pm   0.05 $& $ 0.47  \pm  0.15  $ & $ 0.190  _{-  0.092}^{+    0.091} $ &  56.23/52 \\
3 & $ 1.20  \pm  0.36  $  & $ 2.96 \pm  0.20  $ & $ 2.51\pm   0.08   $ & $ 0.42 \pm   0.14  $ &$ 15.22  _{-  0.68 }^{+   0.81} $& $  1.60 \pm   0.05 $& $ 0.67  \pm  0.22  $ & $ 0.078  _{-  0.061}^{+    0.064} $ &  37.67/52 \\
4 & $ 1.26  \pm  0.38  $  & $ 2.96 \pm  0.21  $ & $ 2.52\pm   0.08   $ & $ 0.44 \pm   0.15  $ &$ 15.26  _{-  0.65 }^{+   0.76} $& $  1.64 \pm   0.05 $& $ 0.78  \pm  0.24  $ & $ 0.063  _{-  0.051}^{+    0.055} $ &  33.10/52 \\
5 & $ 1.32  \pm  0.40  $  & $ 3.03 \pm  0.25  $ & $ 2.49\pm   0.09   $ & $ 0.43 \pm   0.15  $ &$ 15.52  _{-  0.70 }^{+   0.84} $& $  1.64 \pm   0.06 $& $ 0.77  \pm  0.25  $ & $ 0.089  _{-  0.062}^{+    0.064} $ &  43.26/52 \\
6 & $ 1.40  \pm  0.42  $  & $ 2.97 \pm  0.24  $ & $ 2.53\pm   0.09   $ & $ 0.44 \pm   0.15  $ &$ 16.07  _{-  0.67 }^{+   0.81} $& $  1.67 \pm   0.05 $& $ 0.89  \pm  0.28  $ & $ 0.031  _{-  0.031}^{+    0.045} $ &  29.02/52  \\
7 & $ 1.44  \pm  0.43  $  & $ 2.79 \pm  0.10  $ & $ 2.60\pm   0.09   $ & $ 0.45 \pm   0.16  $ &$ 15.92  _{-  0.37 }^{+   0.63} $& $  1.71 \pm   0.05 $& $ 0.98  \pm  0.31  $ & $ - $ &  35.53/52 \\

\enddata
\end{deluxetable}

%

\begin{deluxetable}{llll}
\tabletypesize{\scriptsize}
\tablecaption{Results of timing analysis of Intervals A and B.  \label{tbl-5}}
\tablewidth{0pt}
\tablehead{
 \colhead{ $\nu_{\rm b}$ (Hz)}  &\colhead{$\nu_{\rm HBO}$  (Hz)}&\colhead{ $\Delta $ (Hz) } & \colhead{$\nu_{\rm c}$ (Hz)}
}
\startdata
Interval A  & & & \\
\hline
$3.19    \pm    0.43 $ &   $  12.16  \pm  0.05   $ & $ 3.91 \pm   0.23 $ &   $     12.18  \pm  0.04  $  \\
$2.31    \pm    0.16 $ &   $  12.63  \pm  0.29   $ & $ 1.30 \pm   0.37 $ &   $     12.65  \pm  0.28  $  \\
$3.17    \pm    0.33 $ &   $  17.91  \pm  0.25   $ & $ 10.1 \pm   0.75 $ &   $     18.61  \pm  0.19  $  \\
$4.30    \pm    0.24 $ &   $  23.83  \pm  0.14   $ & $ 7.37 \pm   0.37 $ &   $     24.11  \pm  0.19  $  \\
$4.99    \pm    0.19 $ &   $  28.06  \pm  0.13   $ & $ 5.34 \pm   0.30 $ &   $     28.19  \pm  0.08  $  \\
$5.72    \pm    0.10 $ &   $  31.92  \pm  0.26   $ & $ 9.29 \pm   0.60 $ &   $     32.26  \pm  0.09  $  \\
$6.57    \pm    0.13 $ &   $  37.59  \pm  0.24   $ & $ 8.25 \pm   0.68 $ &   $     37.82  \pm  0.18  $  \\
$6.87    \pm    0.15 $ &   $  42.08  \pm  0.32   $ & $ 8.96 \pm   0.96 $ &   $     42.32  \pm  0.24  $  \\
$7.64    \pm    0.28 $ &   $  43.71  \pm  0.65   $ & $ 10.2 \pm   1.46 $ &   $     44.01  \pm  0.36  $  \\
$8.37    \pm    0.25 $ &   $  46.26  \pm  0.85   $ & $ 9.31 \pm   1.76 $ &   $     46.49  \pm  0.41  $  \\
$8.83    \pm    0.44 $ &   $  45.43  \pm  0.88   $ & $ 8.81 \pm   1.98 $ &   $     45.64  \pm  0.43  $  \\
$9.01    \pm    0.33 $ &   $  48.18  \pm  0.29   $ & $ 13.3 \pm   3.62 $ &   $     48.65  \pm  0.73  $  \\
$7.77    \pm    0.56 $ &   $  52.13  \pm  0.75   $ & $ 16.7 \pm   2.55 $ &   $     52.79  \pm  0.58  $  \\
\hline
Interval B        &                         &                         &                \\
\hline
$6.69    \pm    0.48 $ &   $  27.85  \pm  0.16   $ & $ 5.5 \pm    0.45 $ &   $     27.99  \pm  0.11  $  \\
$6.72    \pm    0.05 $ &   $  34.05  \pm  0.19   $ & $ 5.6 \pm    0.57 $ &   $     34.17  \pm  0.13  $  \\
$8.37    \pm    0.43 $ &   $  43.75  \pm  0.45   $ & $ 3.4 \pm    1.20 $ &   $     43.79  \pm  0.26  $  \\
$9.73    \pm    0.81 $ &   $  46.98  \pm  1.34   $ & $ 5.4 \pm    2.40 $ &   $     47.05  \pm  0.65  $  \\
$9.23    \pm    0.51 $ &   $  49.34  \pm  1.31   $ & $ 5.3 \pm    2.88 $ &   $     49.41  \pm  0.67  $  \\
$9.93    \pm    0.73 $ &   $  53.62  \pm  1.03   $ & $ 5.7 \pm    2.50 $ &   $     53.70  \pm  0.54  $  \\
$8.71    \pm    0.72 $ &   $  58.30  \pm  1.14   $ & $ 3.9 \pm    2.97 $ &   $     58.33  \pm  0.54  $  \\

\enddata
\tablenotemark{-}{$\nu_b$,  and $\nu_c$ are the break frequency and centroid frequency in $\nu P_{\nu}$ presentation of PDS, respectively. $\nu_c=\sqrt{(\nu_{\rm HBO})^2+({\Delta/2})^2}$, where, $\nu_{\rm HBO}$ is the HBO and $\Delta$ is the Full-Width-Half-Maximum of HBO.}
\end{deluxetable}
\clearpage

\begin{deluxetable}{cccc}
\tabletypesize{\scriptsize}
\tablecaption{Results of timing analysis of HB/NB vertices in Stages I and II.  \label{tbl-5}}
\tablewidth{0pt}
\tablehead{
 \colhead{ $\nu_{\rm b}$ (Hz)}  &\colhead{$\nu_{\rm HBO}$  (Hz)}&\colhead{ $\Delta $ (Hz) } & \colhead{$\nu_{\rm c}$ (Hz)}
}
\startdata
Stage I  & & & \\
\hline
$  8.7    \pm     2.1 $ &$ 53.0  \pm  2.6  $  &    $ 10.1  \pm 5.3  $  &   $ 53.2     \pm     1.5  $ \\
$ 10.1    \pm     2.1 $ &$ 50.9  \pm  1.6  $  &    $ 6.3   \pm 4.0  $  &   $ 51.0     \pm     0.9  $ \\
$  9.3    \pm     1.9 $ &$ 57.2  \pm  1.8  $  &    $ 5.3   \pm 3.6  $  &   $ 57.3     \pm     0.9  $ \\
$ 15.7    \pm     5.0 $ &$ 51.3  \pm  1.4  $  &    $ 6.4   \pm 3.5  $  &   $ 51.4     \pm     3.6  $ \\
$  7.7    \pm     2.3 $ &$ 50.0  \pm  4.1  $  &    $ 5.3   \pm 6.4  $  &   $ 50.1     \pm     5.0  $ \\
$ 11.1    \pm     1.9 $ &$ 50.1  \pm  2.0  $  &    $ 5.5   \pm 4.6  $  &   $ 50.2     \pm     2.4  $ \\
$ 18.2    \pm     2.7 $ &$ 51.0  \pm  1.5  $  &    $ 9.2   \pm 3.6  $  &   $ 51.3     \pm     1.0  $ \\
$ 11.6    \pm     1.5 $ &$ 49.4  \pm  2.5  $  &    $ 11.9  \pm 7.1  $  &   $ 49.7     \pm     1.3  $ \\
$  8.9    \pm     1.3 $ &$ 51.8  \pm  2.5  $  &    $ 14.2  \pm 5.6  $  &   $ 52.3     \pm     1.4  $ \\

\hline
Stage II        &                         &                         &                \\
\hline
$ 12.7   \pm   2.4 $    &$48.3   \pm  3.1  $ &  $ 1.8  \pm  0.6  $ &   $   48.4    \pm    1.5 $ \\
$  8.1   \pm   4.4 $    &$44.9   \pm  5.5  $ &  $ 5.8  \pm  1.6  $ &   $   45.0    \pm    0.8 $ \\
$ 10.9   \pm   1.1 $    &$50.5   \pm  1.4  $ &  $ 10.0 \pm  0.6  $ &   $   50.7    \pm    0.9 $ \\
$ 12.4   \pm   1.0 $    &$52.6   \pm  2.2  $ &  $ 13.7 \pm  2.4  $ &   $   53.0    \pm    1.5 $ \\
$ 10.6   \pm   1.2 $    &$52.5   \pm  2.3  $ &  $ 7.2  \pm  2.5  $ &   $   52.6    \pm    3.9 $ \\
$ 11.5   \pm   3.2 $    &$44.1   \pm  0.6  $ &  $ 3.7  \pm  0.4  $ &   $   44.2    \pm    0.4 $ \\
$ 13.2   \pm   2.0 $    &$48.4   \pm  1.3  $ &  $ 4.2  \pm  3.6  $ &   $   48.4    \pm    0.6 $ \\
$ 10.1   \pm   2.2 $    &$49.2   \pm  2.0  $ &  $ 7.7  \pm  5.1  $ &   $   49.3    \pm    1.2 $ \\
$  8.7   \pm   1.8 $    &$46.9   \pm  0.7  $ &  $ 2.9  \pm  2.2  $ &   $   46.9    \pm    0.4 $ \\
$  7.8   \pm   1.0 $    &$43.4   \pm  2.2  $ &  $ 3.9  \pm  1.1  $ &   $   43.5    \pm    1.1 $ \\
$  8.5   \pm   1.0 $    &$51.1   \pm  1.7  $ &  $ 4.9  \pm  2.5  $ &   $   51.2    \pm    0.7 $ \\
$  9.6   \pm   1.1 $    &$47.6   \pm  2.2  $ &  $ 3.3  \pm  1.1  $ &   $   47.7    \pm    0.9 $ \\
\enddata
\end{deluxetable}
\clearpage

\begin{deluxetable}{clllllll}
\tabletypesize{\scriptsize}
\tablecaption{Results of spectral and timing analysis of HB/NB vertex in Stage I. Only data with detected QPO  are listed here.  \label{tbl-6}}
\tablewidth{0pt}
\tablehead{
 \colhead{ $L_{\rm Tot}$ }  &\colhead{$R_{\rm NS}$  }&\colhead{ $kT_{\rm BB}$ }&\colhead{ $L_{\rm BB}$  }&\colhead{ $R_{\rm in}$ }&\colhead{$L_{\rm Disk}$  }&\colhead{ $kT_{\rm Disk }$}&\colhead{Reduced $\chi^2$} \\

 \colhead{($10^{38}$erg/s)}& \colhead{(keV)}& \colhead{($10^{38}$erg/s) }& \colhead{(km)}& \colhead{(keV)}&\colhead{($10^{38}$erg/s)}& \colhead{(Hz)}&\colhead{}
}
\startdata
Stage I & & & & & \\
\hline
$ 2.17  \pm 0.6  $ & $ 3.49 \pm 0.33  $ & $ 2.42 \pm 0.06 $ & $ 0.50 \pm 0.20 $ & $ 21.73 \pm 0.77  $ & $1.67 \pm 0.03 $ & $ 1.65 \pm 0.54 $ & $  49.82/53   $\\
$ 2.16  \pm 0.6  $ & $ 2.93 \pm 0.26  $ & $ 2.54 \pm 0.06 $ & $ 0.43 \pm 0.16 $ & $ 21.25 \pm 0.61  $ & $1.70 \pm 0.02 $ & $ 1.72 \pm 0.55 $ & $  40.19/53   $\\
$ 2.15  \pm 0.6  $ & $ 3.11 \pm 0.27  $ & $ 2.50 \pm 0.06 $ & $ 0.45 \pm 0.17 $ & $ 21.79 \pm 0.66  $ & $1.68 \pm 0.03 $ & $ 1.69 \pm 0.55 $ & $  53.77/53   $\\
$ 2.15  \pm 0.6  $ & $ 2.80 \pm 0.32  $ & $ 2.58 \pm 0.08 $ & $ 0.41 \pm 0.17 $ & $ 20.7  \pm 0.70  $ & $1.72 \pm 0.03 $ & $ 1.73 \pm 0.57 $ & $  57.68/53   $\\
$ 2.15  \pm 0.6  $ & $ 3.33 \pm 0.28  $ & $ 2.46 \pm 0.05 $ & $ 0.48 \pm 0.18 $ & $ 21.61 \pm 0.67  $ & $1.67 \pm 0.03 $ & $ 1.66 \pm 0.54 $ & $  56.00/53   $\\
$ 2.15  \pm 0.6  $ & $ 3.24 \pm 0.27  $ & $ 2.51 \pm 0.05 $ & $ 0.49 \pm 0.18 $ & $ 21.03 \pm 0.65  $ & $1.69 \pm 0.03 $ & $ 1.64 \pm 0.53 $ & $  48.42/53   $\\
$ 2.38  \pm 0.7  $ & $ 3.30 \pm 0.37  $ & $ 2.48 \pm 0.07 $ & $ 0.49 \pm 0.20 $ & $ 22.18 \pm 0.79  $ & $1.70 \pm 0.03 $ & $ 1.88 \pm 0.62 $ & $  58.06/53   $\\
$ 2.38  \pm 0.7  $ & $ 3.54 \pm 0.29  $ & $ 2.45 \pm 0.05 $ & $ 0.54 \pm 0.20 $ & $ 22.71 \pm 0.69  $ & $1.67 \pm 0.03 $ & $ 1.82 \pm 0.59 $ & $  48.40/53   $\\
$ 2.38  \pm 0.7  $ & $ 3.71 \pm 0.33  $ & $ 2.38 \pm 0.05 $ & $ 0.52 \pm 0.20 $ & $ 24.25 \pm 0.78  $ & $1.63 \pm 0.03 $ & $ 1.84 \pm 0.60 $ & $  47.30/53   $\\
\hline
Stage II & & & & & \\
\hline
$ 1.32  \pm 0.39 $ & $ 2.88 \pm 0.2   $ & $ 2.58 \pm 0.05 $ & $ 0.44 \pm 0.16 $ & $ 15.42 \pm 0.63  $ & $1.68 \pm 0.04 $ & $ 0.87 \pm 0.28 $ & $  44.02/53   $\\
$ 1.31  \pm 0.39 $ & $ 2.80 \pm 0.2   $ & $ 2.61 \pm 0.05 $ & $ 0.43 \pm 0.15 $ & $ 15.26 \pm 0.64  $ & $1.69 \pm 0.04 $ & $ 0.86 \pm 0.28 $ & $  38.16/53   $\\
$ 1.35  \pm 0.40 $ & $ 2.78 \pm 0.2   $ & $ 2.61 \pm 0.05 $ & $ 0.43 \pm 0.15 $ & $ 15.28 \pm 0.59  $ & $1.71 \pm 0.04 $ & $ 0.91 \pm 0.30 $ & $  37.12/53   $\\
$ 1.36  \pm 0.41 $ & $ 2.92 \pm 0.2   $ & $ 2.58 \pm 0.05 $ & $ 0.45 \pm 0.16 $ & $ 15.90 \pm 0.65  $ & $1.67 \pm 0.04 $ & $ 0.90 \pm 0.30 $ & $  47.77/53   $\\
$ 1.40  \pm 0.42 $ & $ 2.87 \pm 0.21  $ & $ 2.60 \pm 0.05 $ & $ 0.44 \pm 0.16 $ & $ 15.88 \pm 0.64  $ & $1.69 \pm 0.04 $ & $ 0.94 \pm 0.31 $ & $  33.01/53   $\\
$ 1.19  \pm 0.36 $ & $ 2.75 \pm 0.18  $ & $ 2.62 \pm 0.05 $ & $ 0.42 \pm 0.15 $ & $ 14.59 \pm 0.61  $ & $1.68 \pm 0.04 $ & $ 0.77 \pm 0.25 $ & $  47.03/53   $\\
$ 1.27  \pm 0.38 $ & $ 2.62 \pm 0.19  $ & $ 2.67 \pm 0.05 $ & $ 0.41 \pm 0.15 $ & $ 14.39 \pm 0.58  $ & $1.73 \pm 0.04 $ & $ 0.86 \pm 0.28 $ & $  47.56/53   $\\
$ 1.24  \pm 0.37 $ & $ 2.70 \pm 0.19  $ & $ 2.63 \pm 0.05 $ & $ 0.41 \pm 0.15 $ & $ 14.65 \pm 0.59  $ & $1.70 \pm 0.04 $ & $ 0.83 \pm 0.27 $ & $  36.28/53   $\\
$ 1.22  \pm 0.37 $ & $ 2.89 \pm 0.18  $ & $ 2.58 \pm 0.04 $ & $ 0.44 \pm 0.15 $ & $ 15.27 \pm 0.63  $ & $1.65 \pm 0.03 $ & $ 0.78 \pm 0.26 $ & $  46.56/53   $\\
$ 1.27  \pm 0.38 $ & $ 2.63 \pm 0.21  $ & $ 2.65 \pm 0.06 $ & $ 0.40 \pm 0.15 $ & $ 14.39 \pm 0.61  $ & $1.73 \pm 0.05 $ & $ 0.86 \pm 0.29 $ & $  47.55/53   $\\
$ 0.97  \pm 0.29 $ & $ 2.49 \pm 0.2   $ & $ 2.58 \pm 0.05 $ & $ 0.32 \pm 0.12 $ & $ 13.36 \pm 0.60  $ & $1.68 \pm 0.05 $ & $ 0.65 \pm 0.21 $ & $  36.54/53   $\\
$ 1.24  \pm 0.37 $ & $ 2.75 \pm 0.19  $ & $ 2.61 \pm 0.05 $ & $ 0.41 \pm 0.15 $ & $ 14.89 \pm 0.62  $ & $1.69 \pm 0.04 $ & $ 0.82 \pm 0.27 $ & $  39.24/53   $\\
$ 1.22  \pm 0.37 $ & $ 2.90 \pm 0.18  $ & $ 2.57 \pm 0.04 $ & $ 0.44 \pm 0.15 $ & $ 15.47 \pm 0.65  $ & $1.64 \pm 0.04 $ & $ 0.78 \pm 0.25 $ & $  34.52/53   $\\

\enddata
\end{deluxetable}

\clearpage
\clearpage

\clearpage

\clearpage

\clearpage

\clearpage

\clearpage

\clearpage

\clearpage

\begin{figure}


\includegraphics[width=3.25cm,angle=270]{std2f_grp_nth_hb1_B.ps}
\includegraphics[width=3.25cm,angle=270]{std2f_grp_nth_hb2_B.ps}
\includegraphics[width=3.25cm,angle=270]{std2f_grp_nth_hb3_B.ps}
\includegraphics[width=3.25cm,angle=270]{std2f_grp_nth_hb4_B.ps}
\includegraphics[width=3.25cm,angle=270]{std2f_grp_nth_hb5_B.ps}
\includegraphics[width=3.25cm,angle=270]{std2f_grp_nth_hb6_B.ps}
\includegraphics[width=3.25cm,angle=270]{std2f_grp_nth_ver_B.ps}

\caption{The unfolded spectral of upper HB  in Interval B (Boxes 1-7) fitted by Model II.    \label{fig6-4}}
\end{figure}
\clearpage

\begin{figure}
\epsscale{.8}
\plotone{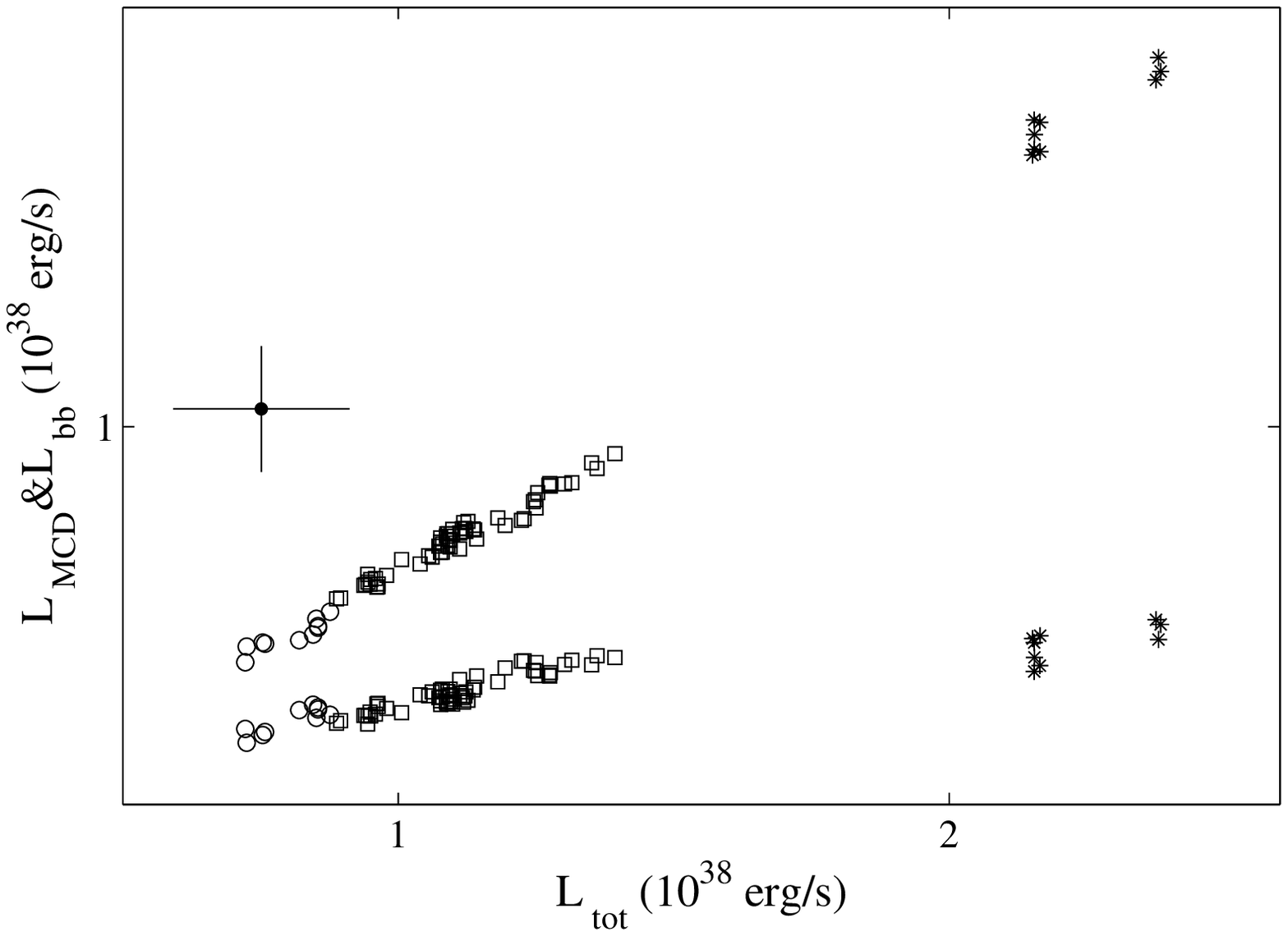}
\plotone{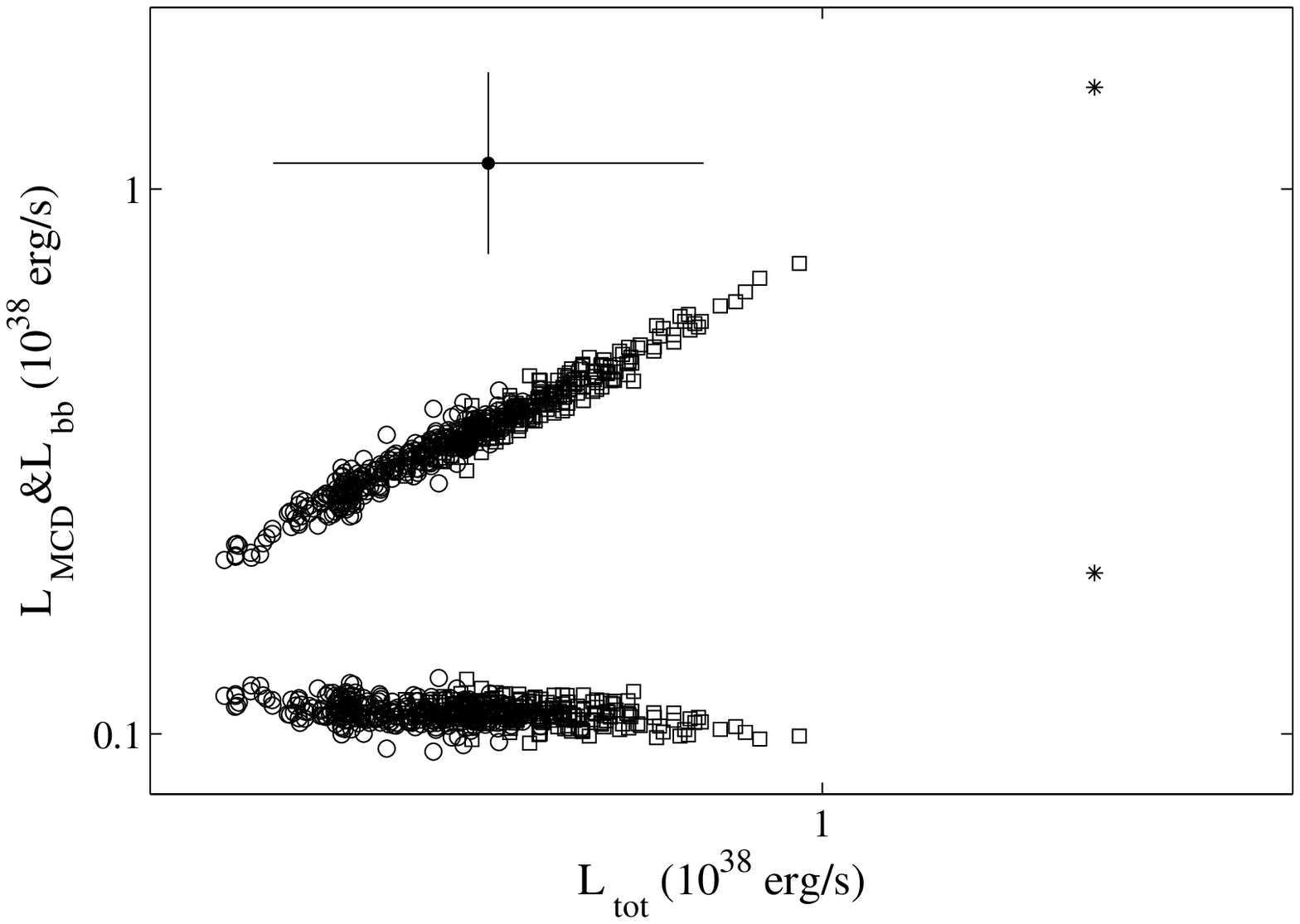}
\caption{The black body and multi-color disk component as a function of total luminosity of the HB/NB vertex (top panel) and the NB/FB vertex (bottom panel). The multi-color disk component has larger luminosity than the black body component which emit from the NS surface. The black dot with error bars are the typical uncertainties of luminosity.  \label{fig7}}
\end{figure}
\clearpage

\begin{figure}
\epsscale{.8}
\plotone{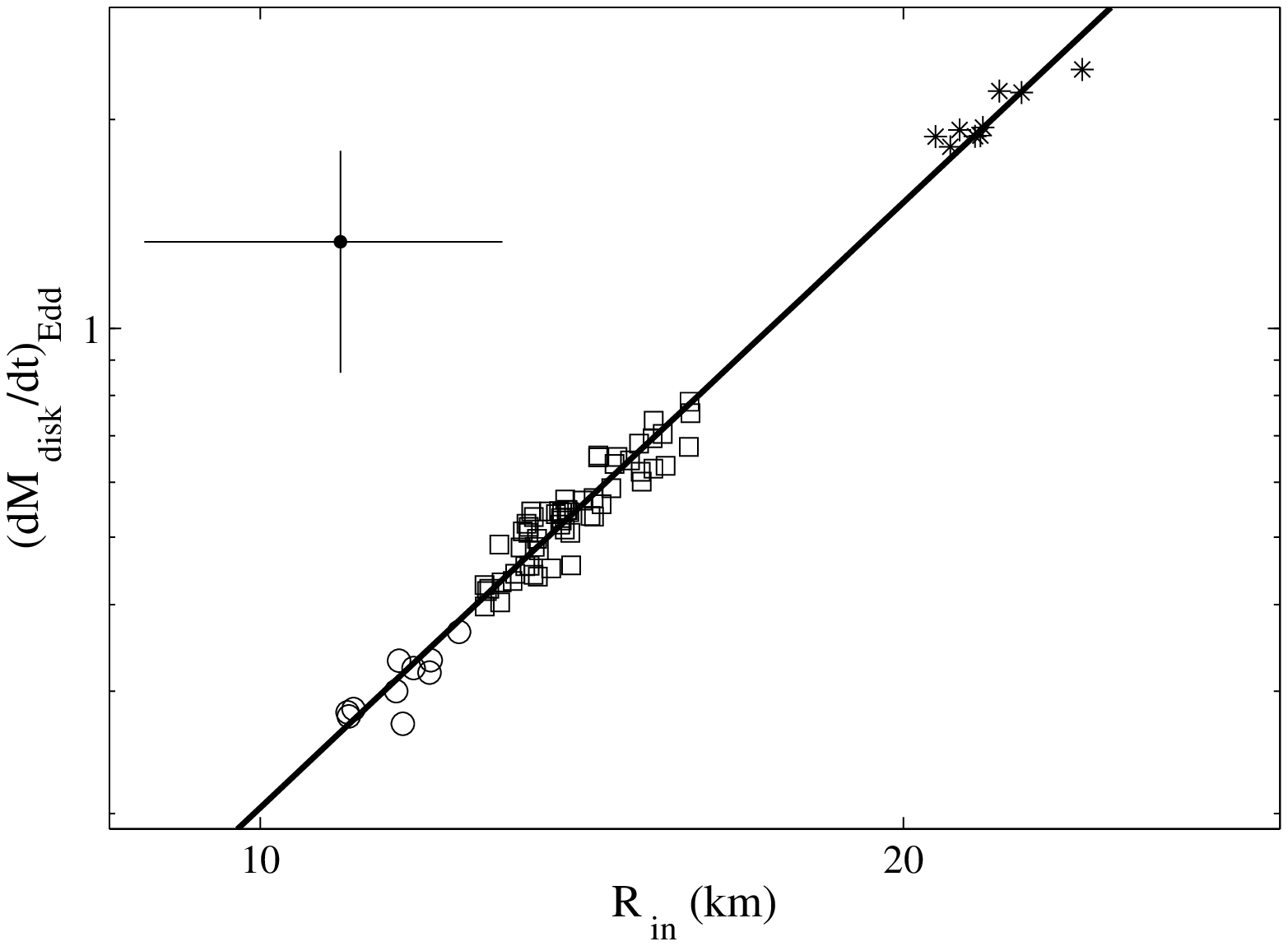}
\plotone{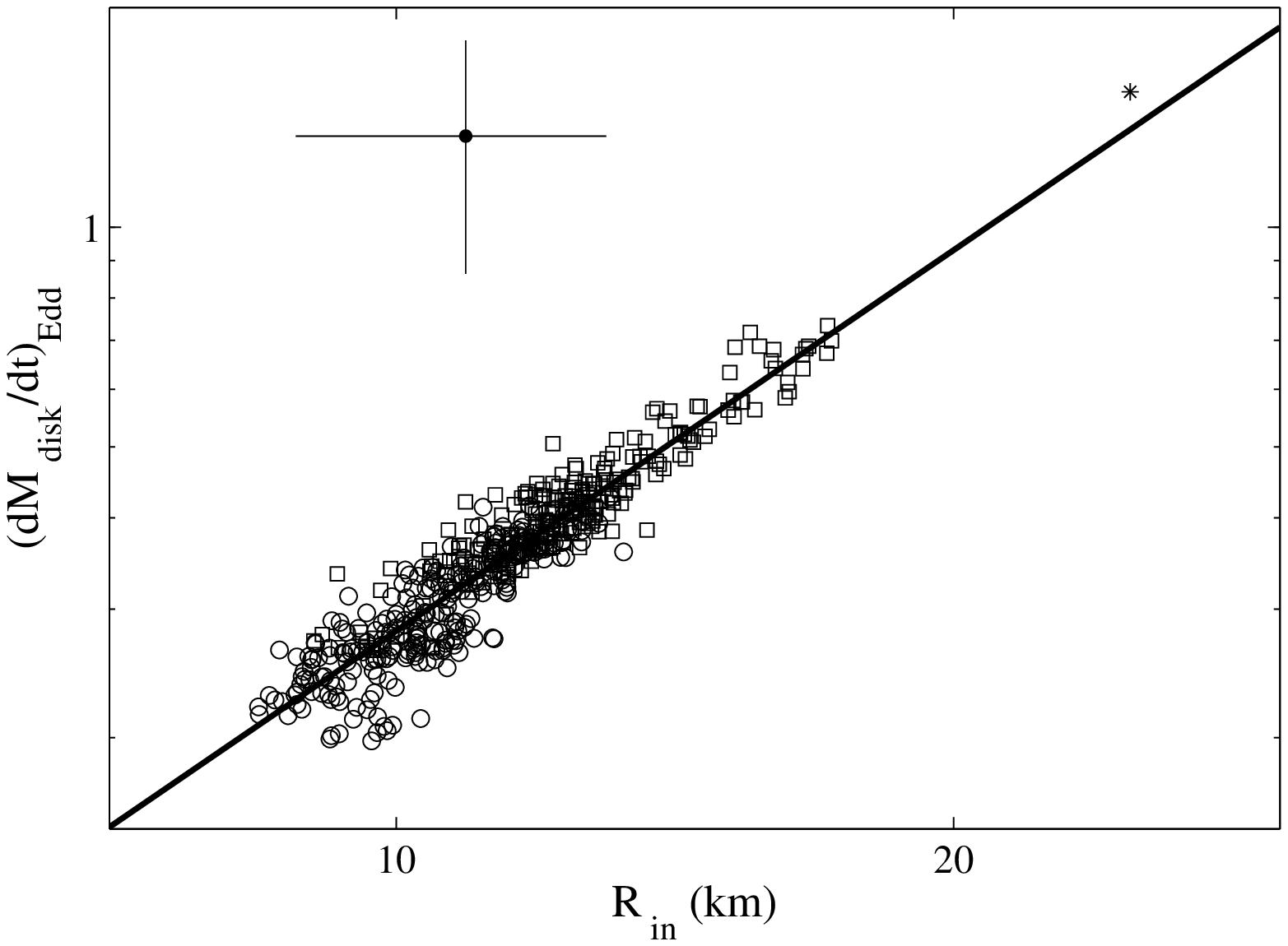}
\caption{Disk accretion rate versus inner disk radius of the HB/NB vertex (\emph{top panel}) and the NB/FB vertex (\emph{bottom panel}). The disk accretion rate is computed from $\dot{M}_{\rm disk}=L_{\rm MCD}R_{\rm in}/2GM$, where the $1.4M_{\odot}$ of NS mass and $3.48\times10^{-8}~ \rm M_{\odot}yr^{-1}$ of Eddinton luminosity are assumed.  The black dot with error bars are the typical uncertainties of disk accretion rate as well as inner disk radius. \label{fig8}}
\end{figure}
\clearpage

\begin{figure}
\epsscale{.8}
\plotone{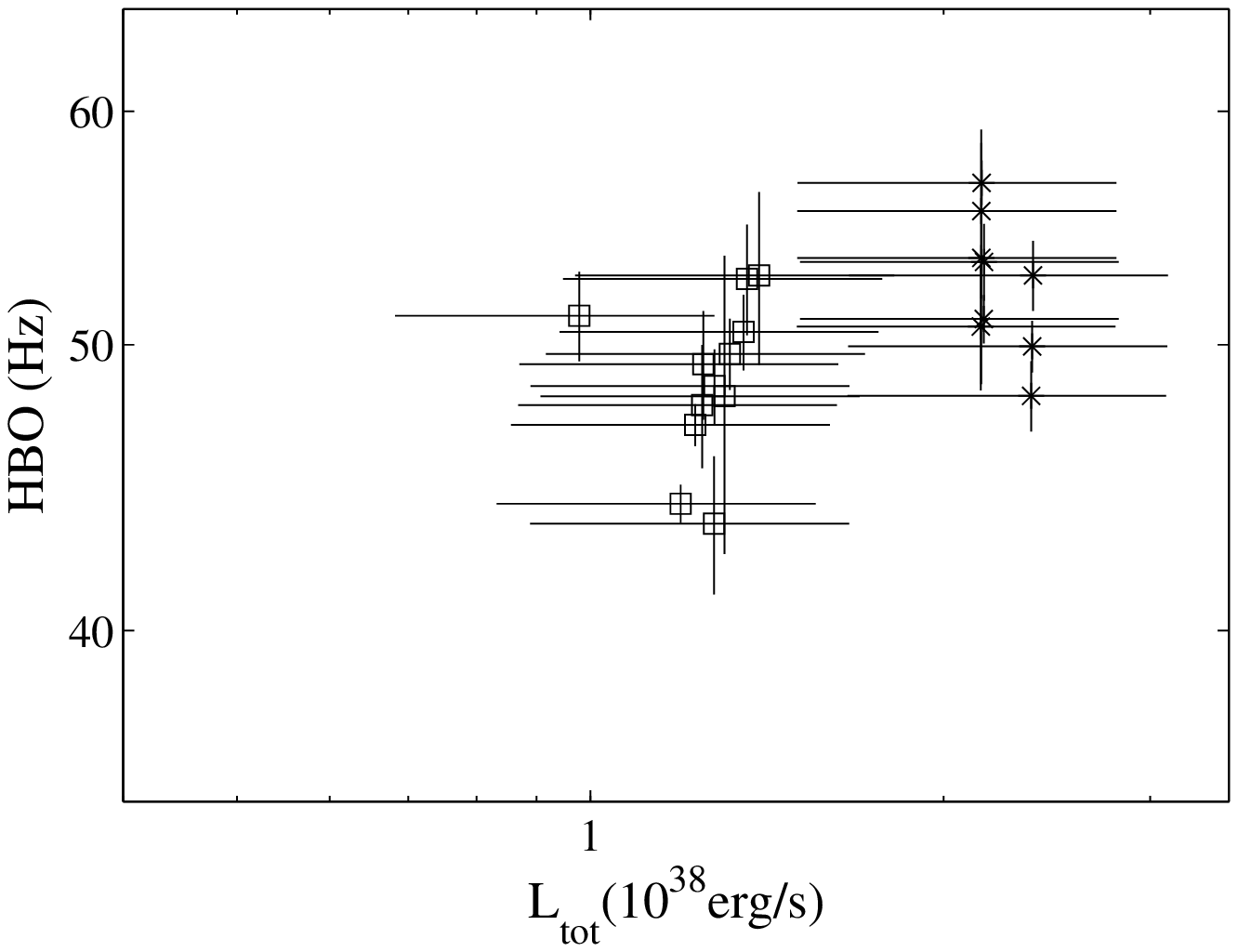}
\plotone{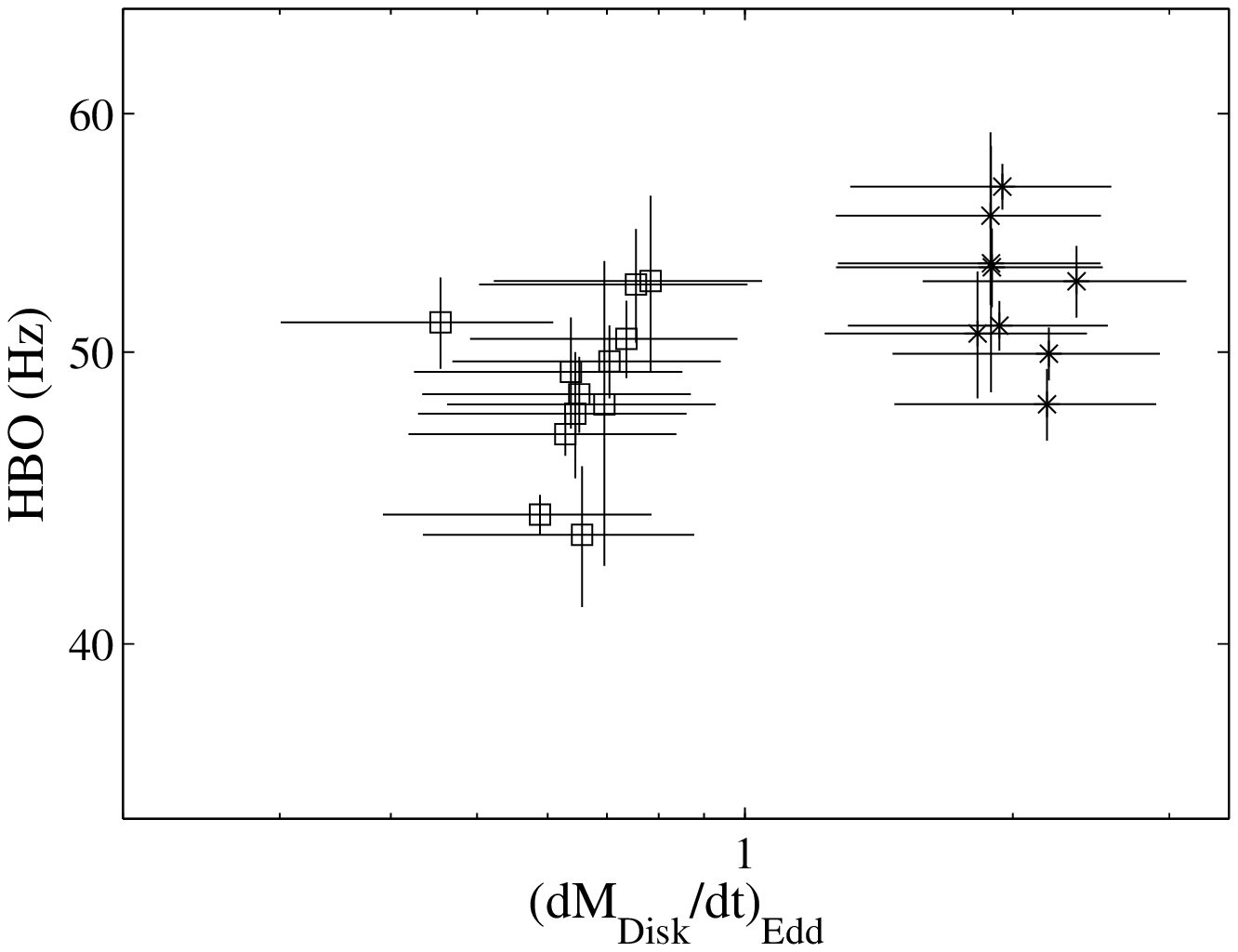}
\caption{The accretion rate dependence of HBO. Top panel, the relation between the $\nu_{\rm{HBO}}$ and the total luminosity of the HB/NB vertex. Bottom panel, the relation between the $\nu_{\rm{HBO}}$ and the disk accretion rate of the HB/NB vertex. Because of lower count rates in Stage III, the HBOs only from Stage I and II are observed. The symbols represent the same as Figure~\ref{fig4}. \label{fig9}}
\end{figure}
\clearpage

\begin{figure}
\epsscale{.8}
\plotone{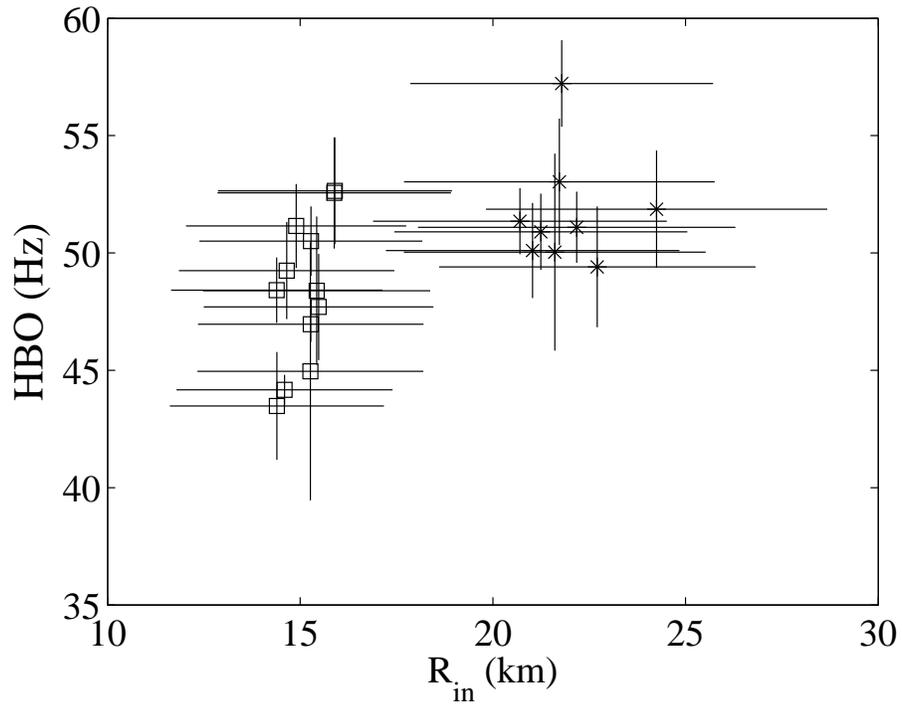}
\caption{The relation between the $\nu_{\rm{HBO}}$ and the inner disk radius of the HB/NB vertex, where the inner disk radius is obtained from the model \textbf{wabs(bbodyrad+diskbb+gaussian)}. The symbols represent the same as Figure~\ref{fig4}. \label{fig10}}
\end{figure}
\clearpage
%

%
%

\begin{figure}
\epsscale{1}
\plotone{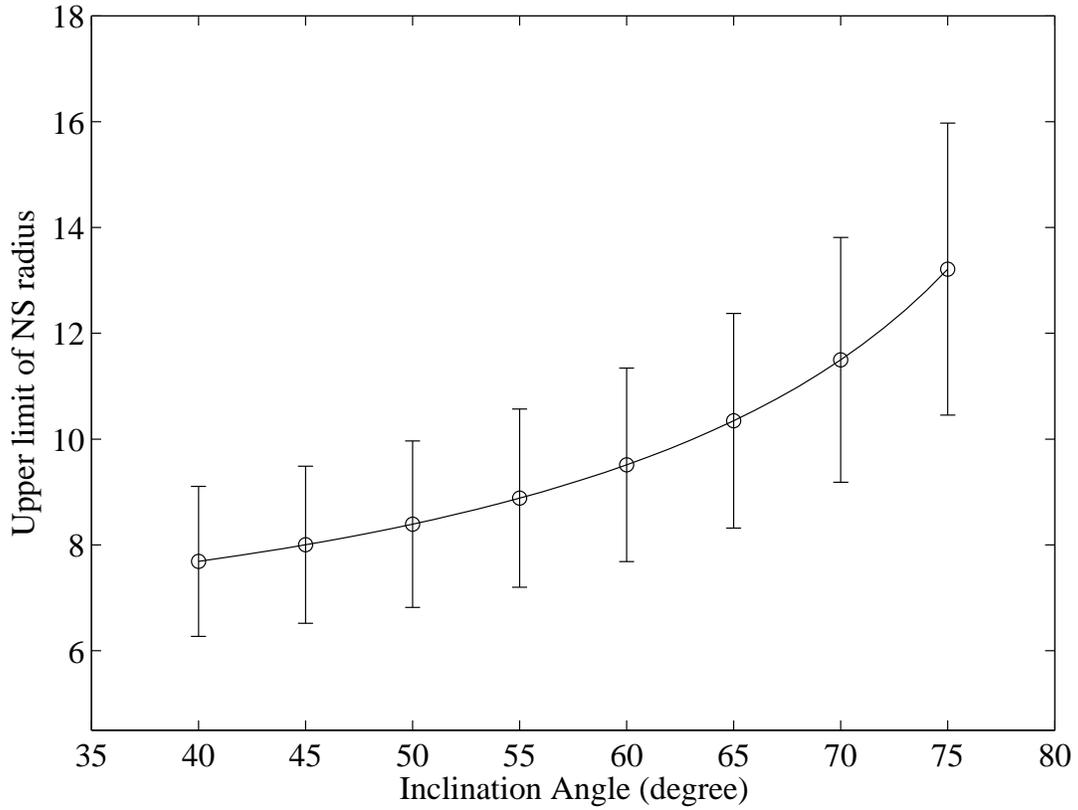}
\caption{The upper limit radius of NS versus inclination angle. The radius is derived from Box 1 in Interval A by Model I. If the inclination angle is set as $70^\circ$, the upper limit radius of NS is $11.5\pm 2.3 ~\rm km$.  \label{fig11}}
\end{figure}
\clearpage

\begin{figure}
\epsscale{.6}
\epsscale{.6}
\plotone{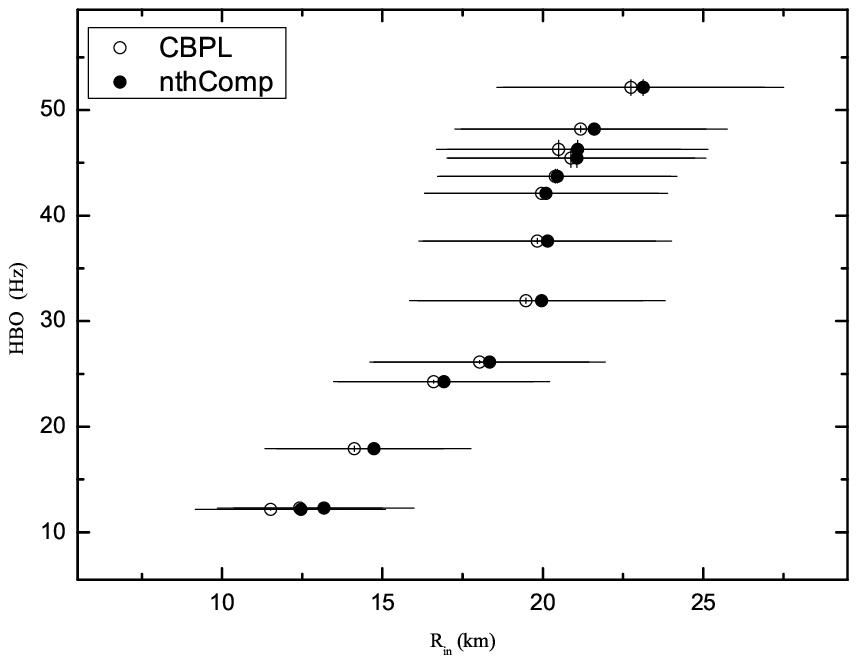}
\plotone{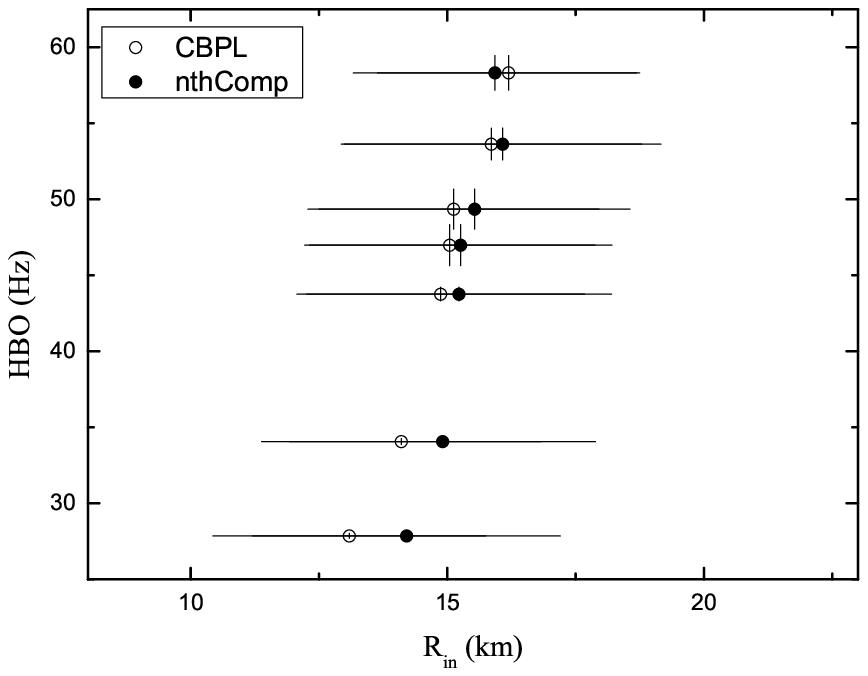}
\caption{The  $\nu_{\rm{HBO}}-R_{\rm{in}}$ relation of Interval A (\emph{top panel}) and Interval B (\emph{bottom panel}). The inner disk radii are derived from Model I (black dots) and Model II (open circles), respectively. \label{fig12}}
\end{figure}
\clearpage

\begin{figure}
\epsscale{.8}
\plotone{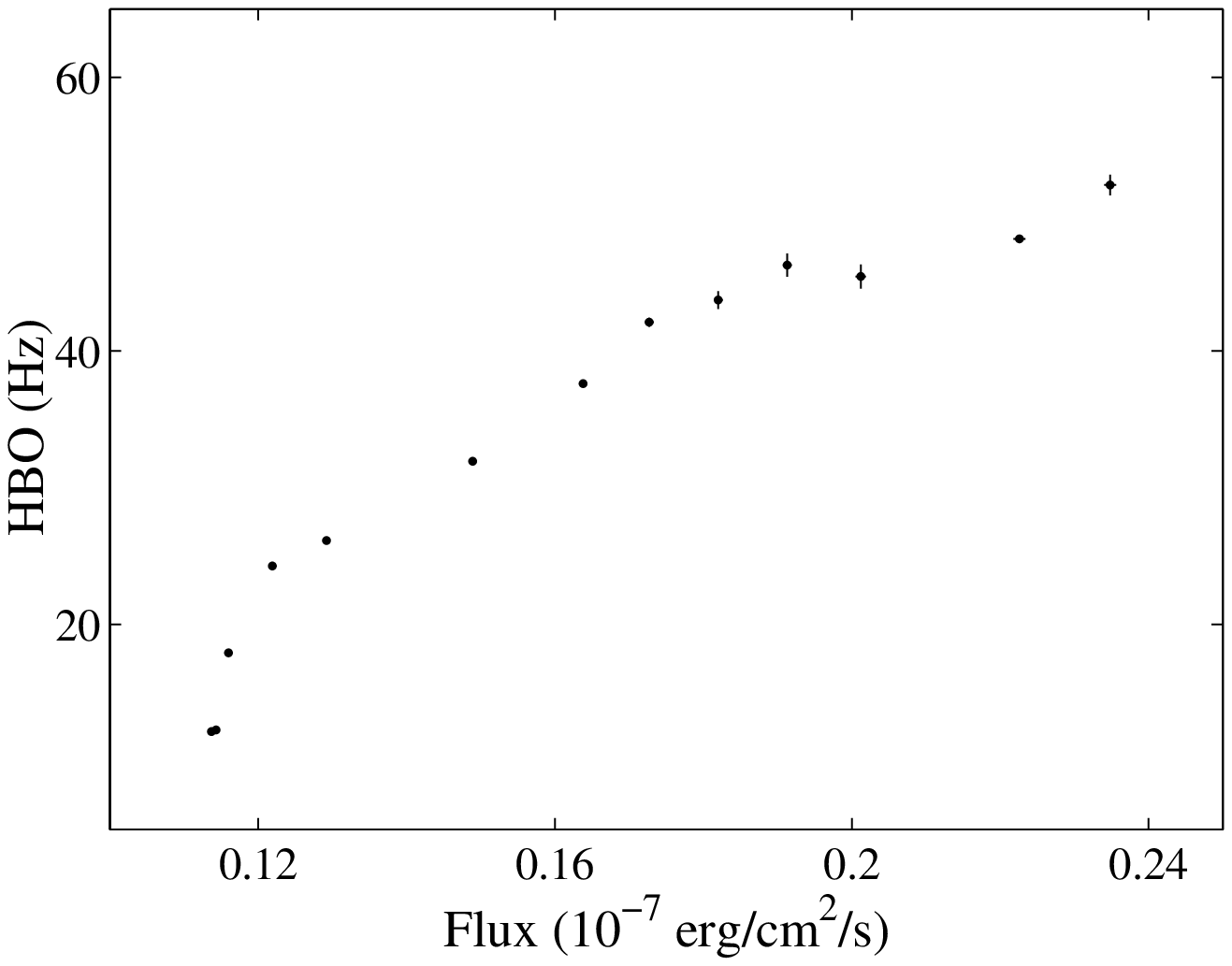}

\plotone{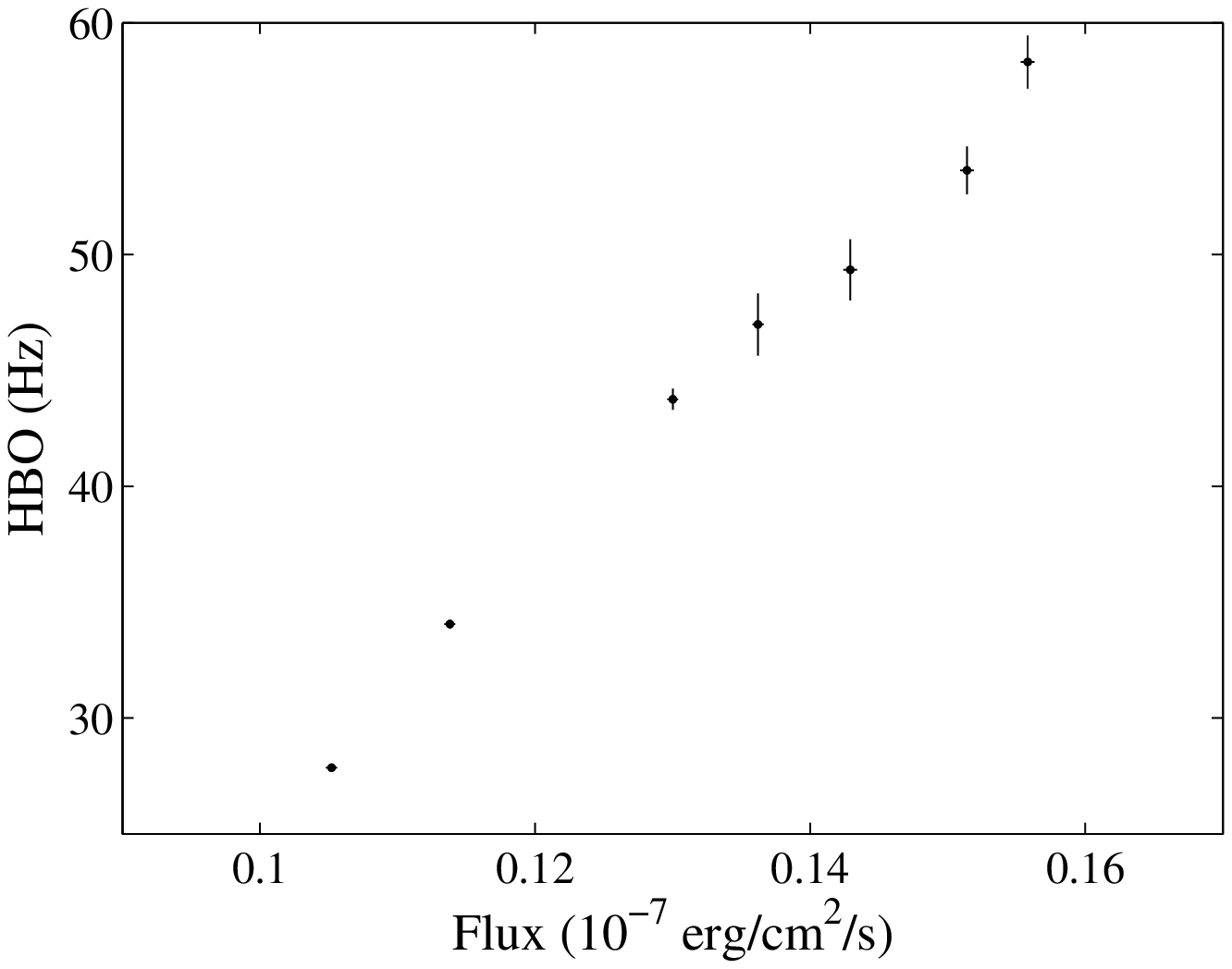}
\caption{The frequency of HBO versus the total flux  of Intervals A (\emph{top panel}) and B (\emph{bottom panel}). The total flux is integrated over the energy range $\sim 3-60 \rm keV$ in units of $10^{-7}~ \rm erg/cm^2/s$.  \label{fig13}}
\end{figure}
\clearpage

\begin{figure}
\epsscale{0.8}
\plotone{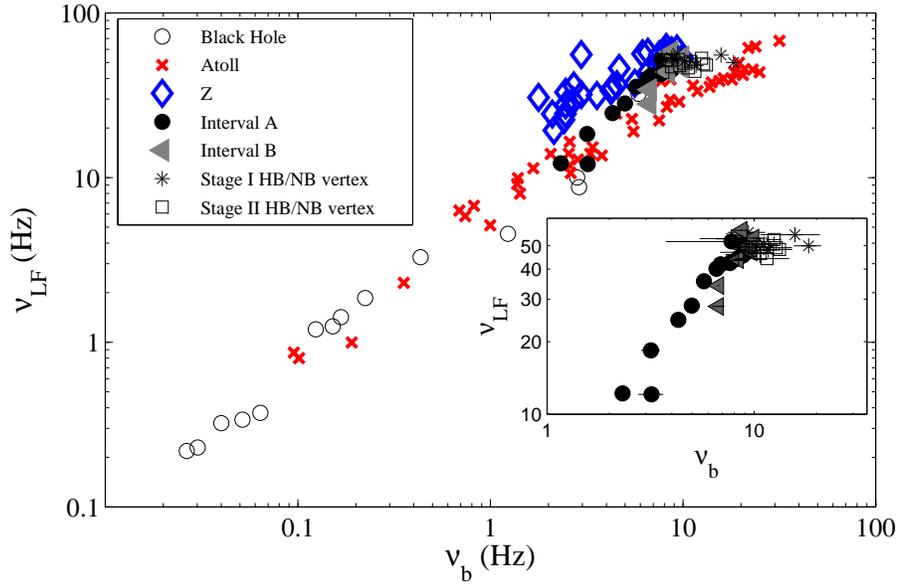}
\caption{The W-K relation of LMXBs. The circle, cross and diamond represent the BHs, the atoll sources and the Z sources from \citet{wij99}, respectively. The black-filled circle, the  grey-filled triangle,  the black star and the open square display the characteristic frequencies of XTE J1701-462 in Interval A, Interval B, Stage I HB/NB vertex, Stage II HB/NB vertex, respectively. The data with error bars are plotted in the insert panel. \label{fig14}}
\end{figure}
\clearpage

\begin{figure}
\epsscale{.8}
\plottwo{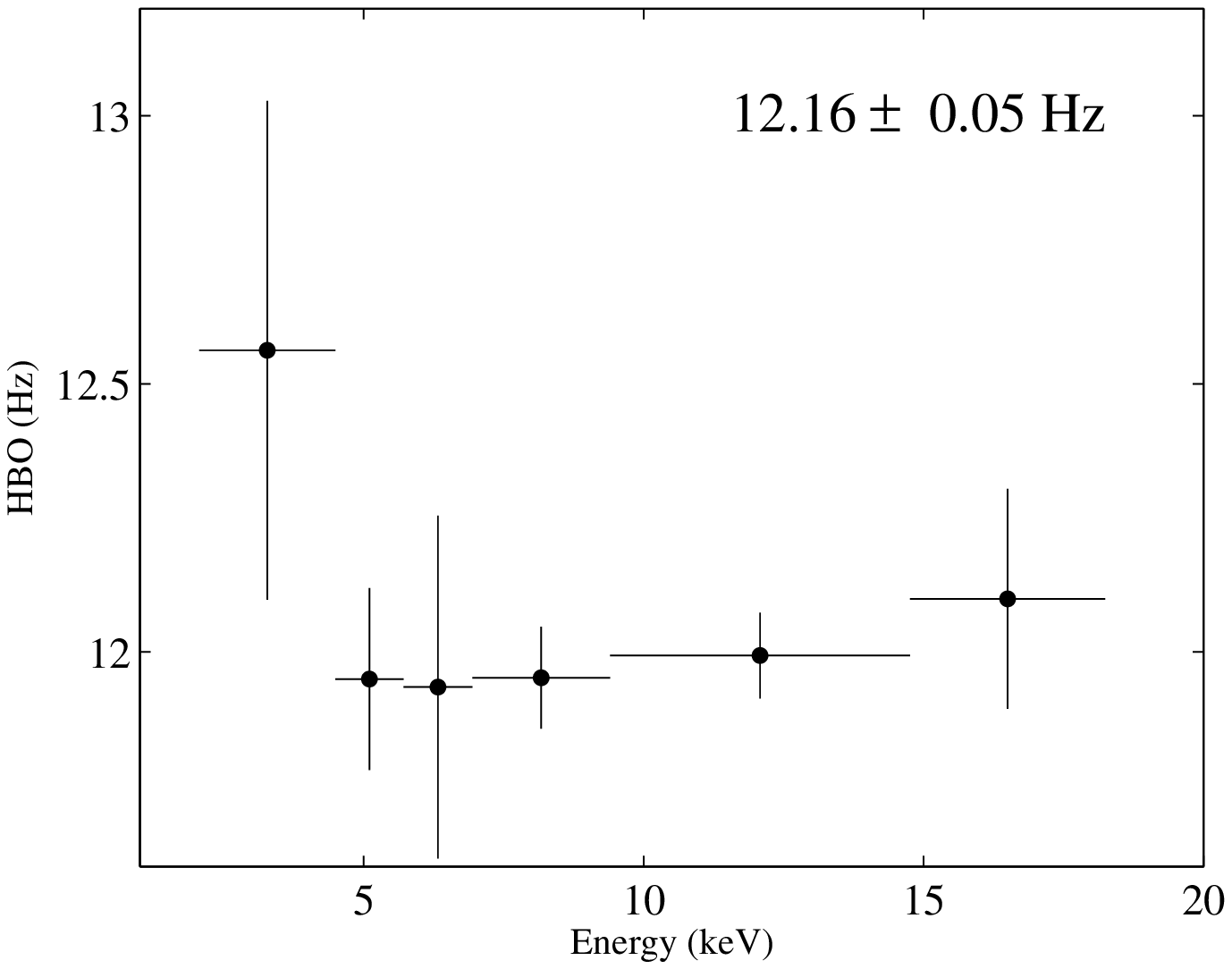}{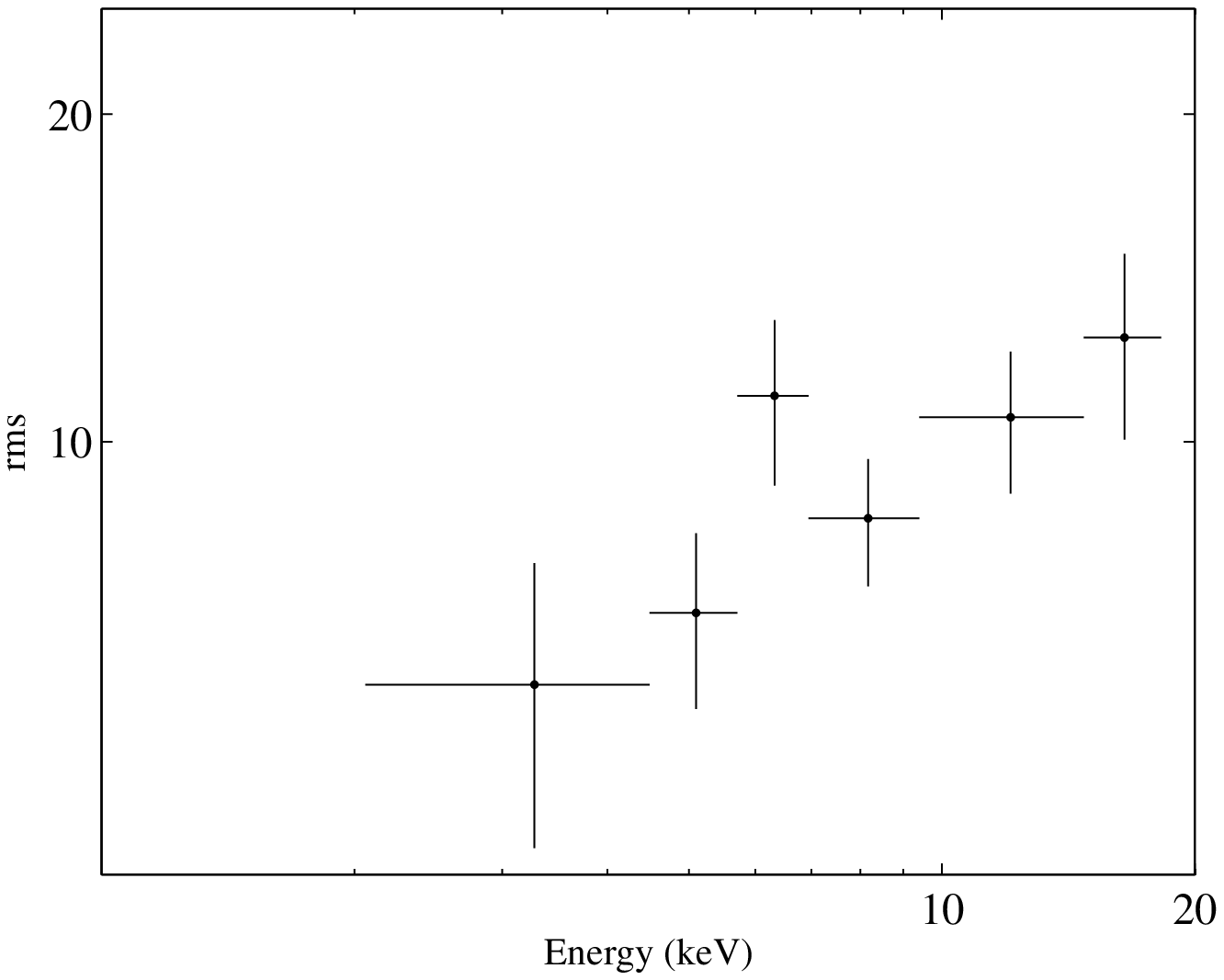}

\plottwo{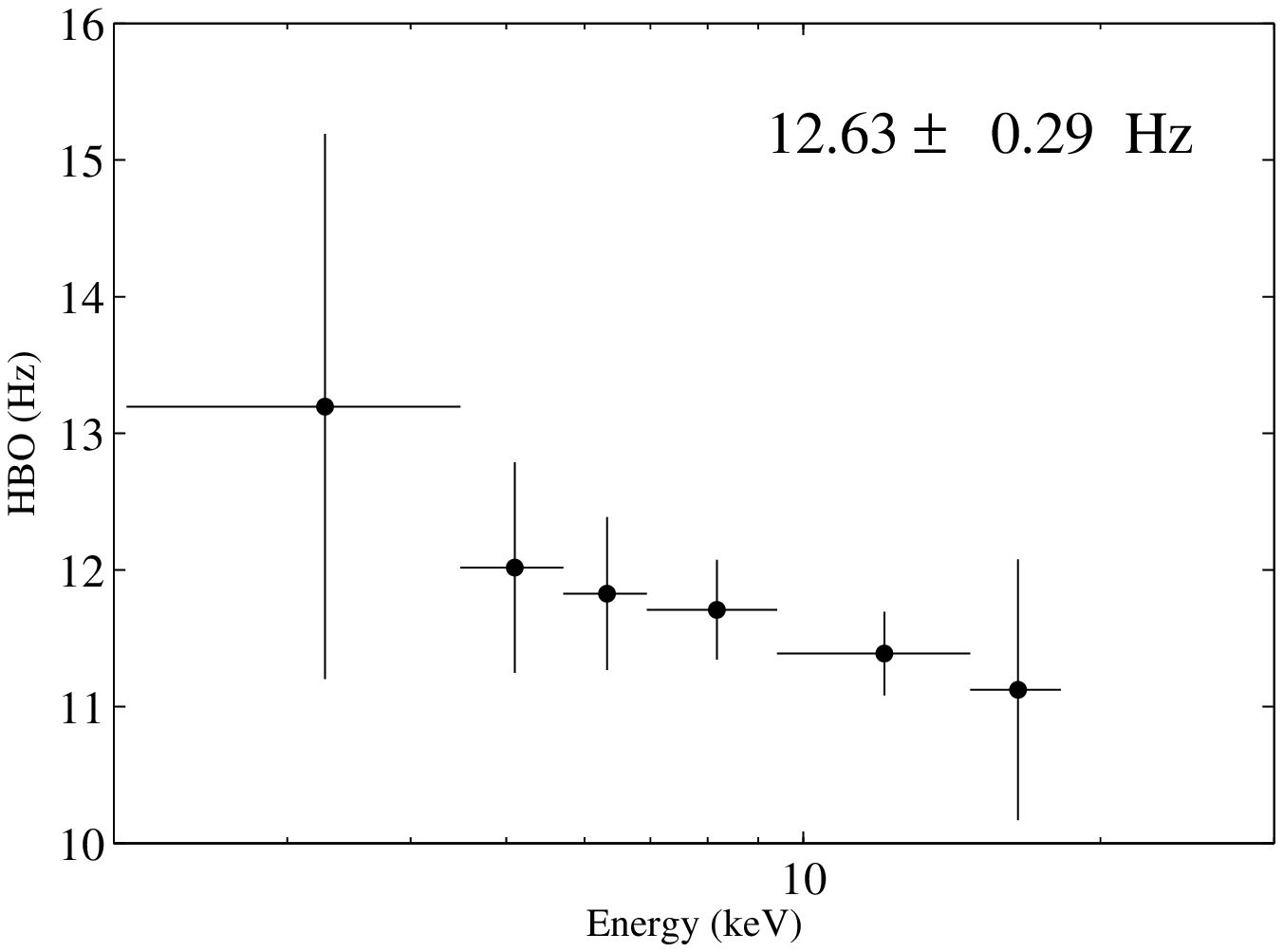}{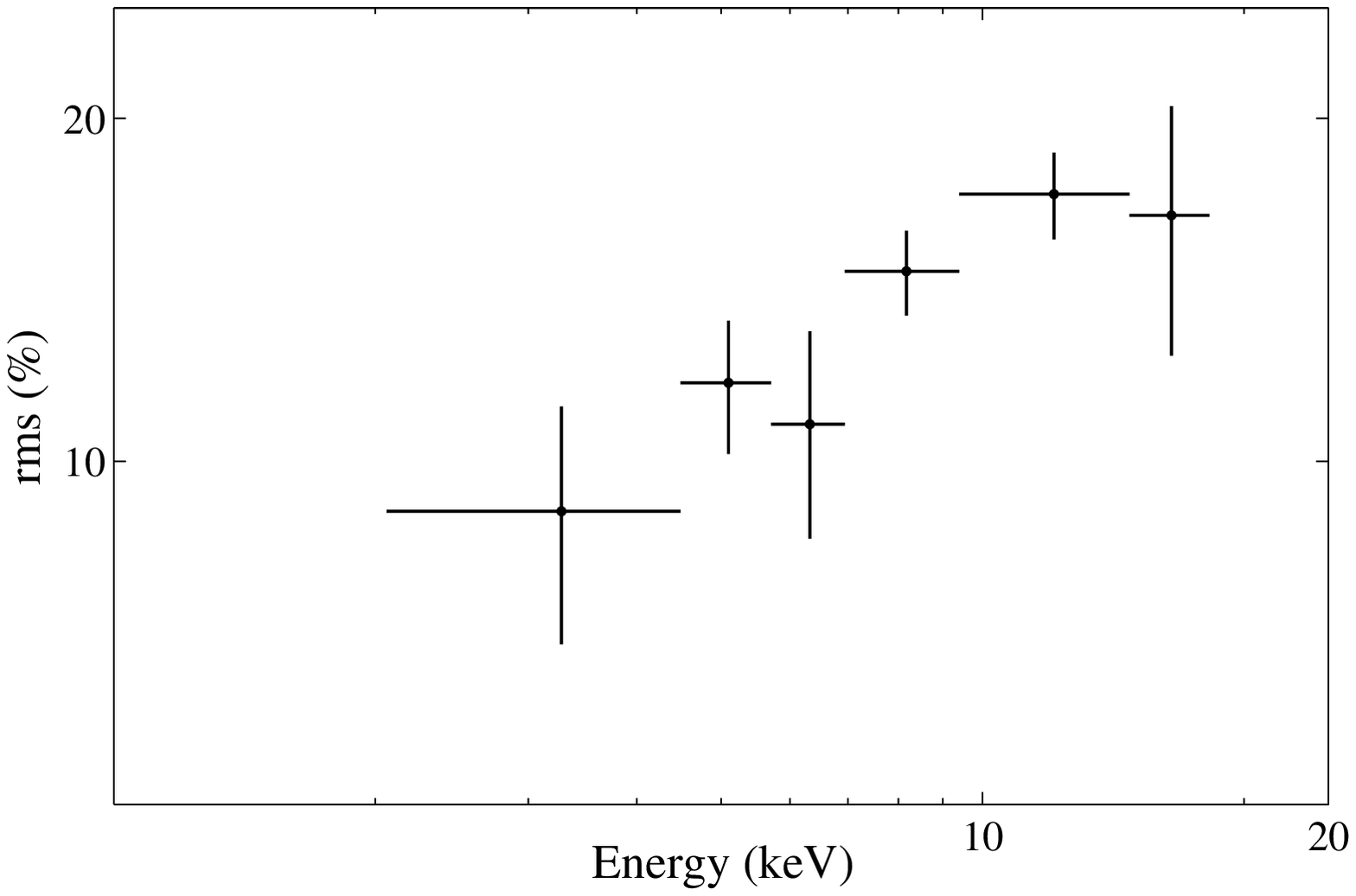}

\plottwo{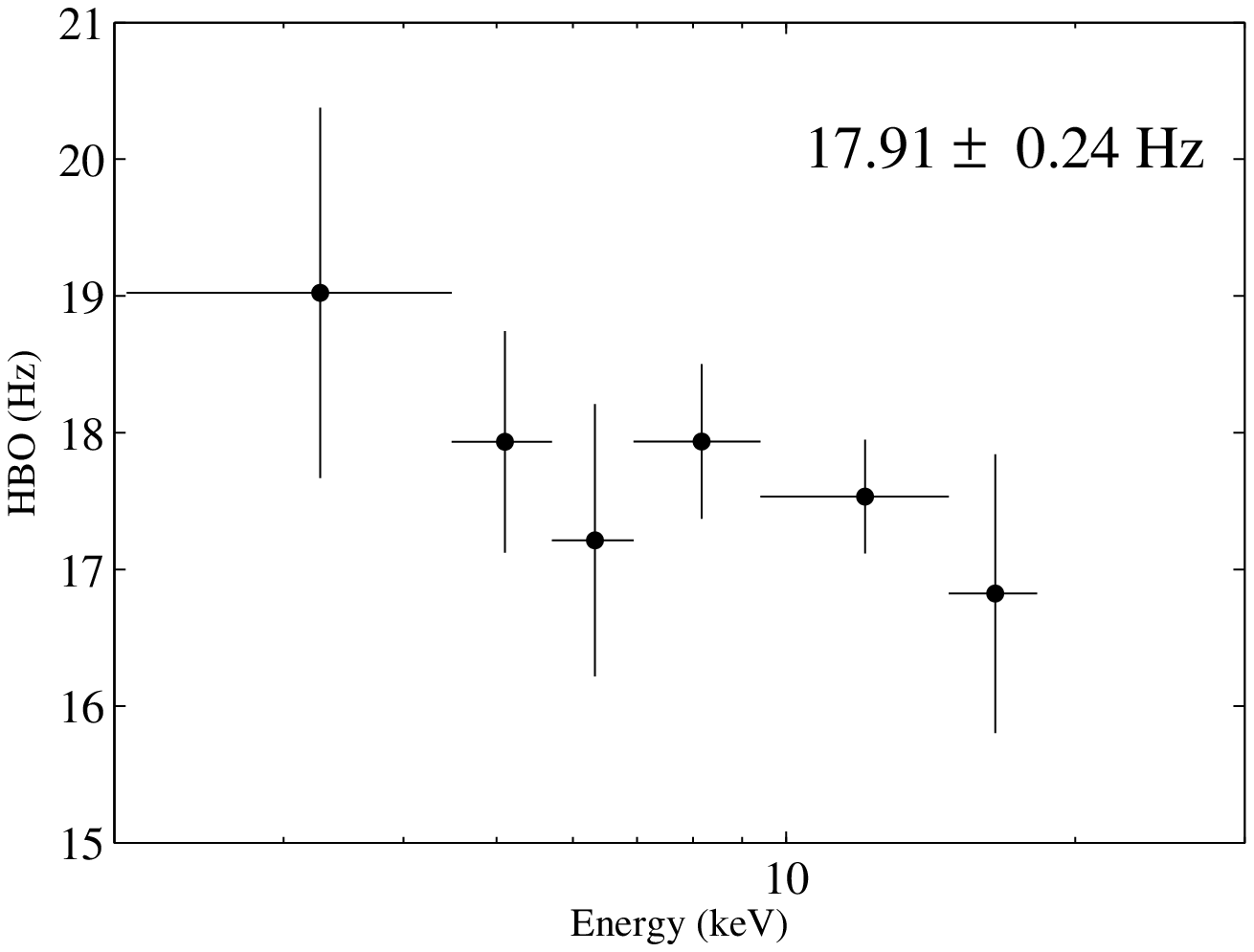}{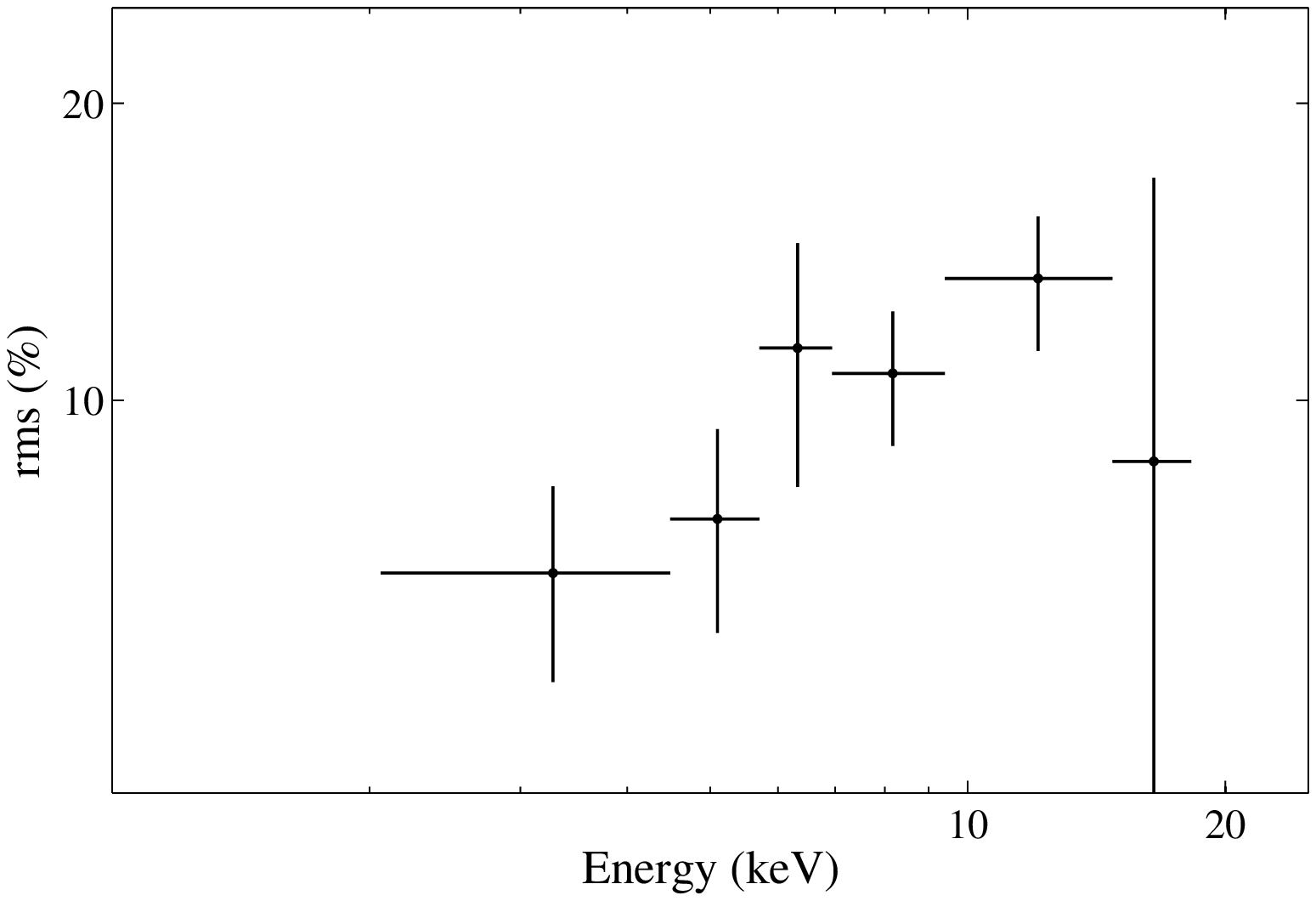}

\plottwo{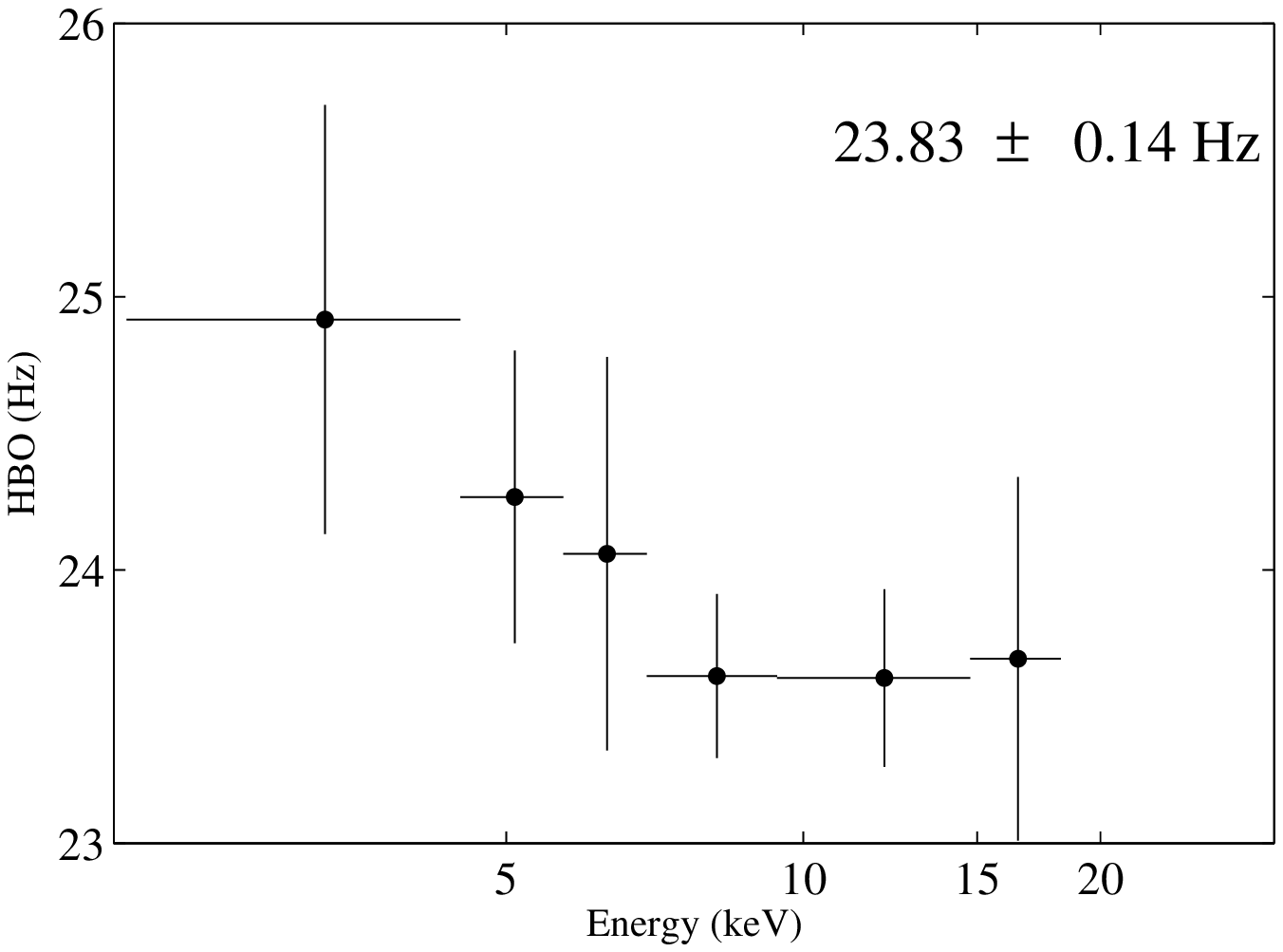}{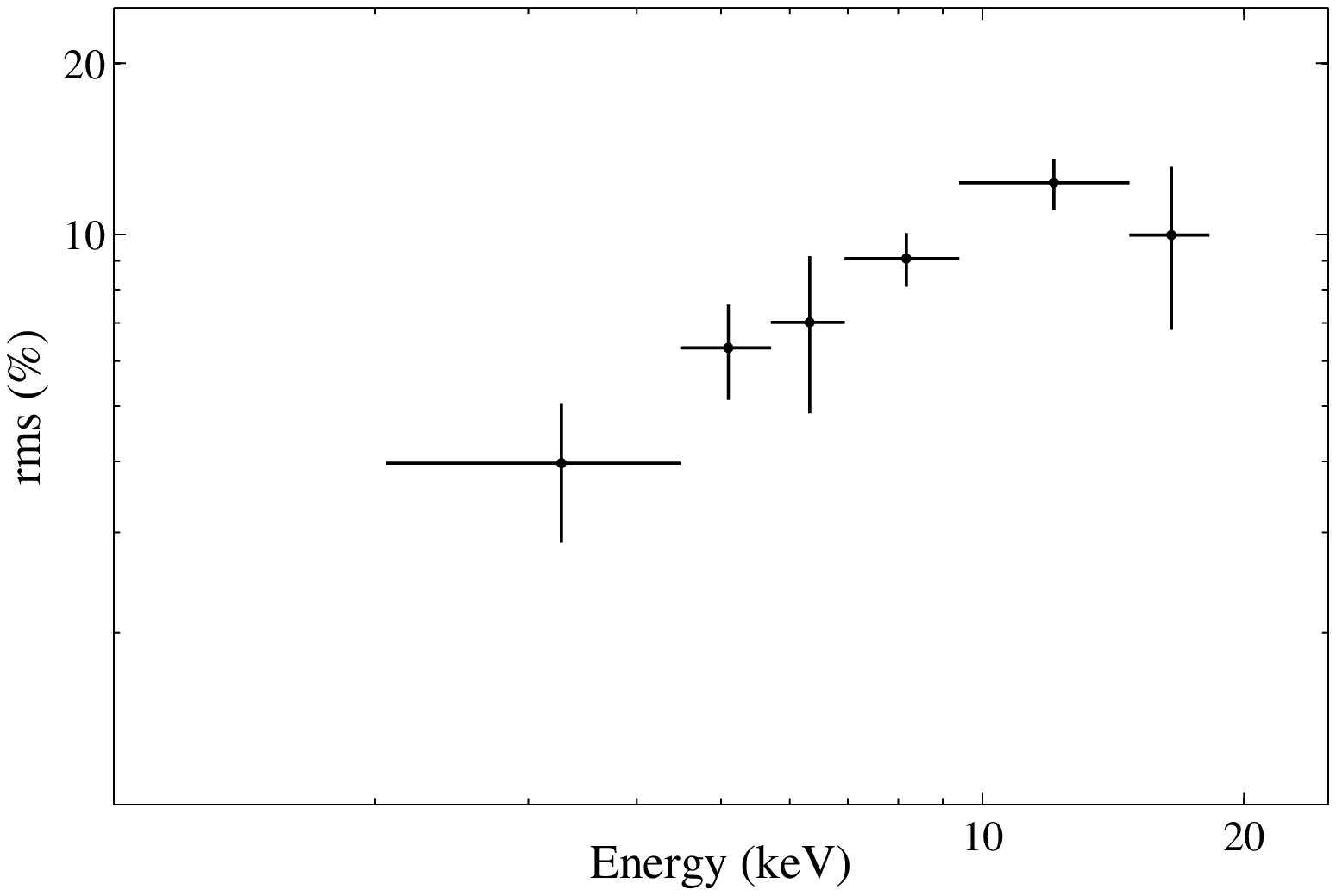}

\plottwo{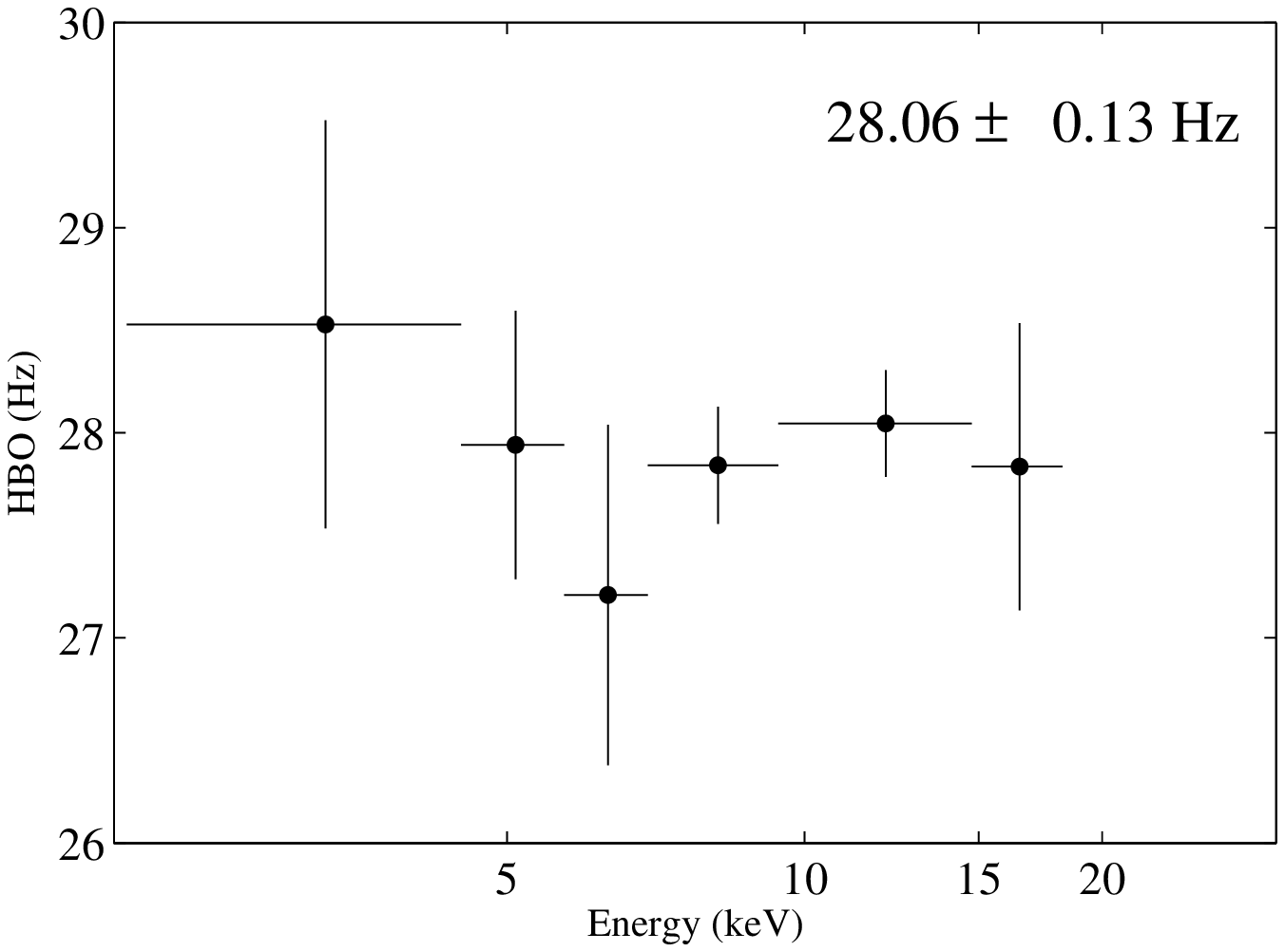}{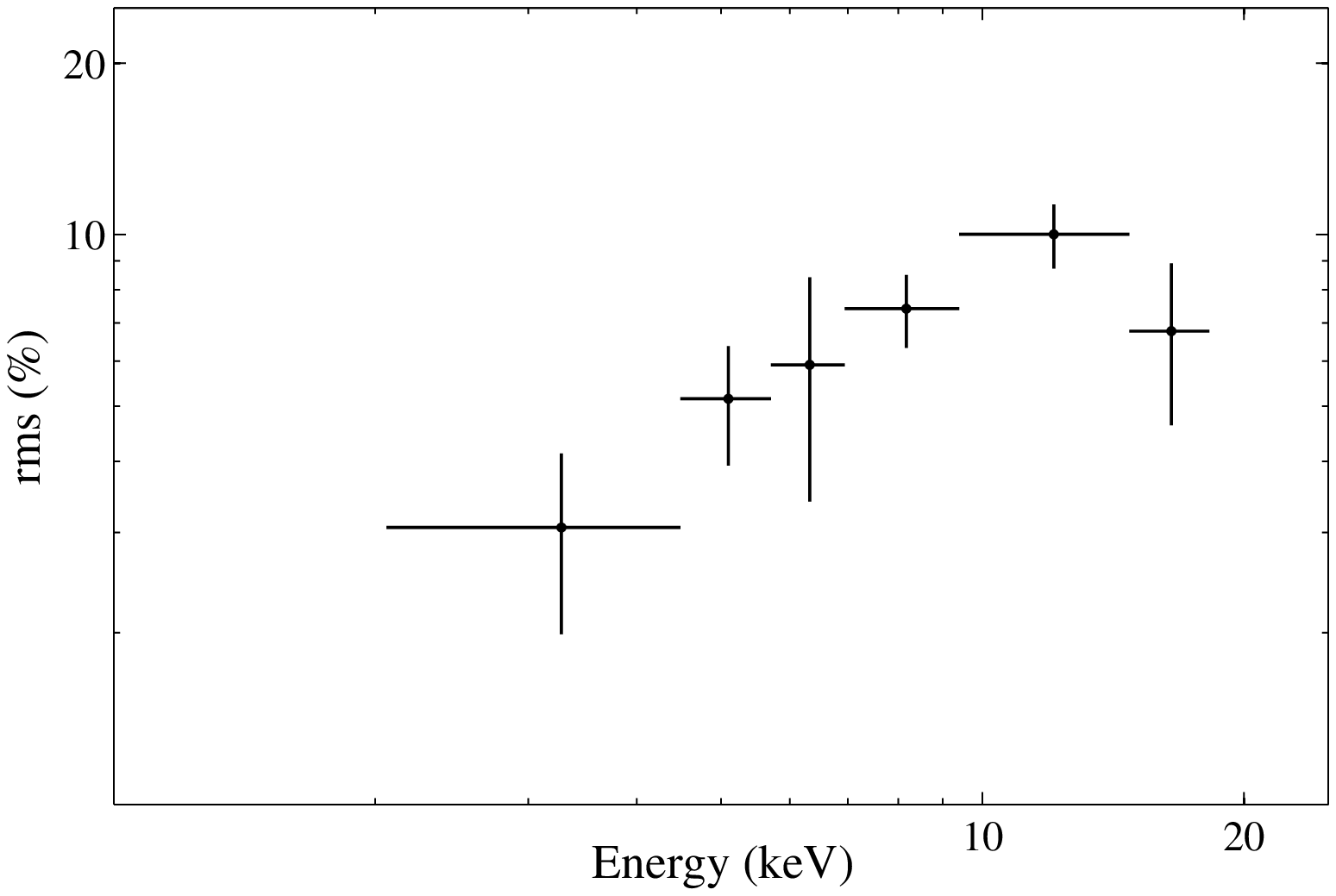}
\caption{Left panels: the energy dependence of HBO in Interval A (Boxes 1-5). Right panels: the corresponding rms of HBO versus photon energy. Left panels offer the full band centroid frequencies of HBO and errors.  \label{fig15}}
\end{figure}

\end{document}